\documentclass[iop,apj]{emulateapj}
\usepackage{amsmath}

\newcommand\msol{{\cal M_{\odot}}}
\newcommand\teff{{T_{\rm eff}}}

\newcommand\pab{$\langle P_\mathrm{ab}\rangle$}
\newcommand\pc{$\langle P_\mathrm{c}\rangle$}

\newcommand\lta{\mathrel{\hbox{\raise 0.6 ex \hbox{$<$}\kern
                   -1.8 ex\lower .5 ex\hbox{$\sim$}}}}
\newcommand\gta{\mathrel{\hbox{\raise 0.6 ex \hbox{$>$}\kern
                   -1.7 ex\lower .5 ex\hbox{$\sim$}}}}
\newcommand{\bea}{\begin{eqnarray}}
\newcommand{\eea}{\end{eqnarray}}

\newcommand\tstrut{\rule{0pt}{2.4ex}}

\usepackage{natbib}
\usepackage{hyperref}

\begin{document}

\epsscale{1.1}

\defcitealias{vbl13}{VBLC13}

\title{CONSTRAINTS ON THE DISTANCE MODULI, HELIUM AND METAL ABUNDANCES, AND
AGES OF GLOBULAR CLUSTERS FROM THEIR RR LYRAE AND NON-VARIABLE HORIZONTAL-BRANCH
STARS.~II.~Multiple Stellar Populations in 47\,Tuc, M\,3, and M\,13}

\shortauthors{Denissenkov et al.}
\shorttitle{HB Stars in 47\,Tuc, M\,3, and M\,13}

\author{
Pavel~A.~Denissenkov\altaffilmark{1,2}, 
Don A.~VandenBerg\altaffilmark{1}, 
Grzegorz~Kopacki\altaffilmark{3},
and
Jason~W.~Ferguson\altaffilmark{4}
}

\altaffiltext{1}{Department of Physics \& Astronomy, University of Victoria, 
P.O.~Box 1700, STN CSC, Victoria, BC, V8W 2Y2, Canada; pavelden@uvic.ca,
vandenbe@uvic.ca}

\altaffiltext{2}{Joint Institute for Nuclear Astrophysics, Center for the Evolution of the Elements, Michigan
       State University, 640 South Shaw Lane, East Lansing, MI 48824, USA}

\altaffiltext{3}{Astronomical~Institute, University~of~Wroc{\l}aw,
ul.~Koperni-\hfil\break
ka 11, 51-622~Wroc{\l}aw, Poland; kopacki@astro.uni.wroc.pl}

\altaffiltext{4}{Department of Physics, Wichita State University, Wichita, KS 67260-0032, USA; jason.ferguson@wichita.edu}

\submitted{Submitted to The Astrophysical Journal}

\begin{abstract}
We present a new set of horizontal-branch (HB) models computed with the
MESA stellar evolution code. The models
adopt $\alpha$-enhanced \cite{ags09} metals mixtures
and include the gravitational settling of He. They are
used in our HB population synthesis tool to generate theoretical distributions of
HB stars in order to describe the multiple stellar populations in the globular
clusters 47\,Tuc, M\,3, and M\,13.  The observed HB in 47\,Tuc is
reproduced very well by our simulations for [Fe/H] $= -0.70$ and [$\alpha$/Fe]
$= +0.4$ if the initial helium mass fraction varies by $\Delta Y_0
\sim 0.03$ and approximately 21\%, 37\%, and 42\% of the stars have $Y_0 = 0.257$,
0.270, and 0.287, respectively.  These simulations yield $(m-M)_V = 13.27$,
implying an age near 13.0 Gyr.  In the case of M\,3 and M\,13, our synthetic HBs
for [Fe/H] $= -1.55$ and [$\alpha$/Fe] $= 0.4$ match the observed ones quite well
if M\,3 has $\Delta Y_0 \sim 0.01$ and $(m-M)_V =  15.02$, resulting in an age of
12.6 Gyr, whereas M\,13 has $\Delta Y_0 \sim 0.08$ and $(m-M)_V = 14.42$,
implying an age of 12.9 Gyr.  Mass loss during giant-branch evolution and $\Delta Y_0$
appear to be the primary second parameters for M\,3 and M\,13.
New observations for 7 of the 9 known RR Lyrae
in M\,13 are also reported. Surprisingly, periods predicted for the $c$-type variables tend to
be too high (by up to $\sim 0.1$~d).

\end{abstract} 


\section{Introduction}
\label{sec:intro}

Based on his comparisons of the color-magnitudes diagrams (CMDs) of the
globular clusters (GCs) M\,3 and M\,92, \cite{sandage:53} concluded that there must
be cluster-to-cluster variations in some intrinsic property in order to explain
the striking differences between them.  He proposed, crediting M.~Schwarzschild
(private conversations) for the initial suggestion, that this property might be
metal abundance; and indeed, subsequent studies 
\citep[e.g.,][]{sandage:60,feast:60,arp:62} revealed that the metals-to-hydrogen
ratio is a distinguishing characteristic of GCs.  As the photometric data
improved \citep[see, e.g.][]{sandage:70}, the increasingly well-defined turnoff (TO)
luminosities and principal photometric sequences suggested that the stars in
each cluster formed with the same metallicity at the same time.  (It was known
early on that $\omega$\,Cen had an unusually broad giant branch
(\citealt{wooley:66}), but this system appeared to be an exception to the rule.)

During the same period of time, \cite{vdbergh:67} and \cite{sandage:67}
independently deduced from the diversity in the horizontal branch (HB)
morphologies of GCs with nearly the same metal abundance that at least two
parameters, one being metallicity, are needed to describe them.  Age and
helium abundance were the most obvious candidates for the (primary) ``second
parameter" \citep[e.g.,][]{rood:73}, though the abundance of nitrogen was also 
considered when \cite{hartwick:72} found that CN is much stronger in NGC\,7006
than in either M\,92 or M\,3.

The first strong evidence of the presence of multiple stellar populations in GCs
was the discovery by \cite{norris:79} of the bimodal distribution of CN
molecular band strengths in 142 red giants in 47\,Tuc.  These results were based
on measurements of the photometric color index C(4142), the same one used by
\cite{hartwick:72}.  At nearly the same time, analyses of low-resolution spectra
showed that differences in CN band strengths could be traced down to upper
main-sequence (MS) stars in 47\,Tuc \citep{hesser:78,hesser:80}, pointing to a
primordial origin of the N abundance variations. 

Soon afterwards, the first
spectroscopic data indicating strong star-to-star variations of the O, Na and
Al abundances in red giants of the GCs M 13 and NGC 6725 were presented by
\cite{pilachowski:80}, \cite{peterson:80} and \cite{norris:81}.
However, what is arguably the most important result from chemical composition
studies of GC stars during the last three decades was the discovery of correlations
between the abundances of C, N, O, Na Mg, and Al
\citep{pilachowski:88,paltoglou:89,kraft:94,norris:95,kraft:97,carretta:05,carretta:09},
with the signs and magnitudes of the abundance variations in those correlations being
indicative of their simultaneous origin in H-burning reactions of the CNO, NeNa,
and MgAl cycles at $T\gta 70\times 10^6$ K \citep{langer:97,denissenkov:98,prantzos:07}.
This conclusion is further supported by the correlations between the Mg isotopic and Al
abundances recently revealed in $\omega$\,Cen and several other clusters
\citep{dacosta:13,denissenkov:14}.  

Both the indications in support of a high-temperature
environment and the fact that most of the proton-capture element abundance anomalies
(star-to-star variations and correlations between them) were found not only along
the upper RGB in GCs, but also in other evolutionary phases of low-mass stars --- including the
MS, lower RGB, HB, \citep{briley:96,briley:02,briley:04,gratton:01,grundahl:02,gratton:11,gratton:13},
and even the asymptotic-giant branch (AGB) \citep{lapenna:16} --- pointed to massive stars as
the most probable origin of the observed abundance trends.
\citep{dantona:83,dantona:16,decressin:07,demink:09,denissenkov:14}. 
This would also explain why the CN-weak and CN-strong stars in GCs define
the same fiducial sequences in the turnoff regions of $(B-V,\,V)$ CMDs (see the
study of 47\,Tuc by \citealt{cannon:98}), which suggests that they have the same age and C$+$N$+$O
abundances.

Hydrogen burning should necessarily result in an enhancement of the He abundance and,
indeed, increased He mass-fraction abundances by $\Delta Y\approx 0.04$ and $\Delta Y\approx 0.09$
from the value of $Y_0\approx 0.25$, which is usually assumed to represent the initial value of
$Y$ in low- to intermediate-metallicity GCs \citep{salaris:04},
were directly measured by spectroscopic methods in blue HB (BHB) stars of the GCs M 4 \citep{villanova:12} and NGC 2808 \citep{marino:14}.
Indirect methods based on a comparison of the observed distribution of HB stars in a GC with that simulated using an HB population synthesis
tool confirm that HB stars in some GCs, especially in GCs with extended BHBs, have a varying He abundance
\citep{dicriscienzo:10,valcarce:16,tailo:16}, with $Y$ increasing towards the
blue tails of HBs. The latter result is consistent with observational data which indicates that
the anti-correlating deviations of the O and Na abundances from their initial values are stronger 
in the bluer HB stars \citep[e.g.,][]{gratton:13}.

Significant progress in the splitting of CMDs of GCs, including the majority of those with the homogeneous Fe abundances, into sets of two to seven well
distinguishable evolutionary sequences (MS, RGB and HB) has been made recently, when the optical photometry in the {\it HST ACS} 
({\it Hubble Space Telescope
Advanced Camera for Surveys}) filter $F435W$ was combined with the UV photometry in the {\it HST ACS} filters $F275W$ and $F336W$ \citep{piotto:15}.
These filters cover parts of stellar spectra that are strongly affected by the presence of CN+CH, NH and OH molecular bands,
respectively \citep{sbordone:11,milone:12}; therefore, the photometric magnitudes and, especially, colors derived from them turn out to be
very sensitive to the O depletion and its correlated N enhancement, i.e. they actually measure the relative amount of material processed
in the CNO cycle that is present in the stellar atmosphere.
Application of these tools has revealed that almost every GC
hosts multiple populations with the fraction of the first-generation stars\footnote{We equate ``first-generation stars'' 
with ``CN-weak stars that have the lowest helium
abundance'' since it is not yet known whether the chemically distinct multiple
stellar populations formed in separate star formation epochs or at the same time \citep{charbonnel:16}.}
varying from 8\% to 67\% \citep{bedin:15,marino:15a,milone:15a,milone:17}.

The goal of the present study is to use our new HB stellar evolution models in conjunction with an HB population synthesis tool
to simulate the observed distributions of HB stars in 47\,Tuc, M\,3 and M\,13. A comparison of the synthetic and
observed HB populations, including the period distributions of their RR Lyrae variables, 
complemented with fits of isochrones to the MS, subgiant-branch (SGB) and RGB CMDs for each of these GCs, can 
constrain the basic parameters of its multiple populations, such as their shared age, metallicity 
([Fe/H]\footnote{We use the standard spectroscopic notation [A/B] = $\log_{10}(N(\mathrm{A})/N(\mathrm{B})) -
\log_{10}(N_\odot(\mathrm{A})/N_\odot(\mathrm{B}))$, where $N(\mathrm{A})$ and
$N(\mathrm{B})$ are the mass fractions or number densities of the nuclides A
and B.}), distance modulus, and reddening, as well as their individual mean
initial He abundances $Y_{0,i}$, fractions $f_i$, mean mass losses by the RGB stars $\Delta {\cal M}_i$, and
their total number $N_\mathrm{pops}$. 

The selection of 47\,Tuc, M\,3 and M\,13 for this paper is motivated by the fact
that multiple populations in these GCs have been extensively studied, using HB models and population synthesis tools, by other researchers
\citep{dantona:02,caloi:05,caloi:08,dantona:08,valcarce:08,dicris:10,dalessandro:13,valcarce:16,salaris:16} whose results
can serve as tests of our computations. Besides, 47\,Tuc is relatively metal rich and one of the most massive GCs, while
M\,3 and M\,13 represent one of the most famous second-parameter pairs of intermediate-metallicity GCs, with M\,13
showing the most extreme abundance anomalies of the $p$-capture elements.
Consequently, they should be especially interesting targets for the first
application of our new HB models and population synthesis tool. 

The paper has the following structure. Section~\ref{sec:MESA} summarizes the input physics and other options
that have been chosen for the MESA stellar evolution code to compute our HB models. Section~\ref{sec:HBtool}
describes the new HB population synthesis tool that we have used to simulate the distributions of HB stars,
including RR Lyrae, in the selected GCs. In Section~\ref{sec:synth}, we first analyze how the mean color
and magnitude of a single generation of HB stars depend on various model parameters; then we compare our
synthetic distributions of HB stars with those observed in the GCs 47\,Tuc, M\,3 and M\,13 to estimate
their distance moduli, reddenings and parameters of multiple stellar populations. The distance moduli are used
in Section~\ref{sec:age} to derive the ages of these GCs by fitting isochrones to their observed turn-off
luminosities. As a consistency check, in Section~\ref{sec:rrl} we compare the periods of RR Lyrae stars in M\,3, M\,13 and 47\,Tuc
with their values predicted by our HB models. In Section~\ref{sec:47TucSMS}, we present a new model of
a supermassive MS star that can reproduce both the O-Na anti-correlation and the correlations between the [Al/Fe]
enhancement and the Mg isotopic ratios in 47\,Tuc stars. The final Section~\ref{sec:concl} contains
summary and discussion.

\section{MESA HB evolution models}
\label{sec:MESA}

\subsection{Input physics}

In this work, we have used revision 7624 of the MESA stellar evolution code
with the MESA default equation of state --- see Section \,4.2 and Fig.\,1 in
\cite{paxton:11} and Section\,A.2 in \cite{paxton:13}.  For the initial mixture
of elements and isotopes, we have adopted the Solar System chemical composition
given by \cite{ags09}, with the abundances of the $\alpha$ elements enhanced to
[$\alpha$/Fe]\,$=+0.4$ and scaled to the [Fe/H] values that have been adopted for the
three GCs considered in this investigation.  As opacities for these abundances are not
included in the MESA library, we had to generate OPAL opacity data \citep{iglesias:93,iglesias:96}
without and with enhanced abundances of C and O (i.e., Type 1 and Type 2 tables,
respectively),\footnote{\href\protect{https://opalopacity.llnl.gov/new.html}} in order to
model both the H- and He-burning evolutionary phases of low-mass stars.
For the same reason, new low-temperature molecular opacity tables \citep{ferguson:05} have
been computed for the \cite{ags09} mixture with [$\alpha$/Fe]\,$=+0.4$. 

We have followed the nucleosynthesis of 30 isotopes from $^1$H to $^{28}$Si in
H-burning reactions of the $pp$-chains, CNO, 
NeNa and MgAl cycles, as well as in the main He-burning reactions. The rates of those
reactions were taken from the {\it JINA Reaclib} database \citep{cyburt:10}. 

To obtain a close match to the input physics of the Victoria code, we have selected
the Krishna-Swamy atmospheric boundary condition \citep{ks66} and
the MESA ``Henyey'' MLT option with its default parameters $y=0.333$ and $\nu = 8$
\citep{henyey:65}, along with the solar-model calibrated mixing length
parameter, $\alpha_\mathrm{MLT} = 2.0$ (in pressure-scale heights $H_P$). 
To reach the best agreement
between the RGB tracks and ZAHB loci computed with the Victoria and MESA codes 
(see Figure~1 in \citealt[][Paper~I]{vandenberg:16})
we had to use higher-order (cubic) interpolation of the opacities in $Z$ and to
include the gravitational settling of He during the MS and subgiant branch (SGB)
stages of evolution.  (The diffusion of the metals was not treated.)

The initial chemical composition for our MESA computations is prepared by
executing a separate Fortran code that can use, as input data,
lists of isotopic abundances corresponding to the solar compositions of \cite{gn93}, \cite{gs98} or \cite{ags09}. The solar-scaled
initial abundances are calculated for specified values of $Y$, [Fe/H], [$\alpha$/Fe] and [O/$\alpha$]. The last parameter is useful
for models with [O/Fe]\,$>$\,[$\alpha$/Fe]. To be able to compute MESA models
for the latter possibility, we have additionally generated Type\,1 and Type\,2
OPAL opacities for the \cite{ags09} mixture with [O/Fe]\,$= 0.6, 0.7, 0.8$ and $1.0$,
assuming [$\alpha$/Fe] $= 0.4$ for all of the other $\alpha$-elements.
At low metallicities ([Fe/H] $\lta -1.3$), low-mass stellar models
with O-enhanced abundances can be computed with MESA using the same low-$T$
opacities that were generated for [O/Fe]\,$=$\,[$\alpha$/Fe]\,$= 0.4$, provided
that they are interpolated for the (lower) value of $Z$ that corresponds to
[O/Fe]\,$=0.4$. This ensures that all of the other heavy elements have the
correct abundances and that their contributions to the low-$T$ opacities are
properly treated, irrespective of the assumed oxygen abundance, which has no
significant influence on the opacity at $\log\,T \lta 4.0$ in relatively warm,
metal-deficient stars.  

To reiterate: the correct O abundance was assumed in the equation of
state, the nuclear reactions, etc.  We have simply opted to use low-$T$ opacities from
the existing set for [$\alpha$/Fe] $=0.4$ for the appropriate value of $Z$,
when such data are essentially independent of the oxygen abundance.  (Clearly,
this strategy cannot be employed once molecules become important sources of opacity.)
In fact, we have verified that O-enhanced MESA stellar evolutionary tracks that
employ this approximation match their counterparts derived from the Victoria
code, which does have low-$T$ opacity tables for O-enhanced mixtures, very well.
Unfortunately, such data are not yet available in the format used by MESA.

It should be appreciated that there are significant differences in the
input physics of our HB models compared with those adopted in published
computations (e.g., the widely used BASTI models by \citealt{pietrinferni:04}).
In particular, our models are the first ones to assume the \cite{ags09} metals
mixture with [$\alpha$/Fe] $= 0.4$, for which we have generated proper OPAL and
low-$T$ opacity tables.  They also incorporate improved nuclear reaction rates
and the gravitational settling of He during the MS and SGB phases, together with
sufficient extra mixing below convective envelopes to provide reasonable
consistency between predicted and spectroscopically derived surface abundances.
As these updates change the He and heavy-element mass-fraction abundances 
of our HB models for the same {\it initial} values of $Y$ and [Fe/H], it is
important to investigate how they affect the derivation of such GC parameters as
their distance moduli, ages, the predicted periods of their RR Lyrae stars, etc.

\subsection{Mixing prescription for the core He burning}

As an HB star burns He in its convective core, thereby producing C and O,
the opacity of stellar material inside the convective core increases. Therefore,
the ratio of the radiative and adiabatic temperature gradients (logarithmic and with respect to pressure),
$\nabla_\mathrm{rad}/\nabla_\mathrm{ad}$, also increases at the convective core
boundary, which, according to the Schwarzschild criterion 
for convective instability, prompts the boundary to advance outwards in mass.
However, because the ratio $\nabla_\mathrm{rad}/\nabla_\mathrm{ad}$ remains below $1.0$ immediately outside the boundary, a stellar
evolution code has to be instructed what to do in this ambiguous situation. Although there are several alternative instructions
(algorithms) available in the literature \cite[e.g., see the review of five mixing models for the core He burning phase in
Section\,2 of][and references therein]{straniero:03}, none of them has been tested in 3D hydrodynamical simulations yet.

Fortunately, \cite{constantino:15} and \cite{constantino:16} have used new
observational data to constrain the mixing prescription for core He burning.
In the first work, they found that the average period spacing
for the asymptotic dipole $g$ modes inferred for HB stars in the {\em Kepler} field
can only be reproduced by models that implement their proposed
``maximal overshoot'' mixing scheme. This algorithm models mixing outside the
convective boundary, usually called
``convective overshooting'' (OS) or ``convective boundary mixing'' (CBM), with
the following diffusion coefficient:
\bea
D_\mathrm{OS}(r) = D_0\exp\left[\frac{-2(r-r_0)}{f_\mathrm{OS}H_P}\right],
\label{eq:DOS}
\eea
where $r_0$ is the radius immediately below the convective core boundary, $D_0=D_\mathrm{MLT}(r_0)$ is the MLT diffusion coefficient at $r=r_0$, 
and $f_\mathrm{OS}$ is a free parameter.
This expression was originaly proposed for CBM by \cite{herwig:97} and it is now implemented in MESA. 
If a large and constant value of $f_\mathrm{OS}$ is used, then the He convective core rapidly grows in mass and a minimum of
the ratio $\nabla_\mathrm{rad}/\nabla_\mathrm{ad}$ in the core moves from its boundary inwards. When the minimum of 
$\nabla_\mathrm{rad}/\nabla_\mathrm{ad}$ inside the core drops below $1.0$, a convective shell detaches from the convective core.
To avoid this situation, \cite{constantino:15} assumed that the amount of mass added to the convective core at each time step was proportional to
the minimum of $(\nabla_\mathrm{rad}/\nabla_\mathrm{ad}-1)$ in the core as long as that minimum remained larger than 0.002, otherwise it was zero. 
With this algorithm, they obtained the largest possible
convective He core throughout the HB evolution, a result that is supported by
the {\em Kepler} data.

When the helium abundance $Y_\mathrm{c}$ decreases below $\sim 0.1$ in the convective core, its boundary can occasionally move rapidly outward
in mass, thus ingesting fresh He. This results in a brief rejuvination of the HB star, during which time its evolutionary track makes
a loop. These so-called ``breathing pulses'' increase the lifetime of the HB evolution.
In their second work, \cite{constantino:16} have found that the number ratio of AGB to HB stars in 48 Galactic globular clusters
favours the shorter evolutionary timescale without ``breathing pulses''.

\begin{figure}[t]
\plotone{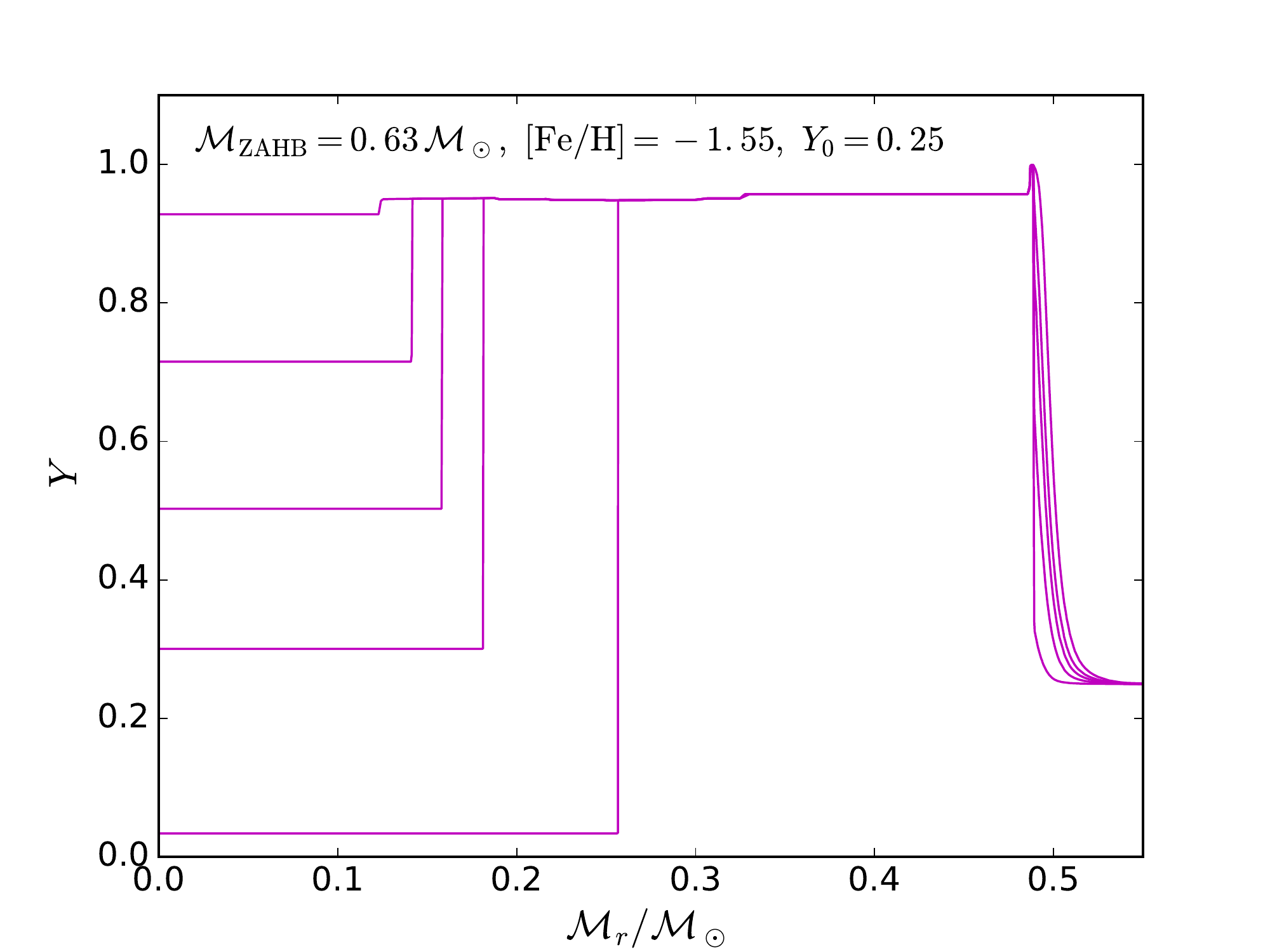}
\caption{The growth of the He convective core and the change of the He abundance profile in
our HB models for the indicated mass and chemical abundances.}
\label{fig:f1}
\end{figure}

To comply with these results, we have designed the following mixing prescription for core He burning. We use the MESA
implementation of Equation (\ref{eq:DOS}) in which $D_0$ is taken at a distance $0.001H_P$ below the Schwarzschild boundary with
$f_\mathrm{OS} = 0.0005$ for $Y_\mathrm{c}\ge 0.05$ and, to eliminate ``breathing pulses'', 
$f_\mathrm{OS} = 0.0005\,(Y_\mathrm{c}/0.05)$ for $Y_\mathrm{c} < 0.05$. To
avoid a possible splitting of the convective core,
we replace $D_\mathrm{MLT}$ with $0.05\,K$ when $D_\mathrm{MLT} < K$, where $K$ is a thermal diffusivity. Our prescription
results in the growth and maximal size of the He convective core (Fig.~\ref{fig:f1}) that are both very similar to those
shown by \cite{constantino:15} in the bottom panel of their Fig.~2 for the ``maximal overshoot'' case. 

\section{HB population synthesis tool}
\label{sec:HBtool}

\subsection{The algorithm and its input parameters}

The Monte Carlo algorithm that we have developed to simulate the distributions
of HB stars in the CMDs of globular clusters is 
not very different from those used by other researchers \citep[e.g.,][]{valcarce:08,sollima:14,salaris:16,tailo:16}.
The input parameters for the algorithm are a total number of HB stars $N_\mathrm{HB}$, the number of different stellar populations in a studied GC
$N_\mathrm{pops}$, where each population is specified with its own mean initial He abundance $Y_{0,i}$, relative fraction $f_i$, the mean mass
$\Delta {\cal M}_i$ lost by the RGB stars, and its dispersion $\sigma_i$. The mean initial masses ${\cal M}_i$ of 
the $N_\mathrm{pops}$ populations, that along with $\Delta {\cal M}_i$ determine the mean masses of the ZAHB stars 
${\cal M}_{\mathrm{HB},i}={\cal M}_i - \Delta {\cal M}_i$, are constrained by their shared age $t_\mathrm{ZAHB}$
and metallicity [Fe/H], but individual $Y_{0,i}$ values.

For each of the $N_i = f_i N_\mathrm{HB}$ stars of the $i$-th population, its
(initial) mass on the ZAHB is randomly selected from
a normal distribution with the mean value ${\cal M}_{\mathrm{HB},i}$ and
dispersion $\sigma_i$. A set of HB evolutionary tracks,
that was generated for the specified values of [Fe/H], $Y_{0,i}$ and for an age close to the one estimated for the GC independently, 
e.g. by an isochrone fitting method, is then interpolated in the selected ZAHB mass 
to identify the two tracks with the closest ZAHB masses.
Then, an age of the synthetic HB star $t_\mathrm{HB}$ is randomly selected from an uniform distribution on the interval [$0,t_\mathrm{HB,max}$],
and its location on a theoretical HR diagram $(\log_{10}T_\mathrm{eff},\log_{10}(L/L_\odot))$
is interpolated in the selected ZAHB mass\footnote{We assume that the stellar mass does not change
during the HB evolution.} and $t_\mathrm{HB}$, using the pair of HB evolutionary
tracks with the nearest ZAHB masses. We adopt the value of $t_\mathrm{HB,max} = 1.5\times 10^8$ yr that
deliberately exceeds the maximum HB ages of the order of $1.3\times 10^8$ yr of our HB models to accomodate
the standard assumption behind synthetic HBs that ``the stars are being fed onto the HB at a constant rate'' \citep{rood:73}.
If the selected age $t_\mathrm{HB}$ happens to be older than the maximum age of the two HB tracks used
for the interpolation, we assume that the star has already left the HB and it is not included
in the synthetic HB populations.  
The effective temperatures and luminosities of our stellar models are transformed
to the $(m_\mathrm{F606W}-m_\mathrm{F814W})_0$ or
$(B-V)_0$ colors and to the $M_\mathrm{F606W}$ or $M_V$ absolute magnitudes using the
bolometric corrections from \cite{cv14} for $\teff$\ values $\le 8000$~K,
and from S.~Cassisi (private communication, but see \citealt{cassisi:04}) for
higher temperatures.  Suitable zero-point adjustments were applied to the latter
transformations in order to ensure continuity with those given by Casagrande \&
VandenBerg.

\begin{figure}[t]
\plotone{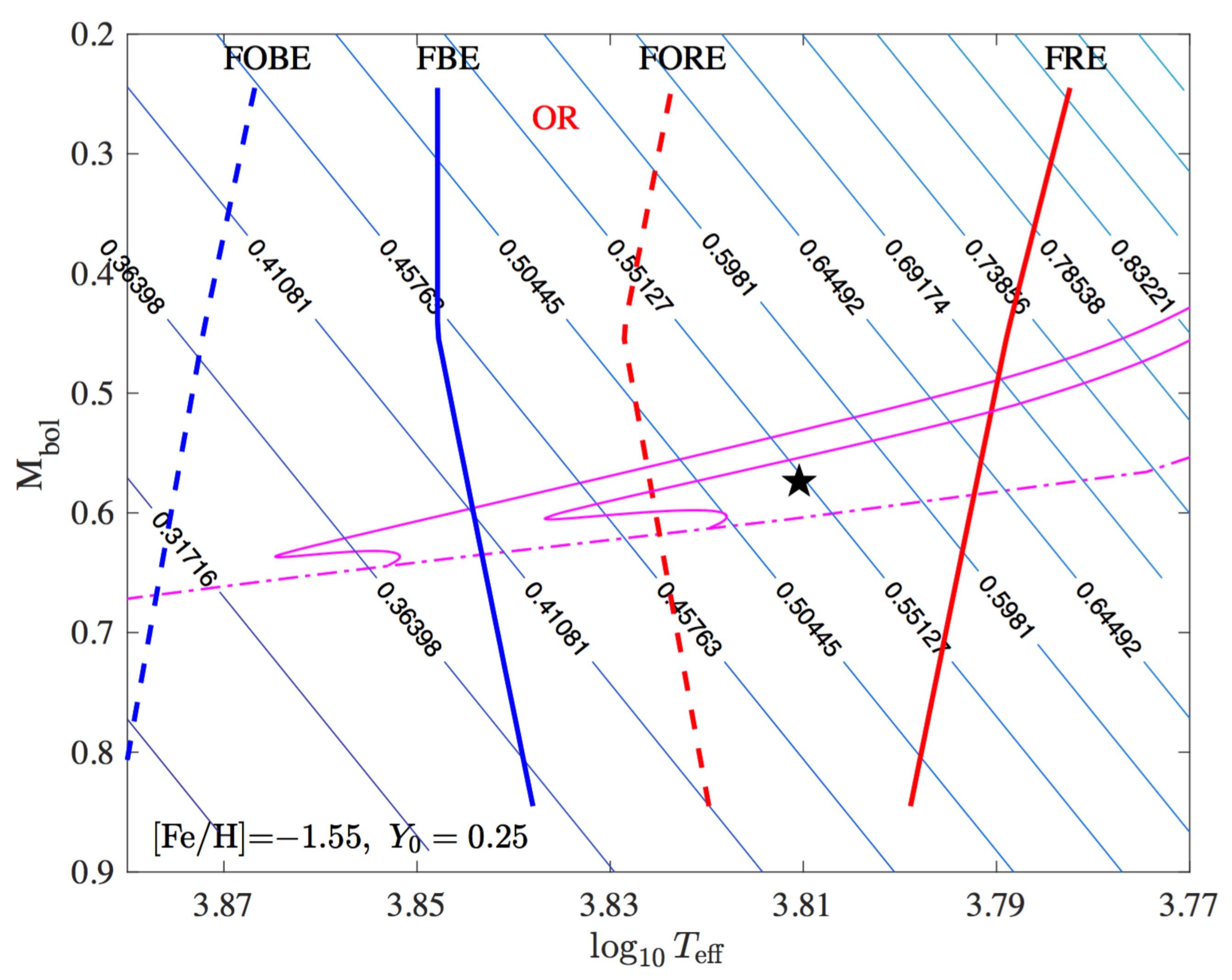}
\caption{The overlapping instability strips for the fundamental (solid vertical curves) and first-overtone (dashed vertical curves) modes, 
   diagonal lines of constant pulsation periods (given in days) for the fundamental mode \citep{mcb15}, 
   the ZAHB (dot-dashed curve), and the mean location of
   our synthetic RRab stars (black star symbol) calculated for [Fe/H]\,$=-1.55$,
   $Y_0=0.25$, [$\alpha$/Fe]\,$=0.4$, ${\cal M}_i=0.8$, $\Delta {\cal M}_i=0.125$ and $\sigma_i=0.01$. The two HB evolutionary
   tracks have ZAHB masses $0.6575\,{\cal M}_\odot$ (left) and $0.6675\,{\cal M}_\odot$ (right). FOBE, FBE, FORE, FRE, and OR stand for
   the first-overtone and fundamental blue and red edges of the instability strips, and the overlap region.}
\label{fig:f2}
\end{figure}

\subsection{The synthetic RR Lyrae stars}

A more sophisticated algorithm is used to determine a pulsation mode of the synthetic HB star when its location happens to be
inside the instability strip (IS), where the RR Lyrae stars are observed. The effective temperatures at the red and blue edges of
the two overlapping instability strips in which the HB star is expected to pulsate either in the fundamental mode (FRE and FBE) or 
in the first overtone mode (FORE and FOBE)
are interpolated in its mass and luminosity (Fig.~\ref{fig:f2}) from the
tables of \cite{bono:95} and \cite{bono:97}. We have found that, in order for all of
the observed RR Lyrae stars in the globular cluster M\,3 to fit inside the IS, both the
interpolated values of $T_\mathrm{eff}^\mathrm{FRE}$ and  
$T_\mathrm{eff}^\mathrm{FOBE}$ have to be increased by $\approx 200$\,K. We
believe that our empirical adjustments of the theoretical data are within
reasonable uncertainty limits, because their magnitudes are comparable to the
corrections that have to be applied to the effective temperatures of
the IS edges when $\alpha_\mathrm{MLT}=1.5$, as used in the models of
\cite{bono:95} and \cite{bono:97}, is replaced with our value of
$\alpha_\mathrm{MLT}=2.0$ \citep{marconi:03}. We note that 
similar adjustments of the M\,3 IS boundaries were made by \cite{dalessandro:13} ($3.80\la\log_{10}T_\mathrm{eff}\la 3.88$,
in good agreement with our Figure~\ref{fig:f2}) and by \cite{caloi:08} ($0.14\la (B-V)_0\la 0.38$,
which compares well with our results; see Section \ref{sec:M3}).

Just the locations of the IS boundaries have been adjusted --- not the evolutionary tracks --- 
as otherwise some of the RR Lyrae in M\,3 would lie outside the IS. Other researchers have found such adjustments
to be necessary, as we have noted. Thus the small $T_\mathrm{eff}$ shift of the tracks that may be necessary
to obtain the best agreement between predicted and RR Lyrae periods is a separate issue.
Note that the small $T_\mathrm{eff}$ offset could instead be due to the assumption of a reddening value or
distance modulus that are not quite right, or small errors in the mean
$\langle B\rangle - \langle V\rangle$ and $\langle V\rangle - \langle I_C\rangle$
colors that are assumed to represent the equivalent static colors of the RR Lyrae (as mentioned on page 19).

An ambiguity in the determining of the pulsation mode occurs when the synthetic HB star 
finds itself in the IS overlap (OR) region between $T_\mathrm{eff}^\mathrm{FBE}$ and $T_\mathrm{eff}^\mathrm{FORE}$. Following
\cite{sollima:14}, we assume that the star belongs to the RRc (RR1) class of the first-overtone pulsators 
if it entered the overlap region crossing its blue boundary (FBE)
and that it is of the RRab (RR0) type (the fundamental-mode pulsator) if it crossed the FORE before 
finding itself in the overlap region (Fig.~\ref{fig:f2}). This ``hysteresis'' effect
was originally proposed by \cite{albada:73}. A small number of synthetic HB
stars that stay in the overlap region until they
leave it, when they cross the FORE, are divided equally into the RRc and RRab
types depending on what edge (FBE or FORE) was closest to their ZAHB locations.

Pulsation periods (in days) of our synthetic RR Lyrae stars are calculated using 
the following metallicity-dependent relations for the fundamental and first-overtone modes given by \cite{mcb15}:
\begin{align}
\log\,&P_\mathrm{ab} = (11.347 \pm 0.006) + (0.860 \pm 0.003)\log(L/L_\odot)\nonumber\\ 
 &- (0.58 \pm 0.02)\log({\cal{M}/\cal{M}_\odot}) - (3.43 \pm
 0.01)\log\teff\nonumber\\
 & + (0.024 \pm 0.002)\log\,Z
\label{eq:pab}
\end{align} 
\noindent and
\begin{align}
\log\,&P_\mathrm{c} = (11.167 \pm 0.002) + (0.822 \pm 0.004)\log(L/L_\odot)\nonumber\\
 &- (0.56 \pm 0.02)\log({\cal{M}/\cal{M}_\odot}) - (3.40 \pm
 0.03)\log\teff\nonumber\\ &+ (0.013 \pm 0.002)\log\,Z.
\label{eq:pc}
\end{align}

\begin{figure}[t]
\plotone{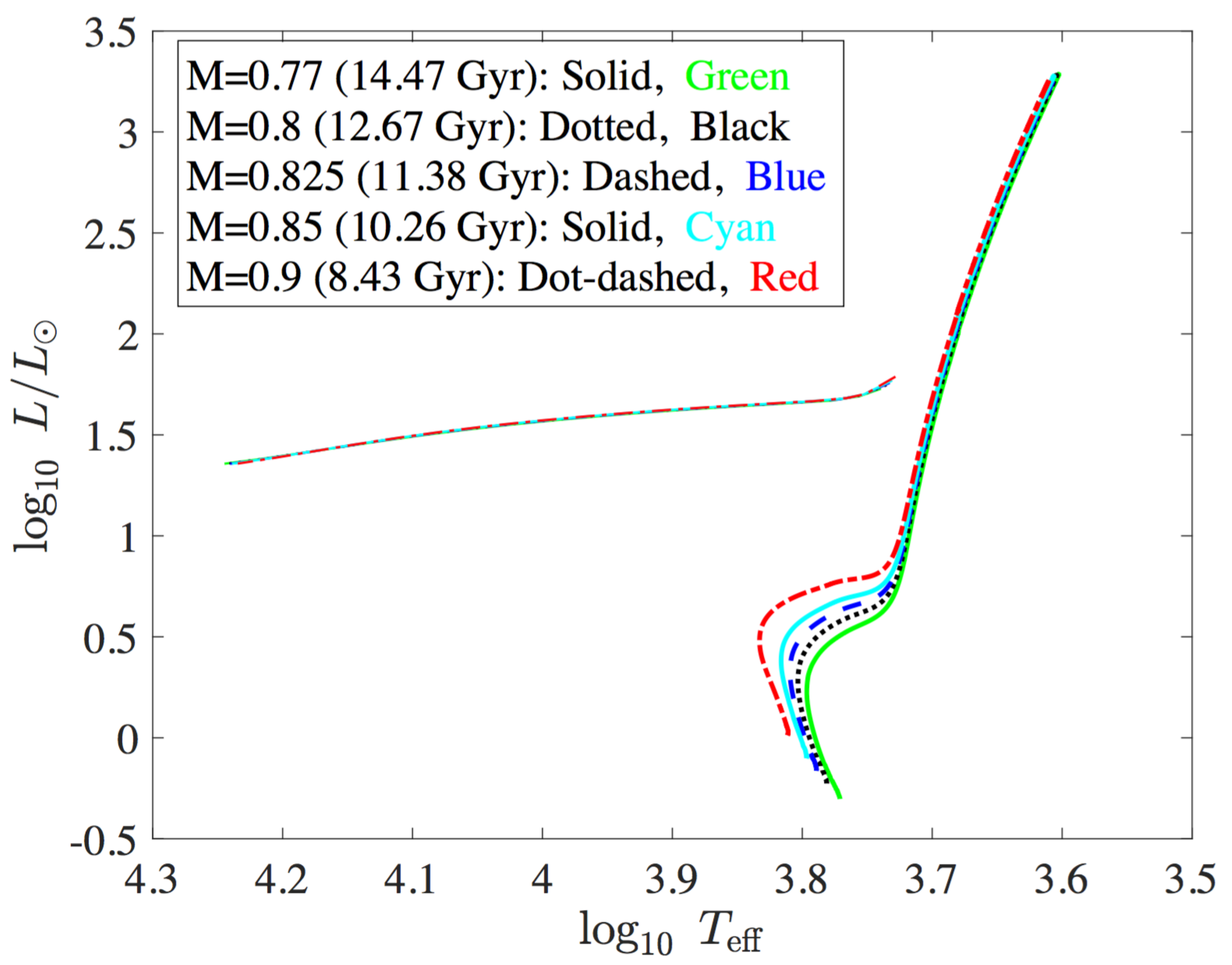}
\caption{A comparison of MS/RGB evolutionary tracks and ZAHBs of different ages for [Fe/H]\,$=-1.55$, $Y_0=0.25$, [$\alpha$/Fe]\,$=0.4$ and [O/Fe]\,$=0.4$.
         The initial masses (in ${\cal M}_\odot$) and ZAHB ages for the tracks are indicated in the legend.}
\label{fig:f3}
\end{figure}

\begin{figure*}[t]
\plotone{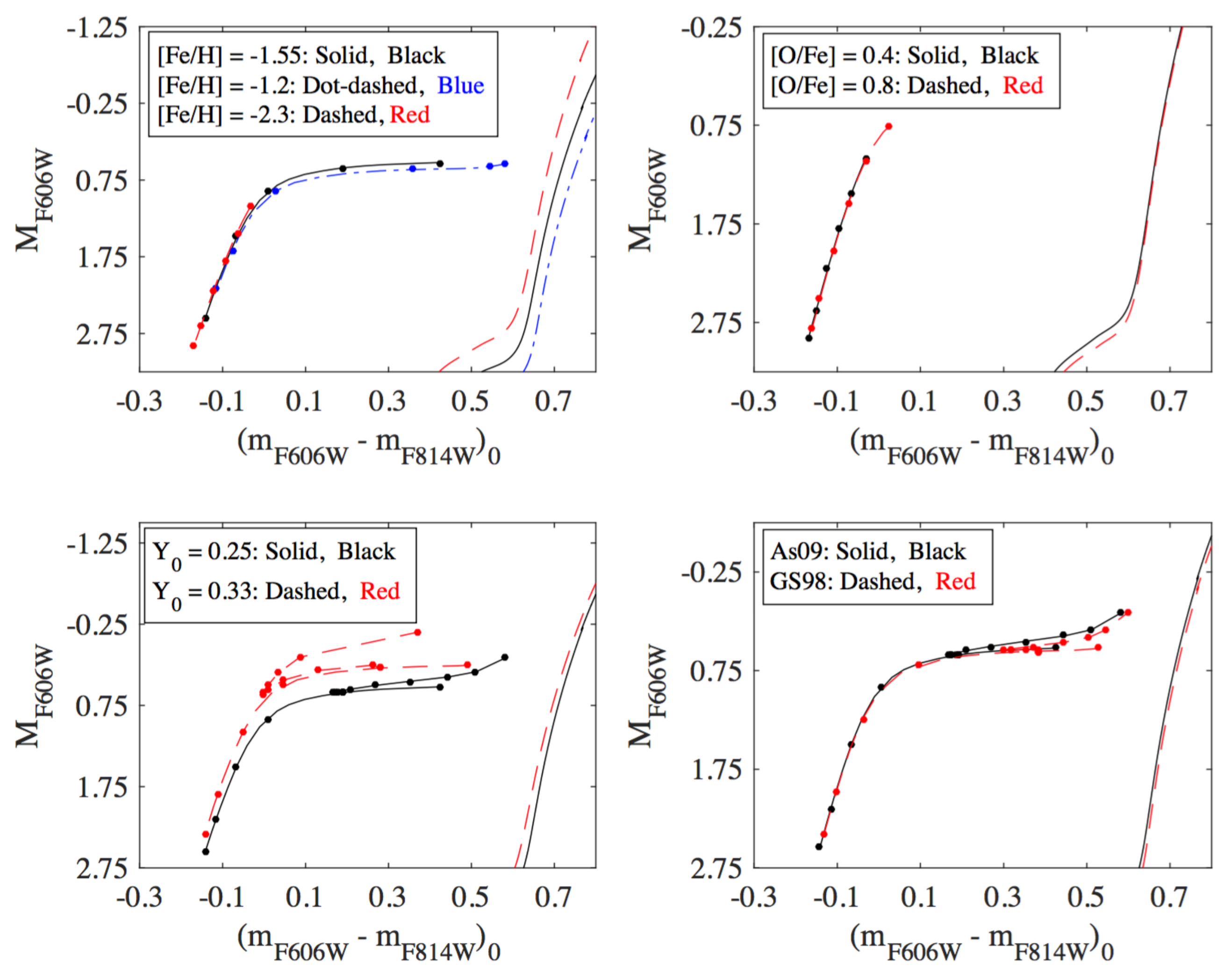}
\caption{The effects of varying [Fe/H], $Y_0$, [O/Fe], and the metals mixture on ZAHB loci and
     (only for the lower panels) selected HB tracks. 
     The masses (in ${\cal M}_\odot$) of six selected ZAHB models are 0.56, 0.575, 0.6, 0.625,
     0.65, and 0.675 (filled circles along the ZAHBs from left to right). The eleven points (some are overlapped)
     along the HB evolutionary tracks are equidistant in age. The template HB model has
     ${\cal M} = 0.8\,\msol$, [Fe/H]\,$= -1.55$, $Y_0 = 0.25$, [$\alpha$/Fe]\,$= 0.4$, [O/Fe]\,$= 0.4$
     and the initial abundances of heavy elements scaled using the solar chemical
     composition of \cite{ags09}. The legends in the panels only specify what has been changed
     in the models compared to the template model, except the top-right panel in which
     [Fe/H] has also been changed to $-2.3$ for both sets of the shown ZAHB models
     and the bottom-left panel in which the HB star model with $Y_0 = 0.33$ has the initial mass ${\cal M} = 0.695\,\msol$ that is
     required for its age to be close to that of the model with $Y_0=0.25$.
     In the bottom-right panel, the acronyms As09 and GS98 indicate the scaling of the initial abundance mix 
     with respect to the solar compositions of \cite{ags09} and \cite{gs98}.}
\label{fig:f4}
\end{figure*}

\begin{center}
\begin{deluxetable*}{cclllcccc}
\tablecolumns{9}
\tablecaption{
Mean Colors, Magnitudes, and Fundamental-Mode Pulsation Periods for Single Populations of Synthetic HB Stars\label{tab:hb_sims}}
\tablehead{
  \colhead{ID$^1$}&\colhead{$t_\mathrm{ZAHB}\ (\mathrm{Gyr})$}&
  \colhead{ ${\cal M}_\mathrm{HB}\ ({\cal M}_\odot) $}&\colhead{ $\langle C\rangle = \langle m_\mathrm{F606W}-m_\mathrm{F814W}\rangle $}
  &\colhead{$\langle M_\mathrm{F606W}\rangle $}& \colhead{$\langle P_\mathrm{ab}\rangle\ (\mathrm{d}) $}&
  \colhead{ $\frac{\Delta\langle C\rangle}{\Delta P}$}& \colhead{ $\frac{\Delta\langle M_\mathrm{F606W}\rangle}{\Delta P}$}&
  \colhead{ $\frac{\Delta t_\mathrm{ZAHB}}{\Delta P}$}}
\startdata
TH & $12.67$ & $0.675$ & $0.4437$ & $0.4486$ & $0.5508$ & $0$ & $0$ & $0$ \\
TL & $12.67$ & $0.6$ & $-0.008251$ & $1.079$ & $0.6838$ & $6.026$ & $-8.405$ & $0$ \\
\hline
MH & $11.38$ & $0.675$ & $0.4484$ & $0.4453$ & $0.5672$ & $0.1880$ & $-0.1320$ & $-51.6$ \tstrut \\
ML & $11.38$ & $0.6$ & $-0.006000$ & $1.065$ & $0.6962$ & $6.059$ & $-8.263$ & $0$ \\
\hline
OH & $12.76$ & $0.675$ & $0.5059$ & $0.4379$ & $0.5879$ & $0.3110$ & $-0.0535$ & $0.45$ \tstrut \\
OL & $12.76$ & $0.6$ & $0.01769$ & $0.9565$ & $0.7039$ & $6.509$ & $-6.915$ & $0$ \\
\hline
FH & $12.37$ & $0.675$ & $0.3414$ & $0.4485$ & $0.5919$ & $0.6820$ & $0.0007$ & $2$ \tstrut \\
FL & $12.37$ & $0.6$ & $-0.04936$ & $1.310$ & $\mathrm{-}$ & $5.210$ & $-11.49$ & $0$ \\
\hline
YH & $11.18$ & $0.675$ & $0.4025$ & $0.3868$ & $0.6054$ & $-2.060$ & $-3.090$ & $-74.5$ \tstrut \\
YL & $11.18$ & $0.6$ & $-0.09014$ & $1.690$ & $\mathrm{-}$ & $6.569$ & $-17.38$ & $0$ \\
\hline
YAH & $12.66$ & $0.675$ & $0.4071$ & $0.3842$ & $0.6058$ & $-1.830$ & $-3.220$ & $-0.5$ \tstrut \\
YAL & $12.66$ & $0.6$ & $-0.02635$ & $1.099$ & $0.6384$ & $5.779$ & $-9.531$ & $0$
\enddata
\tablenotetext{1}{The letter T refers to the template simulations that used ${\cal M} = 0.8\,{\cal M}_\odot$, [Fe/H]\,$=-1.55$, $Y_0=0.25$,
[$\alpha$/Fe]\,$=0.4$, and [O/Fe]\,$=0.4$. 
In the simulations marked with the letters M, O, and F the initial mass was changed to ${\cal M} = 0.825\,{\cal M}_\odot$,
the oxygen abundance was enhanced by $\Delta$[O/Fe]\,$=0.2$, and the metallicity was reduced to [Fe/H]\,$=-1.7$, respectively.
In the simulations Y and YA, the He abundance was increased to $Y_0=0.27$. A difference between the simulations Y and YA
is that in the latter case the initial mass was decreased to ${\cal M} = 0.772\,{\cal M}_\odot$ 
to keep the ZAHB age close to the one in the template simulations. The letters H and L in the simulation IDs signify
the cases with the high and low HB masses. The 7th and 8th columns of the L lines provide the partial derivatives of
the mean color and magnitude with respect to ${\cal M}_\mathrm{HB}$ calculated using the data from the H lines of the same simulations, 
while the last three columns of the H lines contain the partial derivatives of the mean color, magnitude, and age
with respect to the parameter $P$ whose value was changed in those simulations with regard to the TH case. 
All the simulations were run with $\sigma = 0.01\,{\cal M}_\odot$.}
\end{deluxetable*}
\end{center}

\section{Synthetic HB populations}
\label{sec:synth}

\subsection{Parameters that affect the color and magnitude of an HB model star}

\subsubsection{The ZAHB stars}

Fig.~\ref{fig:f3} demonstrates that the luminosity of ZAHB stars does not depend on their age for a fixed initial composition.
Therefore, fitting a theoretical ZAHB to the bottom of an observed distribution
of HB stars can be used to estimate the 
distance modulus of a GC without knowing its age. On the other hand, the luminosity
of the MS turnoff (MSTO) is strongly dependent on age and the morphology
of isochrones in the vicinity of the TO has little sensitivity to the
effective temperature; consequently, these properties can be used to estimate
GC ages in conjunction with distance moduli that are obtained from the ZAHB fits.
This approach was used by \citet[][hereafter VBLC13]{vbl13} and \cite{leaman:13} to
determine the ages of 61 GCs, and a similar method is employed in the present work.
The main difference is that the distance moduli of GCs will instead be based on
the fitting of the entire distributions of observed HB stars by synthetic HB
populations (i.e., not just ZAHBs) to take full advantage of our new tracks for
the core He-burning phase.

When the [Fe/H] and [O/Fe] ratios and the helium mass fraction $Y_0$ in the initial chemical composition are allowed to change,
within their observed or assumed limits appropriate for GCs,
and when the age and RGB mass-loss rate are also varied, 
these changes all affect the location of an HB model star on a CMD as follows.          

When the [Fe/H] value decreases, the ZAHB luminosity increases somewhat and
the color of a ZAHB star becomes significantly bluer
(Fig.~\ref{fig:f4}, the top-left panel). 

An increase in $Y_0$ makes a ZAHB model more luminous and it 
produces an extended blue loop during its evolution (Fig.~\ref{fig:f4}, the bottom-left panel).
For the same mass, a ZAHB star with a higher $Y_0$ has a slightly redder color.
However, if two HB stars in the same GC belong to populations with
different values of $Y_0$ (say, $Y_{0,1}=0.25$ and $Y_{0,2}=0.33$, as shown in
the bottom-left panel), but they have nearly the same age, which is expected if
the second-generation stars in GCs were formed from a mixture of
pristine gas with ejecta from short-lived massive stars, including intermediate-mass and massive AGB stars, then the initial mass of
the He-rich star has to be lower (${\cal M}_2 = 0.695\,{\cal M}_\odot$) 
than the initial mass of the star with the normal He abundance (${\cal M}_1 = 0.8\,{\cal M}_\odot$).
When we apply the Reimers formula for the RGB mass-loss \citep{reimers:75} that
is implemented in MESA, assuming that the same parameter
$\eta_\mathrm{R}=0.5$ applies to both models, they arrive at the HB with the masses ${\cal M}_{\mathrm{HB},2} = 0.507\,{\cal M}_\odot$
and ${\cal M}_{\mathrm{HB},1} = 0.602\,{\cal M}_\odot$. This, according to Fig.~\ref{fig:f4}, means that
He-rich HB stars should be fainter and much bluer than their He-normal counterparts for the same age and RGB mass-loss.

If the trend of increasing [O/Fe] with decreasing [Fe/H] that has been found
for metal-poor field stars in the
Galaxy \citep[e.g.,][]{amarsi:15,dobrovolskas:15} is applicable
to GCs, then stars in a GC with low [Fe/H] may have an excess of oxygen abundance relative to our assumed
value of [$\alpha$/Fe]\,$= 0.4$. The top-right panel of Fig.~\ref{fig:f4} shows that ZAHB stars with
[O/Fe]\,$> 0.4$ are shifted to the red compared to ZAHB stars with [O/Fe]\,$= 0.4$. This is caused by an increase
in the efficiency of the CNO-cycle in the H-burning shell where O plays a role of a catalyst.

Finally, the bottom-right panel of Fig.~\ref{fig:f4} highlights the effect of changing the initial mix of heavy elements
on the location of HB stars. Replacing the GS98 mix with the As09 mix is equivalent to reducing the [Fe/H] ratio
(compare these ZAHBs with those in the top-left panel), because the latter mix has a lower $Z_\mathrm{As09} = 7.6\times 10^{-4}$
than the former $Z_\mathrm{GS98} = 1.1\times 10^{-3}$ for the same metallicity [Fe/H]\,$=-1.55$.

\subsubsection{The evolving HB stars}
\label{sec:evol_HB}

In order to take into account the evolutionary effects on the mean properties
of HB stars, we need to synthesize HB populations.
Some results of such simulations for single populations of HB stars
are presented in Table~\ref{tab:hb_sims}. The parameters of the template case simulations (TH and TL) are representative
of the first generations of stars in the GCs M\,3 and M\,13 that will be discussed in the next section.
The black star symbol in Fig.~\ref{fig:f2} corresponds to the simulation TH. In other simulations,
we have changed the value of only one parameter (two in the YA case) relative to the template case in order
to calculate the partial derivatives of the mean color, magnitude, and age of the synthetic HB stars.
These derivatives and the other data provided in Table~\ref{tab:hb_sims} can be used to estimate changes in
mean properties of HB stars resulting from variations of the parameters that determine the HB morphology
and pulsation periods of RR Lyrae stars. 

Here is an example of such an estimate.
The mean color of a single population of synthetic HB stars is a function of the following seven parameters: 
\begin{align}
\langle m_\mathrm{F606W} - m_\mathrm{F814W}\rangle = \langle C\rangle (\mathrm{[Fe/H]},[\alpha\mathrm{/Fe]},\mathrm{[O/Fe]},\nonumber\\
   Y_0,{\cal M},{\cal M}_\mathrm{HB},\sigma),
\label{eq:hbcolor}
\end{align}
where the mean ZAHB (and HB) mass ${\cal M}_\mathrm{HB}$ can be replaced with the mean RGB mass-loss 
$\Delta{\cal M} = {\cal M} - {\cal M}_\mathrm{HB}$.
It is very difficult to anticipate the reaction of the mean color to simultaneous changes of all 
of the parameters that
it depends on, given that changes in some of the quantities can have the 
opposite effect and thereby partially compensate for changes in other quantities
(Fig.~\ref{fig:f4}). Therefore, it is desirable to first eliminate as many
free parameters as possible
in the above equation by fixing the most plausible values of them.
For example, we assume that the first generations of stars in the GCs M\,3 and M\,13 both had
[Fe/H]\,$=-1.55$, [$\alpha$/Fe]\,$=0.4$, [O/Fe]\,$=0.4$, and $Y_0=0.25$. (These values are close
to the ones that have generally been adopted for M\,3 and M\,13 in the literature.) 

The ZAHB age does not depend on ${\cal M}_\mathrm{HB}$ and $\sigma$
\begin{align}
t_\mathrm{ZAHB} = t_\mathrm{ZAHB}(\mathrm{[Fe/H]},[\alpha\mathrm{/Fe]},\mathrm{[O/Fe]},Y_0,{\cal M});
\label{eq:tzahb}
\end{align}
therefore, once we have chosen the initial chemical composition, $t_\mathrm{ZAHB}$\ depends only on the initial stellar mass.
If we assume that the dispersion of the mean ZAHB mass (or of the mean RGB mass-loss) is nearly the same
for all HB populations, then its value can also be considered fixed. We adopt 
$\sigma = 0.01\,{\cal M}_\odot$ in Table~\ref{tab:hb_sims}.

\begin{figure}[t]
\plotone{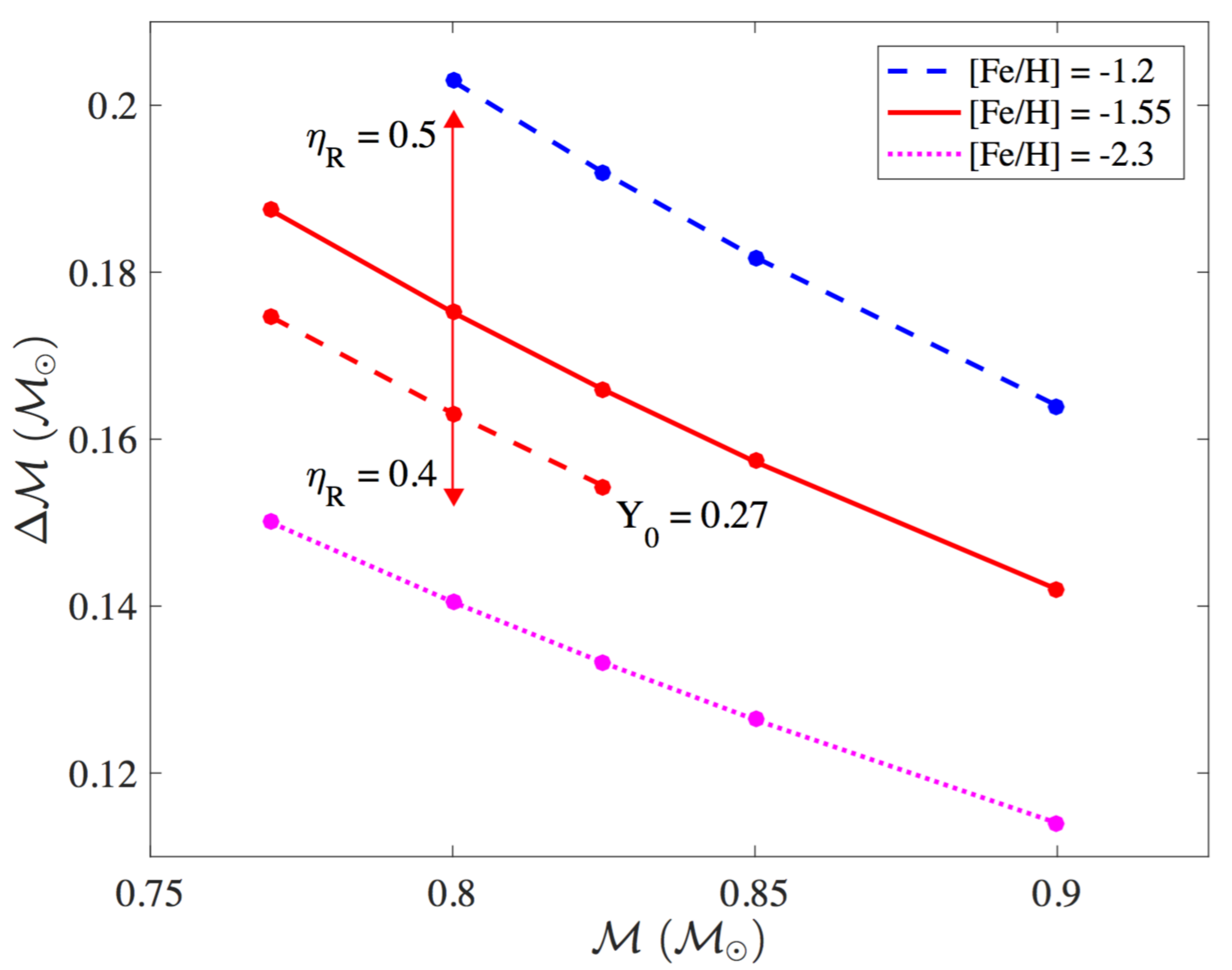}
\caption{The dependences of the mass lost by our benchmark ZAHB models on the RGB 
         on the initial mass, metallicity, and He mass-fraction abundance, assuming
         $\eta_\mathrm{R} = 0.45$ for the Reimers parameter. 
         The results of changing $\eta_\mathrm{R}$ from $0.4$ to $0.5$ for ${\cal M}=0.8\,{\cal M}_\odot$,
         [Fe/H]\,$= -1.55$, and $Y_0 = 0.25$ are indicated by the arrows.}
\label{fig:f5}
\end{figure}

The famous second parameter problem for a pair of GCs with the same metallicity, e.g. for M\,3 and M\,13,
asks which of the other main parameters (i.e., besides [Fe/H]) is primarily
responsbile for the difference in the HB morphologies (usually, in the mean color $\langle C\rangle$ of the HB
stars) of these clusters \citep[e.g.,][and references therein]{dotter:10}. The most likely candidates for
the second parameter are the age ($t_\mathrm{ZAHB}$) and the mean initial He mass fraction ($Y_0$).
To answer this question using the data in Table~\ref{tab:hb_sims}, we will now allow $Y_0$ to vary along with
${\cal M}$ and ${\cal M}_\mathrm{HB}$, while keeping the other parameters fixed. Under these constraints,
we have 
\begin{align}
& \langle C\rangle = \langle C\rangle (Y_0,{\cal M},{\cal M}_\mathrm{HB}),\label{eq:color}\\
& t_\mathrm{ZAHB} = t_\mathrm{ZAHB} (Y_0,{\cal M}), \label{eq:age}
\end{align}
\noindent and
\begin{align}
& {\cal M}_\mathrm{HB} = {\cal M}_\mathrm{HB} (Y_0,{\cal M}). \label{eq:MHB}
\end{align}
We need to add Equation (\ref{eq:MHB}) to this system, because variations of both $Y_0$ and ${\cal M}$ lead
to changes in $\Delta {\cal M}$ for fixed values of [Fe/H] and $\eta_\mathrm{R}$ (Fig.~\ref{fig:f5}).
Differentiating Equations (\ref{eq:color}-\ref{eq:MHB}), it is easy to show that
\begin{align}
& \left(\frac{\partial\langle C\rangle}{\partial Y_0}\right)_{t_\mathrm{ZAHB}} =
\left(\frac{\partial\langle C\rangle}{\partial {\cal M}_\mathrm{HB}}\right)_{Y_0,{\cal M}}
\left(\frac{\partial {\cal M}_\mathrm{HB}}{\partial Y_0}\right)_{\cal M} + \nonumber\\
& \left(\frac{\partial\langle C\rangle}{\partial Y_0}\right)_{{\cal M},{\cal M}_\mathrm{HB}} -
\left(\frac{\partial\langle C\rangle}{\partial {\cal M}}\right)_{Y_0,{\cal M}_\mathrm{HB}}
\frac{\left(\frac{\partial t_\mathrm{ZAHB}}{\partial Y_0}\right)_{\cal M}}{\left(\frac{\partial t_\mathrm{ZAHB}}{\partial {\cal M}}\right)_{Y_0}} - \nonumber\\
& \left(\frac{\partial\langle C\rangle}{\partial {\cal M}_\mathrm{HB}}\right)_{Y_0,{\cal M}}
\left(\frac{\partial {\cal M}_\mathrm{HB}}{\partial {\cal M}}\right)_{Y_0}
\frac{\left(\frac{\partial t_\mathrm{ZAHB}}{\partial Y_0}\right)_{\cal M}}{\left(\frac{\partial t_\mathrm{ZAHB}}{\partial {\cal M}}\right)_{Y_0}},
\label{eq:dCdY}
\end{align}
\noindent and
\begin{align}
& \left(\frac{\partial\langle C\rangle}{\partial t_\mathrm{ZAHB}}\right)_{Y_0} =
\left[
\left(\frac{\partial\langle C\rangle}{\partial {\cal M}}\right)_{Y_0,{\cal M}_\mathrm{HB}} \right. + \nonumber \\
& \left. \left(\frac{\partial\langle C\rangle}{\partial {\cal M}_\mathrm{HB}}\right)_{Y_0,{\cal M}}
\left(\frac{\partial {\cal M}_\mathrm{HB}}{\partial {\cal M}}\right)_{Y_0}
\right]\Big/
\left(\frac{\partial t_\mathrm{ZAHB}}{\partial {\cal M}}\right)_{Y_0}.
\label{eq:dCdt}
\end{align}
After substituting into Equations (\ref{eq:dCdY}) and (\ref{eq:dCdt}) the derivatives
$\left(\partial {\cal M}_\mathrm{HB}/\partial {\cal M}\right)_{Y_0} = 1.391$ and
$\left(\partial {\cal M}_\mathrm{HB}/\partial Y_0\right)_{\cal M} = 0.6050$, calculated
using the data from Fig.~\ref{fig:f5} for $\eta_\mathrm{R} = 0.45$,
and other relevant derivatives from Table~\ref{tab:hb_sims}, we find that,
at fixed [Fe/H] and $Y_0$,
\bea
\Delta \langle C\rangle = -0.1661\,\Delta t_\mathrm{ZAHB}\ (\mathrm{Gyr}),
\label{eq:dcdt}
\eea
whereas, at fixed age and [Fe/H],
\bea
\Delta \langle C\rangle = -0.1079\,(\Delta Y_0/0.01).
\label{eq:dcdy}
\eea
For fixed $Y_0$ and $t_\mathrm{ZAHB}$, a similar analysis with 
$\left(\partial {\cal M}_\mathrm{HB}/\partial \mathrm{[Fe/H]}\right)_{\cal M} = -0.05673$ from Fig.~\ref{fig:f5} provides
\bea
\Delta \langle C\rangle = 0.06723\,(\Delta\mathrm{[Fe/H]}/0.1).
\label{eq:dcdfeh}
\eea
From Equations (\ref{eq:dcdt}) and (\ref{eq:dcdy}), we conclude that the mean initial helium mass-fraction abundance
can be considered as the main second parameter if the ages of the second-parameter pair of GCs differ by less than 1 Gyr
and their helium contents differ by $\Delta Y_0\gta 0.015$, which is probably true in the case of M\,3 and M\,13 (see next section).
It should be noted that the results presented in this example are valid only for narrow ranges of parameters centered on the values of
[Fe/H]\,=$-1.55$, $Y_0=0.25$, ${\cal M}=0.8\,{\cal M}_\odot$, and $\eta_\mathrm{R}=0.45$, around which the partial derivatives
were evaluated.
 
\subsection{Main assumptions in our HB population synthesis simulations}

In this work, we consider the mono-metallic GCs 47\,Tuc (NGC\,104), M\,3 (NGC\,5272) and M\,13 (NGC\,6205).
Specific values of [Fe/H] for these clusters are chosen close to their spectroscopically
measured values (see, e.g., \citealt{ki03}, \citealt{cbg09b}).  To be specific,
we assume [Fe/H]\,$=-1.55$ for both M\,3 and M\,13, and [Fe/H]\,$=-0.7$ for 47\,Tuc.
(The latest analysis of the detached eclipsing binary in 47\,Tuc favors a
slightly higher metallicity than we have adopted for this GC; see \citealt{brogaard:17}.)
In addition, [$\alpha$/Fe]\,$=0.4$ (with [O/Fe]\,$=0.4$ as well) has been
adopted for all three GCs, as this enhancement seems to be a common characteristic of
such systems (e.g., \citealt{car96}, \citealt{cbg09a}). 
We also assume that the first generations of stars in M\,3 and M\,13 had $Y_{0,1}=0.25$, which
is slightly higher than current Big Bang nucleosynthesis determinations of the
primordial abundance ($Y_P \approx 0.248$; \citealt{ksd11}, \citealt{aaa14}).  In the
case of 47\,Tuc, we adopt $Y_0 =0.257$ given the expectation that the gas
out of which this relatively metal-rich system formed will have a somewhat 
higher He abundance due to chemical evolution effects.

When comparing the observed distributions of HB stars with their corresponding
distributions of synthetic HB stars, we consider the reddening, $E(B-V)$, and the
apparent distance modulus, $(m-M)_V$,
to be additional free parameters, with initial estimates close to literature
values, e.g. \cite[2010, edition]{harris:96}, \cite{sf11}, and \cite{sfd98}.
For 47\,Tuc and M\,13 we have selected 549 and 418 HB stars from the final-pass photometry of
Ata Sarajedini\footnote{\href\protect{http://www.astro.ufl.edu/{$\sim$}ata/public\_hstgc/databases.html}} 
in the {\it HST ACS} filters $F606W$ and $F814W$. Given that M\,3 has a large 
population of RR Lyrae stars, for which we use the $B$ and $V$ photometry of \cite{ccc05},
we have decided to investigate the distribution of HB stars in this cluster on the $(V,B-V)$ plane, for which
195 non-variable HB stars have been selected from Peter Stetson's photometric standard 
fields\footnote{\href\protect{http://www.cadc.hia.nrc-cnrc.gc.ca/en/community/\hfil\break STETSON/standards/}}.

To take into account photometric errors,
we assume that they have normal distributions with the standard deviations for the observed magnitude and color
$\sigma_\mathrm{phot} = 0.002$ for 47\,Tuc and M\,3, and $\sigma_\mathrm{phot} = 0.01$ for M\,13.
To estimate the quality of our fits, we compare the magnitude and color histograms of the observed and synthetic
distributions of HB stars using the Kolmogorov-Smirnov (K-S) test. The observational data
on the histograms are shown with their Poisson error bars. 
Our goal is to obtain reasonable values for the free parameters that result in
the highest possible K-S probabilities. When trying to find the best fits,
we use a repeatable pseudo-random sequence of 6000 numbers in our MC simulations. After the best fits are found, 
we check that the K-S probabilities remain high when we switch to non-repeating pseudo-random number sequences
and we normalize the number of synthetic HB stars to the number of observed HB stars in each GC.

From an analysis similar to that presented in the previous section, with the data used there, and from
the relation $\Delta {\cal M} = {\cal M} - {\cal M}_\mathrm{HB}$, we find that
a variation of $Y_0$ at a constant age should be accompanied by the following change in the RGB mass-loss:
$$
[\Delta (\Delta {\cal M})]_{t_\mathrm{ZAHB}} = [\Delta ({\cal M})]_{t_\mathrm{ZAHB}} - (\Delta {\cal M}_\mathrm{HB})_{t_\mathrm{ZAHB}},
$$ 
where
\begin{align}
& \left(\Delta {\cal M}_\mathrm{HB}\right)_{t_\mathrm{ZAHB}} =
\left[\left(\frac{\partial {\cal M}_\mathrm{HB}}{\partial Y_0}\right)_{\cal M} -
\left(\frac{\partial {\cal M}_\mathrm{HB}}{\partial {\cal M}}\right)_{Y_0} \right. \times \nonumber \\
& \left. \frac{\left(\frac{\partial t_\mathrm{ZAHB}}{\partial Y_0}\right)_{\cal M}}{\left(\frac{\partial t_\mathrm{ZAHB}}{\partial {\cal M}}\right)_{Y_0}}
\right]\Delta Y_0 = -1.403\,\Delta Y_0,
\label{eq:dMZAHBdY}
\end{align}
and
$$
[\Delta ({\cal M})]_{t_\mathrm{ZAHB}} = \left(\frac{\partial {\cal M}}{\partial Y_0}\right)_{t_\mathrm{ZAHB}}\Delta Y_0 = -1.4\,\Delta Y_0,
$$
the last number having been obtained by comparing the values of ${\cal M}$ and $Y_0$ in the T and YA raws of Table~\ref{tab:hb_sims}.
From these estimates, we conclude that stellar populations with different $Y_{0}$ values in the same GC should have
nearly the same mean RGB mass-loss, because $[\Delta(\Delta {\cal M})]_{t_\mathrm{ZAHB}}(\Delta Y_0)\approx 0$, 
provided that their ages are not very different and the mass-loss experienced
by their RGB stars obeys the same law --- which, in our benchmark case that was used to derive 
Equations (\ref{eq:dCdY}-\ref{eq:dMZAHBdY}), is prescribed by the Reimers formula with
$\eta_\mathrm{R} = 0.45$. We will take this conclusion into account in our simulations, giving the preference to
an enhancement of $Y_0$ over its compensating increase of $\Delta {\cal M}$, where it is required. 

The assumptions we have made about the RGB mass-loss can be summarized as follows. 
We do not include any mass-loss in the computations of our basic ZAHB models that, therefore, have ${\cal M}_\mathrm{HB} = {\cal M}$.
The basic ZAHB models are used to generate sets of ZAHB models with ${\cal M}_\mathrm{HB} < {\cal M}$ and their HB evolutionary tracks.
This allows us to vary the mass of our synthetic HB models assuming that the difference $\Delta {\cal M} = {\cal M} - {\cal M}_\mathrm{HB}$
represents the mass lost on the RGB. By adjusting $\Delta {\cal M}$ and other parameters, we fit the observed distribution of HB stars.
Then the adjusted mean values of $\Delta {\cal M}$ can be compared with those calculated for the benchmark ZAHB models 
that have the corresponding values of [Fe/H], $Y_0$ and ${\cal M}$ (Fig.~\ref{fig:f5}) to see
how close they are and, hence, how well the mass-loss law adopted for the benchmark ZAHB models reproduces the mean values of $\Delta {\cal M}$
used in our HB population synthesis simulations.

\subsubsection{47\,Tuc}

\begin{center}
\begin{deluxetable*}{cclllll}
\tablecolumns{9}
\tablecaption{Fitted GC Parameters from our HB Population Synthesis Simulations\label{tab:fit_par}}
\tablehead{
  \colhead{GC}&\colhead{Population ($i$)}&\colhead{ $f_i$}&\colhead{ $Y_{0,i}$}
  &\colhead{${\cal M}_i/{\cal M}_\odot$}& \colhead{$\Delta {\cal M}_i/{\cal M}_\odot$}&
  \colhead{ $\sigma_i/{\cal M}_\odot$}}
\startdata
NGC\,104 (47\,Tuc) & $1$ & $0.21$ & $0.257$ & $0.875$ & $0.197$ & $0.009$ \\
\ldots & $2$ & $0.37$ & $0.27$ & $0.85$ & $0.184$ & $0.009$ \\
\ldots & $3$ & $0.42$ & $0.287$ & $0.825$ & $0.190$ & $0.009$ \\
\hline
NGC\,6205 (M\,13) & $1$ & $0.45$ & $0.25$ & $0.8$ & $0.193$ & $0.010$ \tstrut \\
\ldots & $2$ & $0.22$ & $0.285$ & $0.75$ & $0.207$ & $0.010$ \\
\ldots & $3$ & $0.33$ & $0.33$ & $0.695$ & $0.2165$ & $0.010$ \\
\hline
NGC\,5272 (M\,3) & $1$ & $0.45$ & $0.25$ & $0.8$ & $0.126$ & $0.008$ \tstrut \\
\ldots & $2$ & $0.10$ & $0.255$ & $0.792$ & $0.157$ & $0.009$ \\
\ldots & $3$ & $0.45$ & $0.26$ & $0.785$ & $0.160$ & $0.010$ 
\enddata
\end{deluxetable*}
\end{center}

\begin{figure*}[t]
\plotone{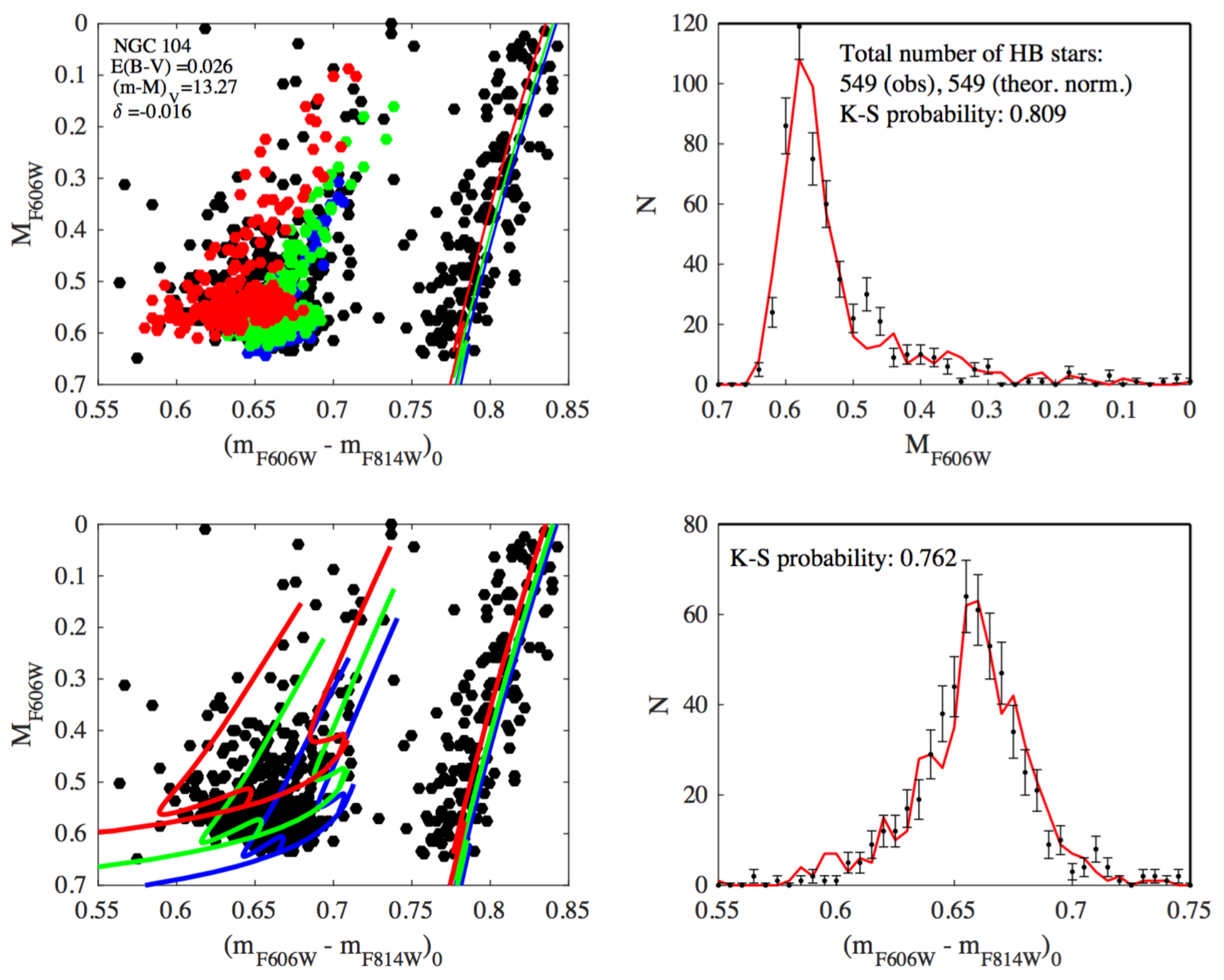}
\caption{The HB population synthesis model of 47\,Tuc 
         ([Fe/H]\,$=-0.7$, [$\alpha$/Fe]\,$=0.4$, and [O/Fe]\,$=0.4$). See text.} 
\label{fig:f6}
\end{figure*}

Fig.~\ref{fig:f6} and the first three raws of Table~\ref{tab:fit_par} show our best fit to the observed
distribution of HB stars in the GC 47\,Tuc and the values of the best-fit parameters. Because of its relatively high metallicity,
all of the HB stars in 47\,Tuc are clustered at the red side of the HB around $(m_\mathrm{F606W}-m_\mathrm{F814W})_0\approx 0.65$.
It is impossible to fit all of them using HB evolutionary tracks with the same initial value of $Y_{0,1} = 0.257$ 
for the stellar population that formed from pristine gas (the blue tracks
in the bottom-left panel of Fig.~\ref{fig:f6} with the ZAHB masses $0.775\,{\cal M}_\odot$ (right) and
$0.66\,{\cal M}_\odot$ (left)). The spread of the observed HB stars above the blue tracks is direct observational evidence that
a significant fraction of them have enhanced He abundances. The wedge-like shape of the distribution of the 47\,Tuc HB
stars can be embraced if we add HB evolutionary tracks with $Y_{0,2} = 0.27$ (the green tracks with the ZAHB masses 
$0.775\,{\cal M}_\odot$ (right) and $0.64\,{\cal M}_\odot$ (left)) and $Y_{0,3} = 0.287$ 
(the red tracks with the ZAHB masses $0.775\,{\cal M}_\odot$ (right) and $0.63\,{\cal M}_\odot$ (left)) representing, respectively,
the intermediate and extreme stellar populations in 47\,Tuc.

The blue, green, and red circles in the top-left panel of Fig.~\ref{fig:f6} are the results of
our HB population synthesis simulations that have used the same sets of HB evolutionary tracks from which the pairs of the blue, green,
and red tracks shown in the bottom-left panel were selected. The positions of the synthetic HB stars have been interpolated
in $Y_0$ assuming that this is a random quantity uniformly distributed between $Y_{0,1}$ and $Y_{0,3}$.
The comparison of the observed and synthetic distributions of HB stars in the magnitude and color histograms
are shown in the top- and bottom-right panels along with their corresponding K-S probabilities.
The K-S probabilities are very sensitive to even small changes in $(m-M)_V$, therefore our estimate of the distance 
modulus for 47\,Tuc $(m-M)_V = 13.27$, which is used for its age estimate in Section~\ref{sec:age},
is quite a robust result (for the assumed chemical abundances). 

The fitted reddening depends on the assumed age of the HB stars; therefore
its estimate should be iterated with the age estimate. We have done one such iteration for 47\,Tuc.
First, we have taken the age of 11.75 Gyr for 47\,Tuc from \citetalias{vbl13} and generated HB evolutionary tracks with
the initial masses $0.9\,{\cal M}_\odot$, $0.877\,{\cal M}_\odot$, and $0.848\,{\cal M}_\odot$ for
the normal, intermediate, and extreme stellar populations having the aforementioned values of $Y_{0,i}$.
Their respective ZAHB ages of 11.58 Gyr, 11.60 Gyr, and 11.61 Gyr are all sufficiently close to 11.75 Gyr.
With these input data, our HB population synthesis tool has provided the best match between the observed and synthetic
distributions of 47\,Tuc HB stars for $E(B-V)=0.015$ and $(m-M)_V=13.28$, while the fitted sets of
$(f_i,\Delta{\cal M}_i,\sigma_i)$ have been very similar to those given in Table~\ref{tab:fit_par}.
However, the age of 47\,Tuc estimated using isochrones with this distance modulus turns out to be close to 13 Gyr,
which is inconsistent with the younger age of the HB models. Also, the inferred reddening is too small
compared with dust-map values\footnote{\href\protect{http://irsa.ipac.caltech.edu/applications/DUST/}} of
$0.0275$ \citep{sf11} and $0.0320$ \citep{sfd98}, though this might be the
consequence of small errors in, e.g., the adopted color--$T_{\rm eff}$ relations or the
photometric zero-points.

Therefore, for the second step, we generated HB evolutionary models with the lower initial masses
$0.875\,{\cal M}_\odot$, $0.85\,{\cal M}_\odot$, and $0.825\,{\cal M}_\odot$
that have the older ages, in turn, of 12.80 Gyr, 12.96 Gyr, and 12.79 Gyr for the same
three stellar populations. These input data have been used to obtain the results presented in Fig.~\ref{fig:f6}
and in the first three raws of Table~\ref{tab:fit_par}. Our final estimates of the reddening and distance modulus
for 47\,Tuc are $E(B-V)=0.026$ and $(m-M)_V=13.27$.  (As alluded to above,
this reddening determination could be in error by as much as $\sim 0.02$\ mag
because the predicted $T_{\rm eff}$\ and color scales, as well as the adopted
metallicity and the photometric data, all involve some uncertainties.  The main points
of this exercise is to illustrate the relationship between the cluster age and the value of
$E(B-V)$\ that is implied by our HB simulations.  The inferred value of $(m-M)_V$ is clearly 
much more secure, as is the cluster age that is derived in Section~\ref{sec:age},
because they do not depend on the adopted reddening or the isochrone colors.)
As explained in the previous section, it is not surprising,
indeed it is expected, that the fitted RGB mass-loss is nearly the same for the populations with different He abundances.
Furthermore, our results are in very good agreement with both the firm lower
limit $\Delta {\cal M} \ge 0.17\,{\cal M}_\odot$ and the maximum value of
$\Delta {\cal M} = 0.21\,{\cal M}_\odot$ obtained for 47\,Tuc RGB stars by \cite{salaris:16}.

\subsubsection{M\,13}

\begin{figure*}[t]
\plotone{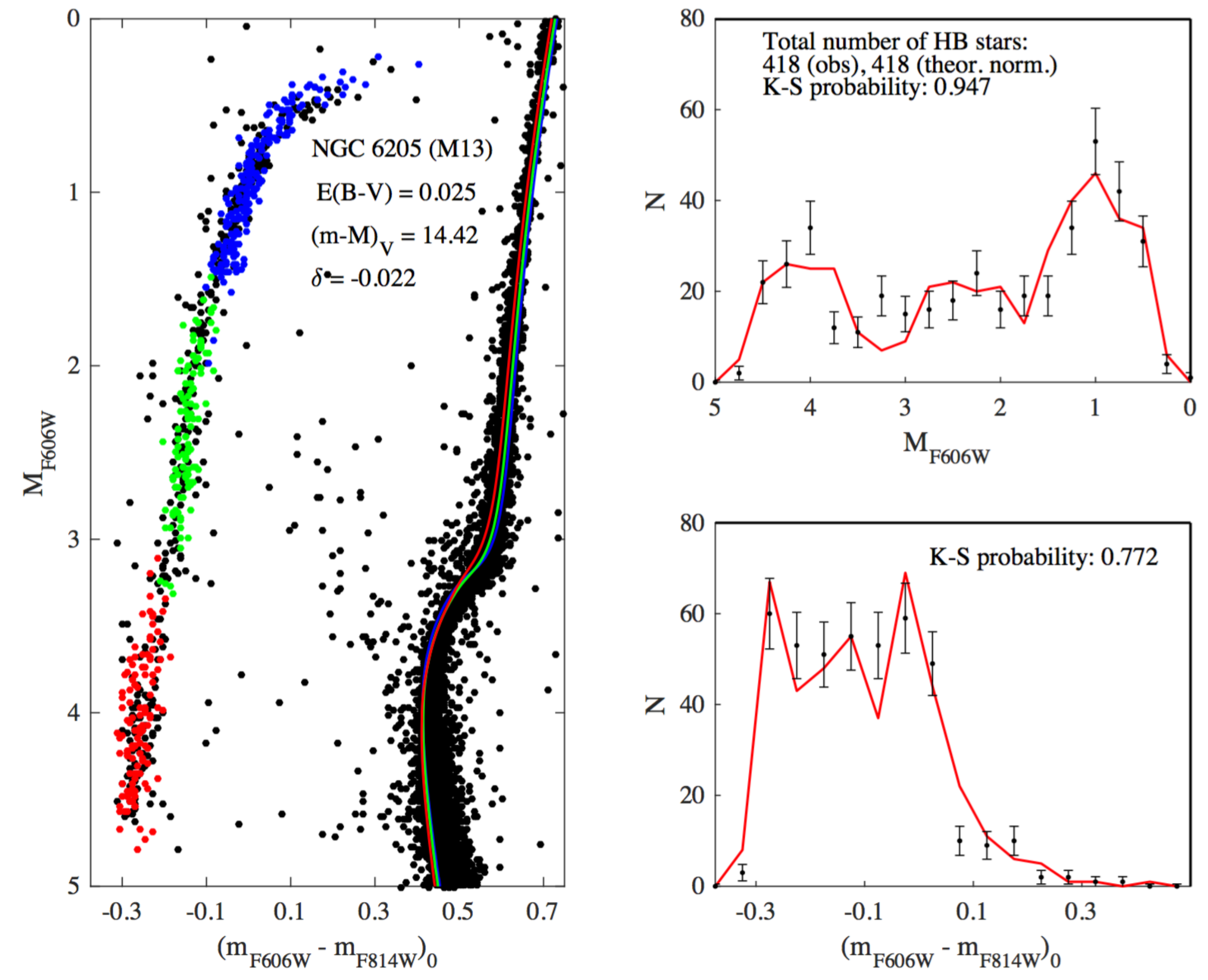}
\caption{The HB population synthesis model of M\,13
         ([Fe/H]\,$=-1.55$, [$\alpha$/Fe]\,$=0.4$, and [O/Fe]\,$=0.4$). See text.}
\label{fig:f7}
\end{figure*}

When the metallicity [Fe/H] decreases, HB stars move to the blue side of the HB (the top-left
panel of Fig.~\ref{fig:f4}).  In addition, at intermediate and low
metallicities, the effect of varying $Y_0$ on the mean color of a population of
HB stars (Equation \ref{eq:dcdy}), and on their mean magnitude when they are
located to the left of the HB ``knee'' at $(m_\mathrm{F606W}-m_\mathrm{F814W})_0 \la -0.05$,
becomes more prominent than at a higher metallicity. Therefore, the HB morphology in the GC M\,13 (the blue, green, and red circles in the left
panel of Fig.~\ref{fig:f7}) is similar to that of 47\,Tuc, except that the normal,
intermediate, and extreme stellar populations have unfolded along the entire HB due to the lower value of [Fe/H] for M\,13.
The large color split is the reason why we do not interpolate positions of HB stars
in GCs with intermediate and low metallicities in $Y_0$ between the three populations, but instead plot them separately for each
of the three values of $Y_{0,i}$.
Another difference between 47\,Tuc and M\,13 is that the latter appears to have a wider range of the helium abundance
$\Delta Y_0 = Y_{0,3} - Y_{0,1} = 0.08$ (the second three raws of Table~\ref{tab:fit_par}) versus $\Delta Y_0 = 0.03$ in 47\,Tuc,
which is required to obtain the best fit of the observed and synthetic distributions of
its HB stars (the right panels in Fig.~\ref{fig:f7}), without making the assumption of a stronger RGB mass-loss
for the extreme population of stars in M\,13. 

Similar to \cite{dalessandro:13}, we have found that the extremely blue extent of
the HB in M\,13 can be explained with a lower value of $Y_{0,3}\approx 0.3$, but only by assuming a significantly higher value of
$\Delta {\cal M}_3 \approx 0.26\,{\cal M}_\odot$, as compared to $\Delta {\cal M}_1$ and $\Delta {\cal M}_2$ (Table~\ref{tab:fit_par}).
As explained in the text associated with Equation (\ref{eq:dMZAHBdY}), we think this is an improper solution, unless independent evidence is
found in favour of higher RGB mass loss by the second-generation stars in GCs. Also, similar to \cite{tailo:16}, who studied   
the GC $\omega$\,Cen, we have used the optical CMD of M\,13 to simulate the distribution of its HB stars. Although, in both cases,
ZAHBs and HB evolutionary tracks converge along the HB blue tails for different values of $Y_0$, it is the extension of
the blue tail towards fainter absolute magnitudes (rather than its width) and its almost vertical shape that allow us to estimate with high accuracy the maximum
value of $Y_{0,i}$ and reddening, respectively.

\subsubsection{M\,3}
\label{sec:M3}

\begin{figure*}[t]
\plotone{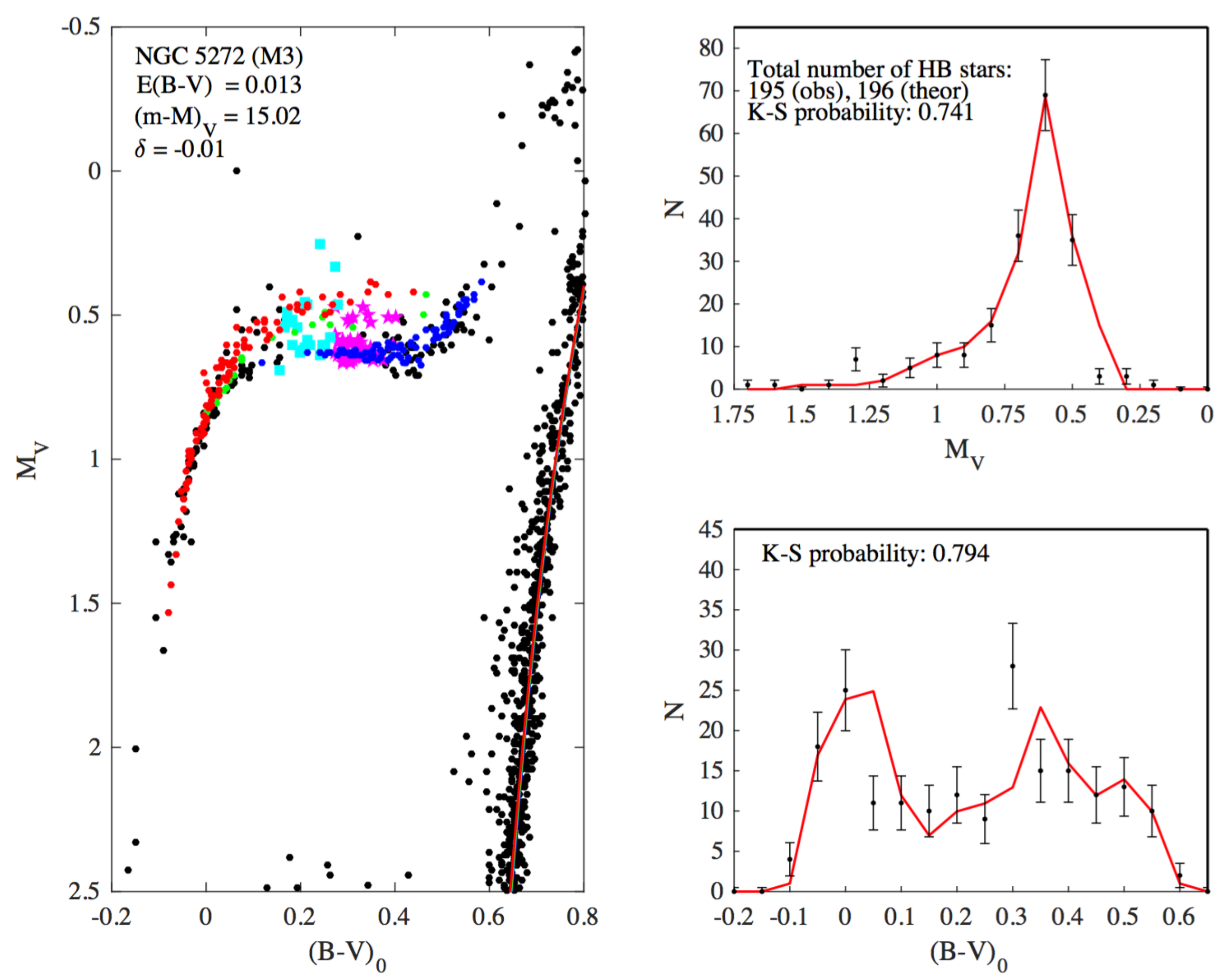}
\caption{The HB population synthesis model of M\,3
         ([Fe/H]\,$=-1.55$, [$\alpha$/Fe]\,$=0.4$, and [O/Fe]\,$=0.4$). The square and star symbols in the left panel
         represent the RRc and RRab Lyrae stars from \cite{ccc05}. See text.}
\label{fig:f8}
\end{figure*}

The HB morphology in the GC M\,3 is very different from the one in M\,13, in spite of the fact that,
like M\,13 and 47\,Tuc, M\,3 is one the most massive GCs (e.g., see Fig.~39 in VBLC13). Our best fit to the distribution of
its HB stars assumes that the star-to-star variations of the He abundance in M\,3 is close to $\Delta Y_0 = 0.01$
(see Fig.~\ref{fig:f8} and the last three raws of Table~\ref{tab:fit_par}). This conclusion is supported by the less extreme abundance
anomalies of the $p$-capture elements in M\,3 stars compared to 
those observed in M\,13 \citep{sneden:04}. It is also surprising that, for our best fit, the RGB stars in M\,3 are required
to have lost less mass, $\Delta {\cal M}_1 = 0.13\,{\cal M}_\odot$ versus $\Delta {\cal M}_1 = 0.20\,{\cal M}_\odot$ in
both M\,13 and 47\,Tuc. The direct signature of this is that the HB in M\,3 extends to $(m_\mathrm{F606W}-m_\mathrm{F814W})_0 \approx 0.6$
(the left panel of Fig.~\ref{fig:f8}), while the normal population of HB stars in M\,13 with 
the same initial values of ${\cal M}_1 = 0.8\,{\cal M}_\odot$ and $Y_{0,1} = 0.25$ does not extend to redder colors than
$(m_\mathrm{F606W}-m_\mathrm{F814W})_0 \approx 0.4$. In fact, when we apply Equation (\ref{eq:dcdt}) with the difference 
in the mean color between the normal ($Y_{0,1}=0.25$) HB populations in M\,13 and M\,3, it gives us the age difference of
2.6 Gyr between the two GCs, i.e. M\,3 has to be 2.6 Gyr younger if its first-generation RGB stars have lost as much mass as
their M\,13 counterparts. As reported in the next section (also see VBLC13), the ages of these
two GCs appear to differ by $\lta 0.3$\ Gyr, which is very much less than the difference in age that would be
needed to explain their very different HB populations if age were the sole second parameter.  We note
that our conclusion concerning the lower RGB mass-loss efficiency
in M\,3 agrees with the findings reported by \cite{mcdonald:15}.

\begin{figure}[b]
\plotone{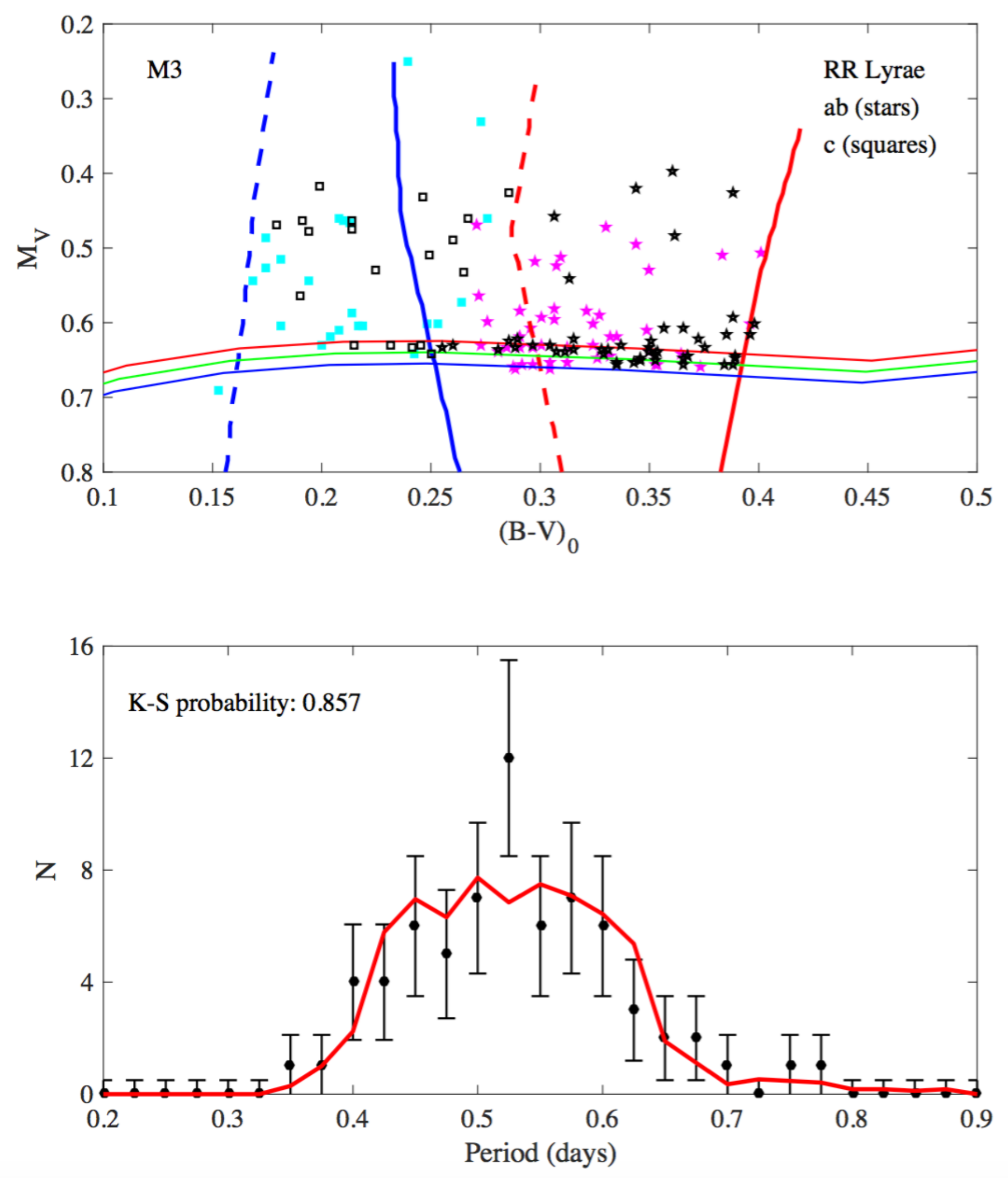}
\caption{Top panel: the observed (filled symbols) and simulated (open symbols) distributions of RR Lyrae stars in M\,3.
         Bottom panel: their observed (black circles are the data of \citealt{ccc05} plotted with their corresponding Poisson error bars) 
         and simulated (red curve) period distributions. The classical relation $\log P_\mathrm{ab} = \log P_\mathrm{c} + 0.127$ has been  used 
         here for the period fundamentalization \citep{mcb15}.
         }
\label{fig:f9}
\end{figure}

We have used Equations (\ref{eq:pab}) and (\ref{eq:pc}) to calculate the pulsation periods for those of our
synthetic HB stars in M\,3 that are inside the instability strip (the top panel of Fig.~\ref{fig:f9}) and
we have compared them with the observed period distribution of RR Lyrae stars in M\,3 from \cite{ccc05} (the bottom panel).
When the periods of the first-overtone pulsators are fundamentalized using the classical relation between
the periods of $ab$- and $c$-type variables \citep{mcb15}, there is no need to assume a very small dispersion of
the RGB mass-loss, as \cite{caloi:08} did, to fit the observed period distribution of RR Lyrae stars in M\,3 using our HB models.
Note that we still get a relatively high K-S probability ($P_\mathrm{K-S} = 0.36$) for the assumption that the observed and synthetic 
periods of M\,3 RR Lyrae stars
are drawn from the same distribution when we separate the fundamental and first-overtone pulsators.
Encouragingly, our predictions for
the mean period of the RRab Lyrae stars ($\langle P_\mathrm{ab}\rangle = 0.558$ days), the HB type (${\cal L} = 0.25$), and
the proportions of the blue, variable, and red HB stars
(${\cal B}:{\cal V}:{\cal R} = 46:33:21$) in M\,3 are very close 
to those reported by \cite{catelan:04}.

A closer inspection of the left-hand panel of Fig~\ref{fig:f8} gives us a hint that we should adopt
a slightly longer distance modulus for M\,3, probably $(m-M)_V = 15.04$, in which case our ZAHB models would provide a somewhat
better fit to the bottom of the observed HB, especially at the reddest colors. The maximum K-S probabilities that we have been able
to obtain with this value of the distance modulus are 0.36 and 0.84 for the magnitude and color histograms, respectively,
while the other characteristics of the fitted synthetic HB have not changed by more than a few percent.
This exercise indicates that the typical uncertainty of a distance modulus that is estimated using our method
is $\sim 0.02$\ mag, which translates into $\sim 1$\% uncertainty in the actual distance.

\section{Cluster Ages}
\label{sec:age}

Having determined the basic properties of 47\,Tuc, M\,3, and M\,13 from comparisons of our
synthetic HBs with the observed HB populations, their ages can be derived
simply by identifying which isochrone for the adopted chemical
abundances provides the best fit to the observed turnoff (TO) luminosity.  The
most objective way of doing this is to follow the procedure outlined by
VBLC13, wherein isochrones for different ages are each shifted, as
necessary, to match the observed TO color until one is found that simultaneously
reproduces the location of the cluster subgiants just past the TO.  Because the shapes of
isochrones near the TO are essentially independent of age, any isochrone (within
reason) will provide exactly the same fit to the observed TO morphology {\it
provided} that the correct distance modulus for that isochrone is assumed.
Conversely, for a given value of $(m-M)_V$, only one isochrone can provide a
simultaneous fit to the TO color and to the stars immediately above (and below)
the TO.  (A very limited color range is considered in deriving the cluster age,
rather than the entire color-magnitude diagram (CMD), because errors in, e.g.,
the adopted color transformations, the atmospheric boundary condition, and the
treatment of convection can easily affect the location of the red-giant branch
(RGB) relative to the turnoff as well as the predicted MS,
SGB, and RGB slopes.  Reference may be made to the VandenBerg et
al.~study for a full discussion of this method.) We have used the Victoria-Regina isochrones
instead of generating and using MESA isochrones because, unlike the Victoria code, the MESA code has not
implemented yet properly an algorithm taking into account the turbulent extra mixing that
reduces the efficiency of atomic diffusion in stellar envelopes.

\begin{figure}[t]
\plotone{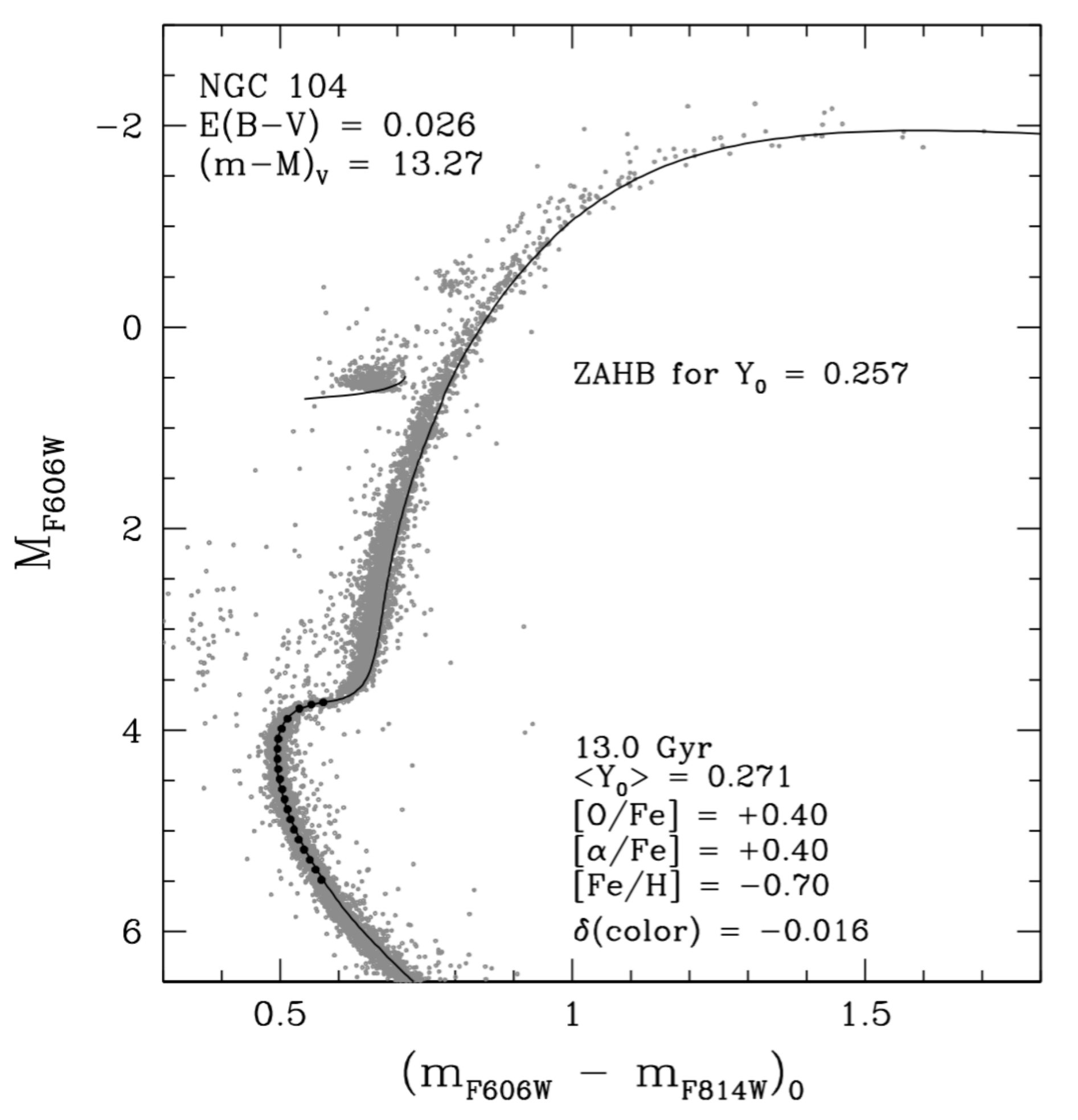}
\caption{Fit of a 13.0 Gyr isochrone for the indicated chemical abundances to
the turnoff and subgiant photometry of 47\,Tuc obtained by \citet{sbc07},
assuming $E(B-V) = 0.026$ and $(m-M)_V = 13.27$.  (In the $F606W$ magnitude,
$(m-M)_{F606W} = (m-M)_V - 0.246E(B-V)$; see \citealt{cv14}.)  The adopted
helium abundance corresponds to the mean initial He content, $\langle Y_0\rangle$, that
is implied by our HB simulations, which indicate that $\Delta Y_0 \approx 0.03$;
see Fig.~\ref{fig:f6}.  The solid curve near the bottom of the distribution of the cluster
HB stars represents a ZAHB for $Y_0 = 0.257$, which is assumed to be initial He
abundance of the lowest-$Y$ population in this cluster.  The small filled
circles near the turnoff indicate the location of the median fiducial sequence
through the photometric data.  To reproduce the TO color, the isochrone that
provides the best fit to the subgiant stars had to be adjusted by 0.016 mag to
the blue.}
\label{fig:f10}
\end{figure}

According to our HB simulations, stars in 47\,Tuc had $0.257 \le Y_0 \le 0.287$,
with a mean value $\langle Y_0\rangle = 0.271$.  We have therefore generated
a set of isochrones from the grids made available by \citet{vbf14} for the mean
He abundance and for metal abundances corresponding to [Fe/H] $= -0.70$ and
[$\alpha$/Fe] $= +0.40$.  As shown in Figure~\ref{fig:f10}, the application
of the isochrone-fitting procedure described above to the {\it HST} observations
that were obtained by \citet{sbc07} yields an age
of 13.0 Gyr.  The isochrone for this age clearly provides an excellent fit to
the cluster SGB just past the TO, which is the primary age indicator, once it has been shifted
horizontally in color by $-0.016$ mag in order to match the observed TO color.
Note that the small black filled circles represent only part of the median
fiducial sequence (just for the vicinity of the TO) that was obtained using the
methods discussed in Section 3.3 of Paper~I. Except along the SGB, median
colors were derived for stars that were sorted into 0.1 mag bins in $F606W$.
Each bin contained up to a maximum of 100 stars with the smallest photometric
uncertainties.  Because the transition from the TO to the RGB is so flat in
47\,Tuc, approximately 3000 of the best observed subgiants (not plotted) were
first sorted into 0.02 mag color bins and then the median magnitude in each
bin was evaluated.  Using this completely objective method resulted in very
smooth fiducial sequences, not only for 47\,Tuc, but also for the other two
clusters considered in this investigation. 

An age of 13.0 Gyr is 1.25 Gyr older than the age found by VBLC13
because of our adoption of a smaller value of $(m-M)_V$ by 0.08 mag, due in part
to the assumption of a higher [Fe/H] value, and a significantly reduced oxygen
abundance. (Rather than try to match a ZAHB to the HB population of 47\,Tuc,
which depends quite sensitively on color uncertainties, VandenBerg et al.~decided
to adopt the distance modulus that was derived by \citealt{thompson:10},
$(m-M)_V = 13.35 \pm 0.08$, from their analysis of the eclipsing binary member known
as V69.  The problem with fitting ZAHBs to the faintest HB stars in relatively
metal-rich GCs is that they follow a curve or sloped line, rather than a nearly
horizontal line as in the case of more metal-deficient clusters like M\,3.  In
47\,Tuc and other GCs of similar or higher metallicity, distance moduli are much
better constrained by comparisons of simulated HB populations with the observed
distributions of stars.)  With regard to the oxygen abundance difference: the
isochrones computed by \citet{vbf14}, which are used in the present work, adopted
the solar abundances reported by \citet{ags09} as the reference mixture whereas
the computations employed by VBLC13 assumed the solar mix of heavy elements
given by \citet{gs98}.  As a consequence, the $\alpha$-element enhanced models
plotted in Fig.~\ref{fig:f10} were generated for a lower absolute O abundance
by $\Delta\log N_{\rm O} = 0.18$ dex than those considered in the 2013 study.
The differences in the apparent distance modulus and the adopted metallicity
(primarily the oxygen abundance) are entirely responsible for the different age
determinations.  (Because both sets of stellar models were generated using
exactly the same evolutionary code; the different age determinations can only
be a consequence of differences in the assumed distances and chemical abundances.)

\begin{figure}[t]
\plotone{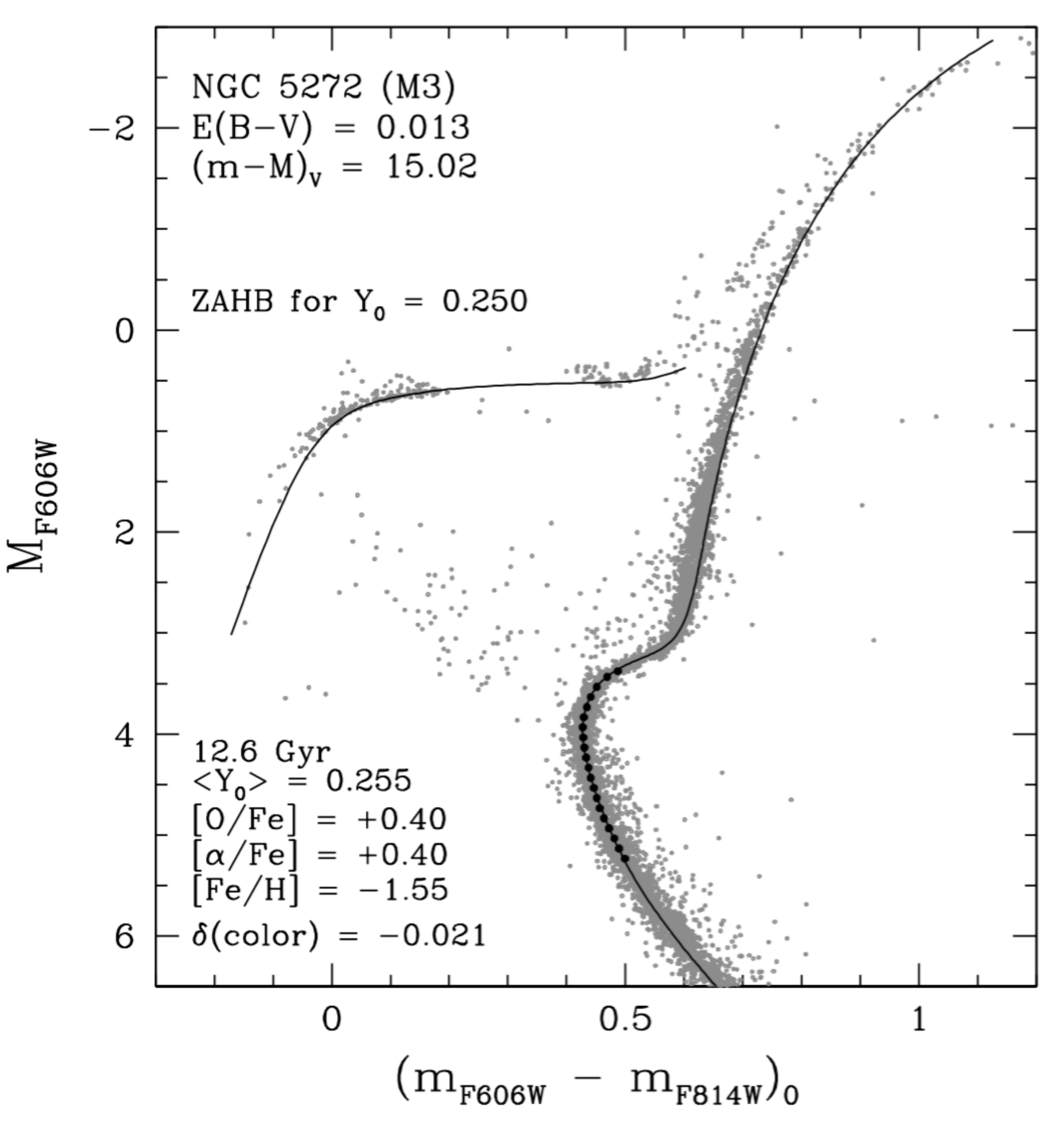}
\caption{Similar to the previous figure, except that a 12.6 Gyr isochrone for
the indicated metal abundances and the mean value of $Y$ found in our HB
simulations, along with a ZAHB for the presumed lowest-$Y$ population, has been
fitted to observations of M\,3 (from \citealt{sbc07}).  As shown in Fig.~\ref{fig:f8}, a
synthetic HB population for the adopted cluster parameters provides a very good
match to the observed HB if $\Delta Y_0 \sim 0.01$.  Approximately 1500
subgiants (not shown) with the smallest photometric errors were used to
determine the median points along the SGB.}
\label{fig:f11}
\end{figure}

Figure~\ref{fig:f11} illustrates the fit of a ZAHB for $Y_0 = 0.250$, [Fe/H] 
$= -1.55$, and [$\alpha$/Fe] $= +0.4$ to the lower bound of the distribution of
HB stars in M\,3, assuming the value of $(m-M)_V$ favored by our HB
simulations.  This is 0.02 mag smaller than the ZAHB-based distance modulus
derived in Paper I.  A small difference can be expected because the modeling of
an entire HB involves many assumptions that are constrained via standard Monte
Carlo methods while the fit of a ZAHB for a single value of $Y_0$ will, in
general, apply only to the faintest HB stars in a limited color range.  As shown in \S~4.2,
our synthetic HB provides a good
match to the observed HB population if M\,3 has $0.250 \le Y_0 \le 0.260$, with
a mean value $\langle Y_0\rangle = 0.255$.  A 12.6 Gyr isochrone for this He
abundance and the aforementioned values of [Fe/H] and [$\alpha$/Fe] provides a
very good fit to the median fiducial for cluster stars that lie close to the TO
(see Fig.~\ref{fig:f11}).  To obtain this fit to the turnoff morphology, the predicted
isochrone colors had to be adjusted by $-0.021$ mag, as indicated.  It is worth
mentioning that the age reported by VBLC13, 11.75 Gyr, is younger than
our determination primarily because of their assumption of a higher oxygen
abundance by $\Delta\log N_{\rm O}$ = 0.29 dex, which takes into account their
adoption of a higher metallicity ([Fe/H] $= -1.50$), a higher value of [O/Fe]
($= 0.50$), and the differences in the adopted solar mixtures that we have
already described.

\begin{figure}[t]
\plotone{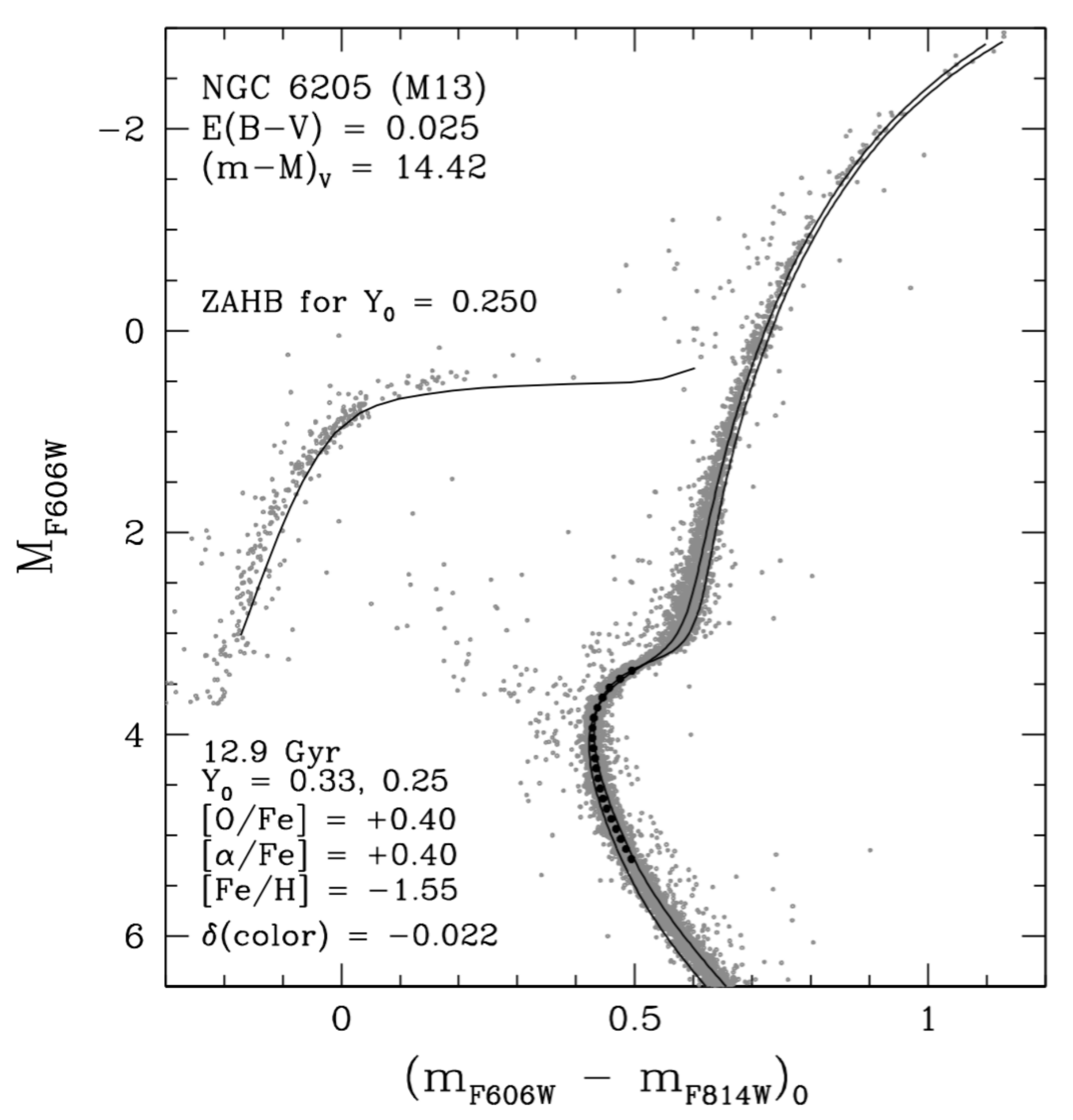}
\caption{Similar to the previous figure, except that 12.9 Gyr isochrones
for $Y_0 = 0.33$ and 0.25 (in the direction from left to right) and a ZAHB for
the presumed lowest-$Y$ population are compared with observations of M\,13
(from \citealt{sbc07}).  As shown in Fig.~\ref{fig:f7}, a synthetic HB population for the
adopted cluster parameters is able to reproduce the observed distribution of HB
stars if $\Delta Y_0 \approx 0.08$.  Approximately 2200 subgiants (not shown)
with the smallest photometric errors were used to determine the median points
along the SGB.} 
\label{fig:f12}
\end{figure}

Figure~\ref{fig:f12} depicts a very similar fit of stellar models for the
same chemical abundances to the CMD of M\,13, which is predicted to be about 0.3
Gyr older than M\,3 if the reddening and apparent distance modulus that are
specified in the top left-hand corner are assumed.  In this case, we have opted
to plot isochrones for $Y_0 = 0.250$ and 0.330 that span the full range in He
abundance that seems to be necessary to explain the observed HB (in particular,
its extended blue tail) rather than a single isochrone for the mean value
$\langle Y_0\rangle = 0.287$ that is implied by our synthetic HB.  In order
for a $Y_0 = 0.287$ isochrone to match the median cluster fiducial for near
turnoff stars, a color shift of $-0.022$ mag was needed.  The same offset was
applied to the isochrones for $Y_0 = 0.25$ and 0.33 so that their relative CMD
locations are maintained.  Thus, we see that the difference in color between
these two isochrones, at a fixed magnitude along the MS, is $\lta$\ the
observed MS width.  In addition, the beginning of their SGB segments are
essentially coincident, just as theory predicts for metal-poor isochrones of
the same age but different $Y$ (see \citealt{vcs12}, \citealt{vbf14}).  As in
the case of M\,3, the younger age that was obtained by VBLC13, 12.0 Gyr,
is mostly due to their assumption of a higher oxygen abundance.

The main purpose of this section is to evaluate the cluster ages when the
helium and metal abundances, the reddenings, and the distance moduli that have
been adopted in, or derived from, our HB simulations are assumed.  It is clear
from the above discussion that ages depend quite critically on the abundance
of oxygen, which is generally derived for giants, rather than dwarf stars, in
globular clusters due to their large distances (see, e.g., \citealt{cm05},
\citealt{km08}).  Such studies are complicated by the presence of the O--Na
anticorrelation (see \citealt{cbg09a}), the occurrence of deep mixing (e.g.,
\citealt{dv03}), and the uncertainties associated with the $\teff$\ and [Fe/H]
scales.  However, the CN-weak populations in these systems appear to be
chemically very similar to halo field stars (see the discussion by \citealt{cohen:05},
for which recent non-LTE abundance analyses generally find [O/Fe]
$= 0.5$--0.6 for [Fe/H] values $\lta -0.8$ (e.g., \citealt{rmc12},
\citealt{zmy16}).  Thus, it is quite possible, if not probable, that our age
determinations are on the high side, since any increase in the assumed O
abundance will cause a reduction in the age at a given TO luminosity.  (In
fact, if 47\,Tuc has [Fe/H] $\le -0.70$, it would be difficult to obtain a satisfactory
match to the properties of the eclipsing binary, V69, unless it has [O/Fe] $\gta 0.6$; 
see \citealt{brogaard:17}.)

\begin{figure}[ht]
\plotone{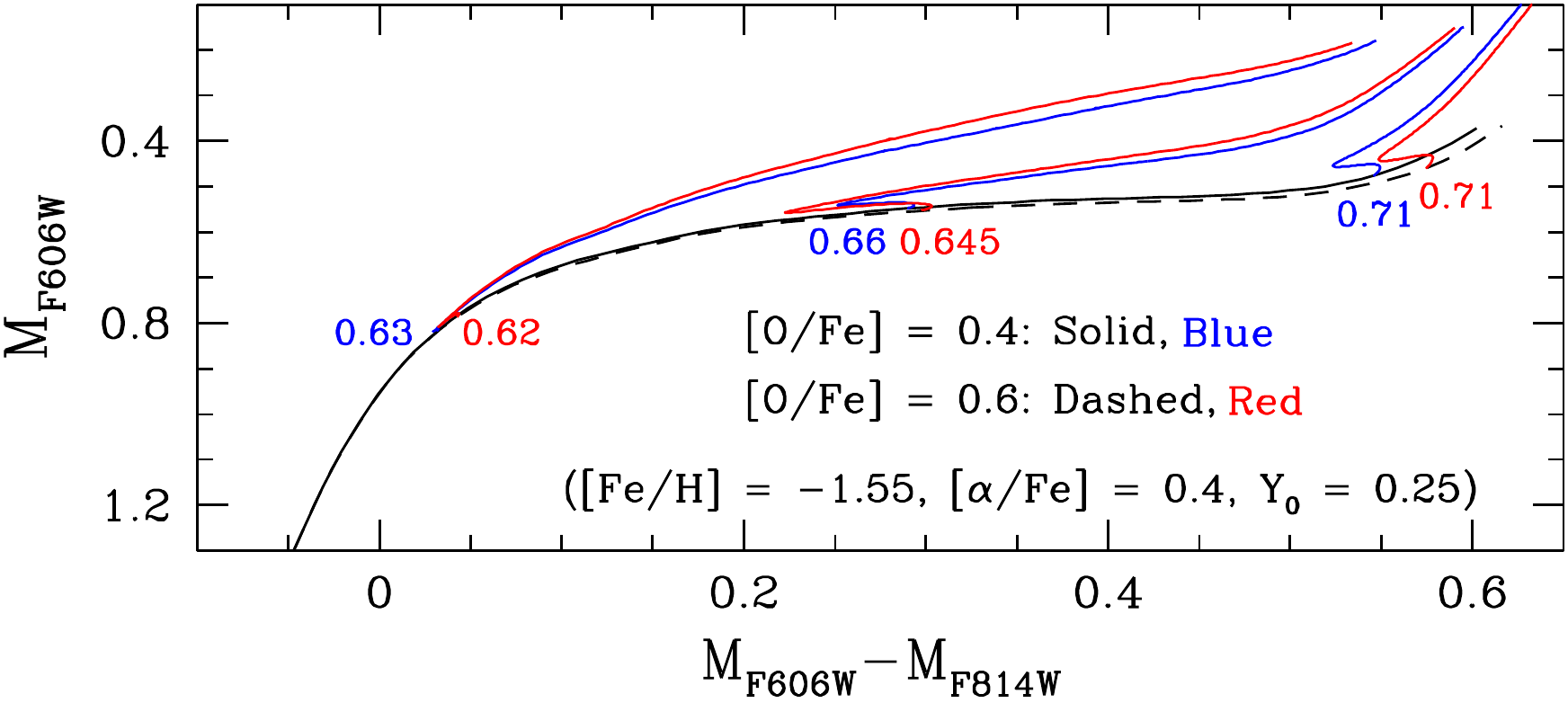}
\caption{Comparison of ZAHB loci (in black) and selected evolutionary tracks
(in blue or red) for the indicated chemical abundances to illustrate the effects
of increasing the assumed value of [O/Fe] by 0.2 dex.  Masses in solar units are
specified in blue, for [O/Fe] $= 0.4$, or in red, for [O/Fe] $= 0.6$, near the
ZAHB locations of the respective tracks.}
\label{fig:f13}
\end{figure}

However, this uncertainty should not affect the results based on our
HB simulations very much.  As shown in Figure~\ref{fig:f13}, the effect of
higher oxygen on a ZAHB is to stretch it to somewhat redder colors and fainter
magnitudes, especially at the red end.  (The differences are more pronounced at
lower [Fe/H] values than we have considered in this investigation.)  Thus, HB
models for [O/Fe] $= 0.6$ would predict essentially the same distance moduli as
those for [O/Fe] $= 0.4$.  Increasing the oxygen abundance by 0.2 dex does have
minor consequences for the predicted masses along a ZAHB, since this leads to
the displacement of a fixed mass to somewhat cooler temperatures and hence
redder colors (see Fig.~\ref{fig:f13}).  In addition, core He-burning tracks
at a given color are predicted to have somewhat longer, post-ZAHB blue loops if
O is enhanced.  For instance, the ZAHB location of a model for
$0.645\,\msol$\ and [O/Fe] $= 0.6$ is slightly redder than that of the
$0.66\,\msol$\ model with [O/Fe] $= 0.4$, but the evolutionary track for the
former extends to bluer colors before reversing direction than the track for
the latter.  However, because the HB is far more dependent on $Y$ than on
[O/Fe] (see Table~\ref{tab:hb_sims} and the next section), it can be expected
that He abundance variations that are derived from HB simulations will be quite
similar regardless of whether they are based on tracks for [O/Fe] $= 0.4$ or 0.6.

\section{RR Lyrae Periods}
\label{sec:rrl}

The distance moduli derived in this study are based primarily on the
evolutionary characteristics of our HB models.  An important consistency check
of these distances is afforded by the periods of individual RR Lyrae stars in
clusters that contain such variables.  If their mean magnitudes and colors
(i.e., the properties of equivalent ``static stars") have been determined,
interpolations can be performed within a relevant grid of HB tracks that is
overlaid onto the observed HB populations to derive the luminosities,
temperatures, and masses of the variables.  These quantities, along with the
value of $Z$ that was assumed in the model computations, can then be used to
predict the periods, in units of days, of the fundamental ($ab$-type) and
first-overtone ($c$-type) pulsators using equations (\ref{eq:pab}) and (\ref{eq:pc}).
If the HB tracks provide good representations of the evolutionary paths that are
followed by cluster stars, and the adopted distances and chemical abundances
are reasonably accurate, the periods given by the above equations should 
reproduce the observed periods of individual RR Lyrae quite well.  Indeed, this
was found to be the case in Paper I, which analyzed the variable (and
non-variable) HB stars in M\,3, M\,15, and M\,92.  

Since the RR Lyrae stars in M\,3 have already been considered in Paper I, only
a brief summary of those findings are provided in the next section, along with
some discussion of the impact of a slightly decreased distance modulus, which
is favored by our comparisons of synthetic and observed HB populations.   This
will be followed by an examination of how well our models are able to explain
the observed periods of the M\,13 and 47\,Tuc RR Lyrae.

\subsection{M\,3}
\label{subsec:m3rr}

Paper I showed that periods calculated using equations (\ref{eq:pab}) and (\ref{eq:pc}) for the 
predicted properties of M\,3 variables from HB models for $Y_0 = 0.25$, [Fe/H]
$= -1.55$, and [$\alpha$/Fe] $= 0.4$ agree with the measured periods to well
within the uncertainties of the computed values of $\log\,P$.  In fact, the
predicted mean period of the sample of $c$-type variables that were considered
reproduced the observed value, \pc $= 0.336$~d, to three decimal places.
Similar consistency was obtained for the $ab$-type pulsators in our sample,
which have \pab $= 0.568$~d, but only if the model temperatures (the most
uncertain of the stellar properties) are reduced by $\delta\log\teff =
-0.0055$.  However, this offset could easily be partly, or even mostly, due to
errors in the color transformations that were used, or in the adopted mean colors
for the equivalent static stars, $(B-V)_S$ (from \citealt{ccc05}), rather than
problems with the model $\teff$\ scale.  We have used $\delta\log\teff$\ simply
as a free parameter that can be adjusted in order to force the predicted and
observed values of \pab\ to agree.

If the value of $(m-M)_V = 15.04$ used in Paper I should be decreased by
0.02 mag, as suggested by the present work, the variables would have reduced
luminosities by $\delta\log\,(L/L_\odot) \approx 0.008$ (since the change in
$M_V$ will be nearly identical to the change in $M_{\rm bol}$).  Models that 
have lower luminosities by this amount would predict shorter periods, unless,
e.g., their temperatures are decreased by 0.002 in $\log\teff$.  (Note that the
ratio of the luminosity and temperature coefficients in equations (\ref{eq:pab}) and (\ref{eq:pc}) is
very close to 1:4.)  In other words, one would recover the comparisons between
the predicted and observed periods for the M\,3 RR Lyrae, as well as the
ensemble averages, that are reported in Paper I, but for $(m-M)_V = 15.02$
instead of 15.04, if $\delta\log\teff = -0.0075$ and $-0.002$, for the
$ab$- and $c$-type variables, respectively.

Given the uncertainties associated with the reddening, distance, [Fe/H], and
$\teff$ scales, this level of consistency seems quite satisfactory.  For
instance, the above temperature offsets would be reduced in an absolute sense
if a slightly lower
reddening were adopted or if a lower metallicity, implying a brighter HB and
therefore increased periods, were assumed.  Although the HB simulations
presented in this study argue for a variation in $\delta\,Y_0 \approx 0.01$\ in
M\,3, which was not taken into account in Paper I, this would have only very
small effects on the predicted masses and periods of the variables, in view of
the relatively weak dependence of $\log\,P$ on mass; see equations
(\ref{eq:pab}) and (\ref{eq:pc}).  Indeed, such effects are overwhelmed by the
consequences of uncertainties in $(B-V)_S$ for the corresponding temperatures
and periods.

\subsection{M\,13}
\label{subsec:m13rr}

The most extensive work to date on the variable stars in M\,13 was carried
out by \citet{kkp03}, whose paper provides quite a thorough review of the
efforts made over the preceding century to discover such stars in this system;
also see the latest version of the ``Catalogue of Variable 
Stars in Globular Clusters"
(\citealt{cmd01}),\footnote{\href\protect{http://www.astro.utoronto.ca/{$\sim$}cclement/read.html}}
for updated information on some of the variables.  The intensity-weighted
mean magnitudes ($\langle V\rangle$) that were derived by Kopacki et
al.~for the 9 known RR Lyrae in M\,13 are listed in this catalogue, as well
as the best available determinations of their periods.  Only one of them (V8)
is an $ab$-type (fundamental mode) pulsator. 

In 2014, new $BVI_{\rm C}$ CCD photometry of
M\,13 was obtained during the nights of July 1--7 at the Bia\l{}k\'ow
Observatory of the University of Wroc\l{}aw using a 60-cm Cassegrain telescope
equipped with an Andor DW432-BV back-illuminated CCD camera.  The CCD frames
were corrected for instrumental effects (bias, dark current) and flat-fielded in
the usual way (see the description by \citealt{jpk96}).  Instrumental magnitudes
were then computed for all stars in the observed field using the DAOPHOT
profile-fitting software described by \citet{st87}. The differential photometry
was derived on a frame-to-frame basis.  In this approach, the instrumental
photometry for each frame is shifted to the magnitude scale of a reference
frame by the average offset in brightness between them.  The average offset is
determined from a large number of bright unsaturated stars.

Average differential magnitudes and colors were transformed to the standard
system using the photometric data of M\,13 taken from \citet{st00} database.
From 410 stars in common with \citet{st00}, we obtained the following
transformation equations:
%
%
\begin{equation}
\setlength{\arraycolsep}{0pt}
 \begin{array}{rll} 
  V-v& {}= -0.047\times(v-i)-0.753,& \hskip0.7cm\sigma=0.020,\\
   B-V& {}=  +1.235\times(b-v)-0.413,& \hskip0.7cm\sigma=0.014,\\
   V-I_{\rm C}& {}= +0.945\times(v-i)+0.506,& \hskip0.7cm\sigma=0.012,
 \end{array}  
\label{eq:stdtrf}
\end{equation}
where uppercase and lowercase letters denote standard and instrumental
magnitudes, respectively, and $\sigma$ is the standard deviation of the fit.
Individual instrumental $BVI_{\rm C}$ magnitudes of RR Lyrae stars were 
transformed to the standard system using the above relations, but with
instantaneous instrumental colors derived from the phase diagrams for the
appropriate passbands.

The new $BVI_{\rm C}$ observations of M\,13 were supplemented with the
$VI_{\rm C}$ data of \citet{kkp03}, which were taken at the same observatory
but using different equipment.  These older observations, obtained in 2001,
were also tied to the standard system using Stetson's data.  The differences
between the mean $VI_{\rm C}$ magnitudes that were derived for the RR Lyrae
stars from both datasets were found to be small, amounting to about 0.03 mag.
Therefore, we decided to merge the RR Lyrae light curves from these two sets.

Table~\ref{tab:meanBVI} gives the mean magnitudes and colors for those RR Lyrae
stars in M\,13 for which reliable photometry could be derived.  (Observations
were not obtained for three of the known RRc variables; namely V25, V31, and
V35.)  For each star in this table, the mean $V$ magnitudes and the mean $B-V$
and $V-I_C$ colors are given as intensity-weighted, $\langle{ }\rangle$,
and magnitude-weighted, $({ })$, averages.  (This notation
follows the convention that is often used; see, e.g., \citealt{bcs95}.)
These mean values were determined from the fit of a truncated Fourier series to
the light curves.  Average colors are defined as the arithmetic difference
between the appropriate mean magnitudes.

It should be noted that complete light curves in all passbands were obtained for
the observed RRc stars, so their mean magnitudes should be very reliable, with
the possible exceptions of those for V5 and V36.  V5 exhibits some deviations
from regularity in its $B$ light curve, which could be an indication of some
instrumental problems (due perhaps to close stars) or possibly the Blazhko
effect (e.g., \citealt{bk11}), whereas V36 appears to be undergoing
multi-periodic non-radial pulsations (see the discussion by \citealt{kkp03}).
The only star of the RRab type is V8.  Due to the short timespan of the 2014
observations, its $B$-filter light curve has a few gaps, but the maximum and
minimum phases, and some intermediate phases, are very well covered.  Although
we believe that the tabulated mean $B$ magnitude and mean $B-V$ colors for V8
are accurate, they should nevertheless be used with caution.

\begin{table}[t]
\centering
\caption{Mean $BVI_{\rm C}$ magnitudes and colors of RR Lyrae stars in M\,13.}
\label{tab:meanBVI}
\setlength{\tabcolsep}{1.7pt}
\begin{tabular}{ccccccc}
\hline
\hline
Var.& $\langle{}V\rangle$& $(V)$& 
$\langle{}B$--$V\rangle$& $(B$--$V)$&
$\langle{}V$--$I_{\rm C}\rangle$& $(V$--$I_{\rm C})$\\
& [mag]& [mag]& [mag]& [mag]& [mag]& [mag]\\ [0.5ex]
\hline
 V5& 14.717& 14.728& 0.293& 0.298& 0.437&  0.445\\
 V7& 14.895& 14.900& 0.247& 0.250& 0.352&  0.355\\
 V8& 14.822& 14.853& 0.379& 0.399& 0.535&  0.554\\
 V9& 14.786& 14.797& 0.302& 0.307& 0.430&  0.437\\
V34& 14.803& 14.810& 0.317& 0.320& 0.439&  0.444\\
V36& 14.782& 14.783& 0.271& 0.271& 0.362&  0.363\\
\hline
\end{tabular}
\end{table}

Insofar as the colors of the equivalent static stars are concerned, the
magnitude-weighted $B-V$ color has often been thought to be a better $\teff$
indicator than the intensity-weighted color (e.g., \citealt{san90}).  However,
\citet{bcs95} found from their analysis of nonlinear, nonlocal, time-dependent
convective models of RR Lyrae stars that static colors should be quite well
approximated by $\langle B\rangle - \langle V\rangle$, especially in the case
of $c$-type variables.  (Their work also justifies the common practice of
equating intensity-weighted mean $V$ magnitudes with static $V$ magnitudes.)
The differences between the colors listed in fourth and fifth columns of
Table~\ref{tab:meanBVI}, or between those in the sixth and seventh columns, are
sufficiently small, in any case, that the choice of which color to use is
inconsequential.  In what follows, intensity-weighted mean magnitudes and colors
have been adopted in CMDs that plot the locations of the M\,13 RR Lyrae.

\begin{figure}[t]
\plotone{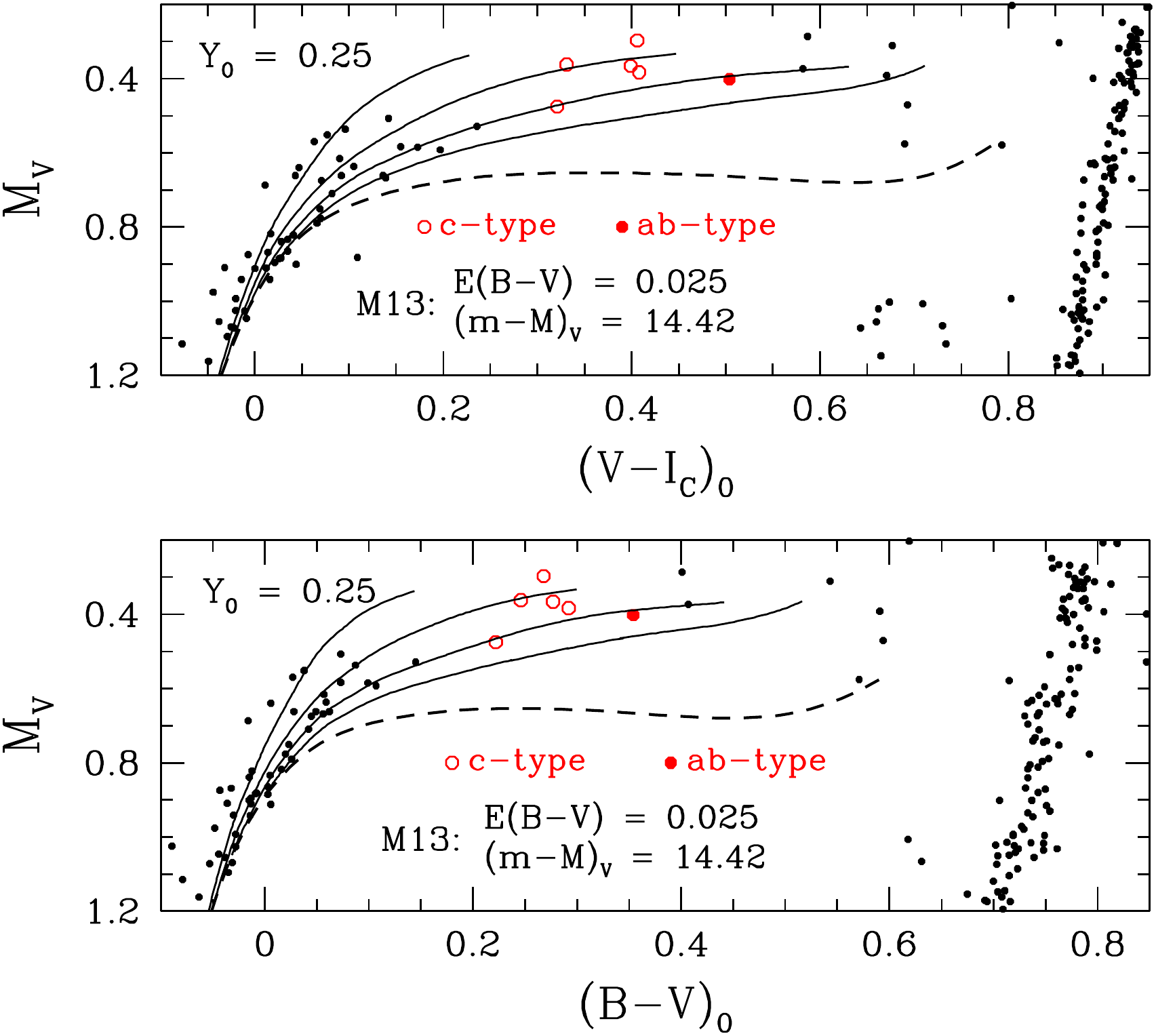}
\caption{Overlay of a ZAHB (dashed curve) and evolutionary tracks for masses of
0.60, 0.61, 0.62, and $0.63\,\msol$, in the direction from left to right
(solid curves), for [Fe/H] $= -1.55$, [$\alpha$/Fe] $= 0.4$, and $Y_0 = 0.25$ onto
the CMD for the HB and RGB populations of M\,13 that have $0.2 \le M_V \le 1.2$. 
Small black filled circles indicate the locations of non-variable HB stars and
giants, while filled and open circles (in red) are used to identify the $ab$-
and $c$-type RR Lyrae, respectively.  Note that the HB tracks provide very
similar fits to the $VI_C$ observations (top panel) and the $BV$ data (bottom
panel).}
\label{fig:f14}
\end{figure}

Figure~\ref{fig:f14} superimposes on the part of the M\,13 CMD that contains
its HB stars with $M_V \le 1.2$ a subset of the same HB tracks (for [Fe/H] $=
-1.55$, [$\alpha$/Fe] $= 0.4$ and $Y_0 = 0.25$) that were compared with
observations of M\,3 in Paper I.  (The $BVI_C$ photometry was taken from the
``Photometric Standard Fields" archive developed and maintained by P.~Stetson;
see footnote 10.)  According to these tracks, the RR Lyrae variables have masses
in the range of 0.60 to $0.63\,\msol$ and they evolved from ZAHB locations well
to the blue of the instability strip where there are substantial numbers of
non-variable HB stars.  Each of these variables is $\gta 0.2$ mag brighter 
than the ZAHB at the same color, in stark contrast with the M\,3 RR Lyrae,
which are mostly found in close proximity to the ZAHB, but with a few outliers
at brighter magnitudes (see Paper I).  Note that the locations of the M\,13
variables relative to the evolutionary tracks are very similar on both the
$[(V-I)_0,\,M_V]$- and $[(B-V)_0,\,M_V]$-diagrams, which indicates that there
is good consistency between the predicted and observed colors.

\begin{figure}[b]
\plotone{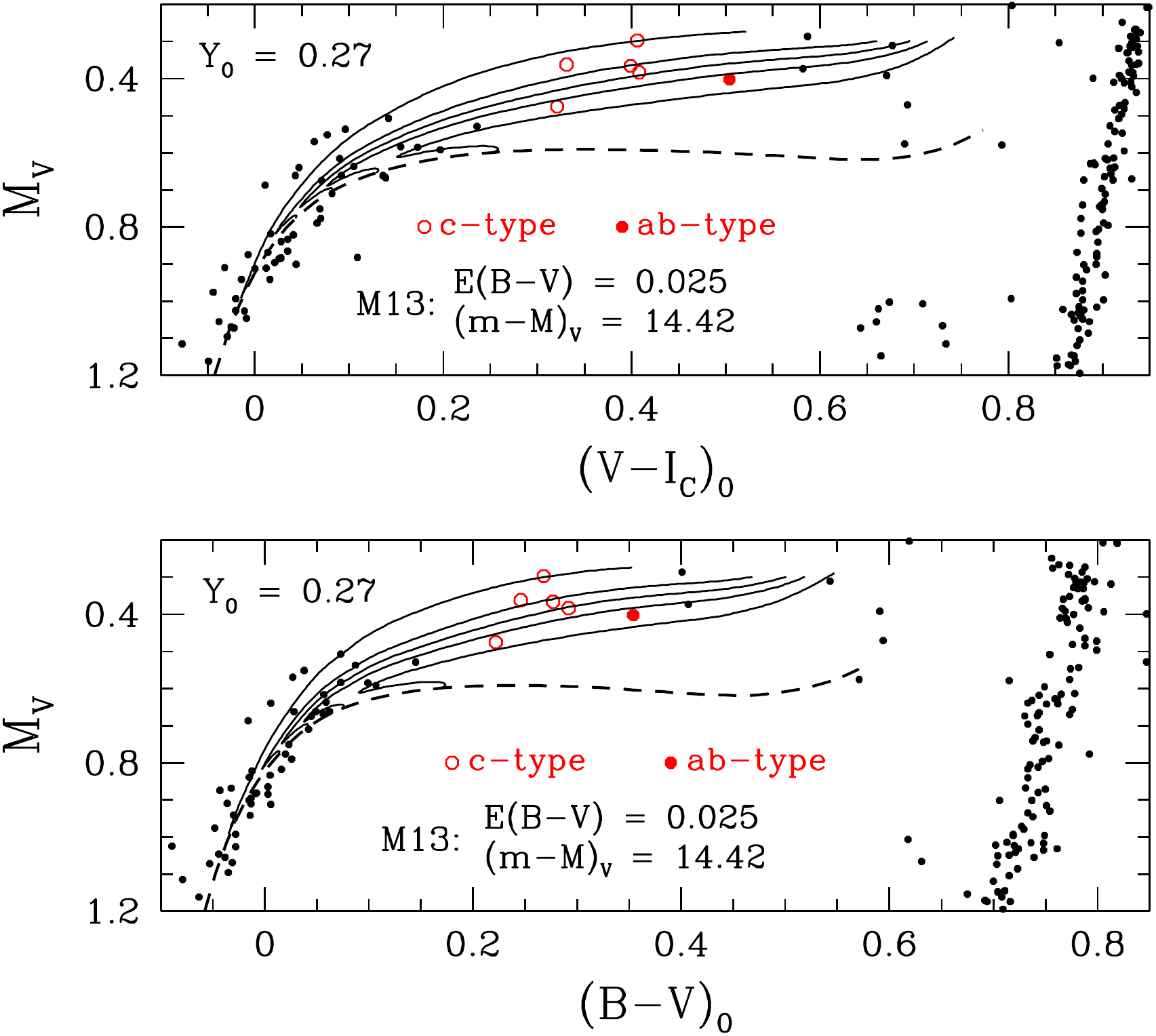}
\caption{Similar to the previous figure, except that the models assume $Y_0 =
0.27$.  The tracks were computed for masses of 0.62, 0.63, 0.635, 0.64, and
$0.65\,\msol$, in the direction from left to right.}
\label{fig:f15}
\end{figure}

\begin{figure}[t]
\plotone{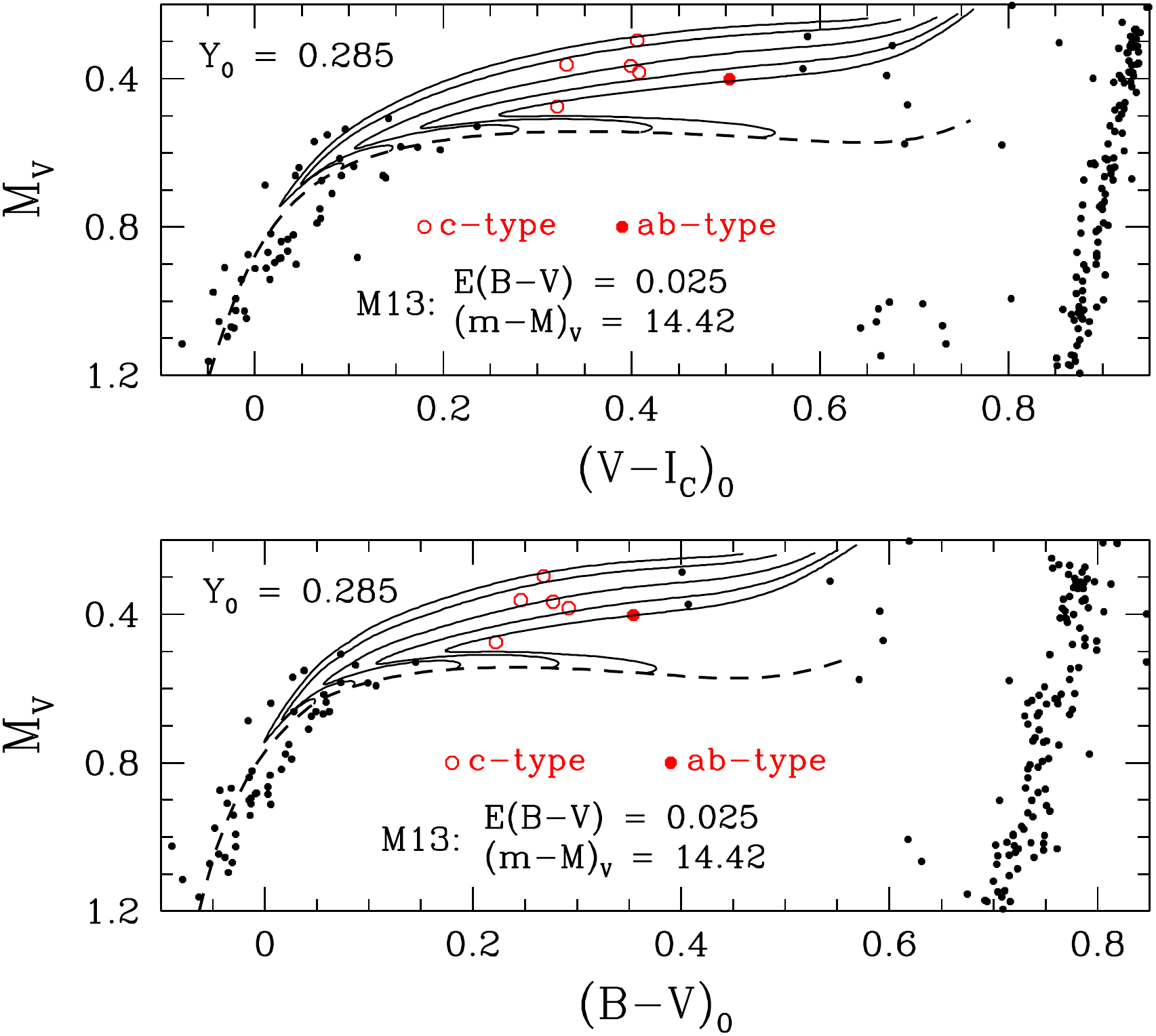}
\caption{Similar to Fig.~\ref{fig:f14}, except that the models assume
$Y_0 = 0.285$.  The tracks were computed for masses of 0.635, 0.64, 0.65, 0.66,
and $0.67\,\msol$, in the direction from left to right.}
\label{fig:f16}
\end{figure}

Very similar comparisons between theory and observations are shown in
Figures~\ref{fig:f15} and \ref{fig:f16},
the only differences being the assumption of $Y_0 = 0.27$ and 0.285 in the
stellar models that are compared with the observations (as indicated) and small
adjustments to the masses that were assumed in the HB tracks.  The higher the
helium abundance, the higher the masses of the variable stars that are implied
by the superposition of the evolutionary tracks onto their CMD locations.  The
variation in mass is not large; e.g., the brightest of the $c$-type RR Lyrae
is predicted to have a mass near $0.60\,\msol$, $0.62\,\msol$, or $0.637\,\msol$
if it has $Y_0 = 0.25$, 0.27, or 0.285, respectively.  

Figs.~\ref{fig:f14}--\ref{fig:f16} also illustrate the strong sensitivity of
the luminosities of the models to $Y$, as well as the increased prominence
of blue loops in the
post-ZAHB evolutionary tracks, which have important consequences for the
interpretation of the data.  For instance, the reddest track in
Fig.~\ref{fig:f16}, for a mass of $0.67\,\msol$,
makes an especially long excursion to the blue before
bending back to the red and passing through the location of the one $ab$-type
variable in M\,13 (V8).  The fact that M\,13 does not contain any stars in the
vicinity of the ZAHB location where this evolutionary sequence originates makes
it unlikely that V8 has a helium abundance of $Y_0 = 0.285$ (or higher).  The
same can be said of the faintest $c$-type variable.  Indeed, $Y_0 = 0.285$ would
appear to be slightly too high for the next one or two (more luminous)
first-overtone pulsators as well because there are no stars near the ZAHB in M\,13
at colors where the high-$Y$\ tracks most relevant to these RR Lyrae begin. 

Due, in particular, to the strong dependence of $\log\,P$ on $\log\,\teff$, which
is affected by errors in the photometry, the reddening, the determination of the
colors of equivalent static stars, the adopted color transformations, and many
physics ingredients of stellar models (such as the treatment of convection and
the atmospheric boundary condition), the periods that are predicted for
RR Lyrae stars necessarily involve fairly large uncertainties.  If, e.g.,
the $\teff$\ of a 6500~K fundamental-mode pulsator has an errorbar of
$\pm 70$~K, the associated uncertainty in the value of $\log\,P_\mathrm{ab}$ given
by equation (\ref{eq:pab}) would be $\pm 0.016$, which corresponds to $\pm 0.023$~d if it
has a 0.65~d period.  As Victoria models satisfy empirical constraints on the
temperatures of turnoff and subgiant stars by \citet{crm10} quite well (see
\citealt{vcs10}), they should fare equally well, if not better, in matching
the properties of warm HB stars.  
We conclude from these considerations that satisfactory consistency between
the predicted and observed periods of a given RR Lyrae star is obtained if
they agree to within $\sim \pm 0.03$~d.

Interpolations in the HB tracks yield the luminosities, temperatures, and
masses at the CMD locations of the M\,13 RR Lyrae variables.  In the case of
the models for $Y_0 = 0.25$, the mass-fraction metal abundance $Z = 
7.623 \times 10^{-4}$.  Using this information, we can calculate their periods,
and in addition, we can determine how much the interpolated $\teff$\ values
would have to be adjusted in order for the predicted and measured periods to
agree to, say, three decimal places (assuming that the
discrepancies are primarily due to errors in the inferred temperatures).  

A summary of our results for V8, the only RRab-type variable in M\,13, is
provided in Table~\ref{tab:v8}.  If no $\teff$\ offset is applied to the models,
the HB tracks for $Y_0 = 0.25$ predict periods that are too low by 0.023~d or by
0.054~d, depending on whether the $BV$ or $VI_C$ observations are considered.
Not unexpectedly, the results from the two color planes are not in perfect
agreement, likely due in errors in the derived static colors and/or in the
adopted $(B-V)$--$\teff$, $(V-I_C)$--$\teff$\ relations (from \citealt{cv14}).
In order to reproduce the observed period (0.750~d), the temperatures inferred
from the $\langle B-V\rangle$ and $\langle V-I_C\rangle$ colors need to be
revised, in turn, by $\delta\log\teff = -0.0040$ and $-0.0095$ (see the second
row of numbers).  Thus, the temperatures derived from $B-V$ colors are
somewhat cooler than those based on $V-I_C$ colors.  Finally, as discussed at
the end of the preceding section, there is some justification for correcting
the model temperatures by $\delta\log\teff = -0.0075$, which results in a
period from $BV$ photometry that is too high by 0.021~d or one from $VI_C$
observations that is too low by 0.012~d.

\begin{table}[ht]
\centering
\caption{Predicted Periods of V8 ($P_{\rm obs} = 0.750$~d)}
\label{tab:v8}
\setlength{\tabcolsep}{7.0pt}
\begin{tabular}{ccccc}
\hline
\hline
Y &  $\delta\log\teff$ & $P$ & $\delta\log\teff$ & $P$ \\
 &  \multicolumn{2}{c}{[from $\langle B-V\rangle$]}  &
   \multicolumn{2}{c}{[from $\langle V-I_C\rangle$]} \\ [0.5ex]
\hline
 0.25 & $+0.0000$ & 0.727 & $+0.0000$ & 0.696 \\
      & $-0.0040$ & 0.750 & $-0.0095$ & 0.750 \\
      & $-0.0075$ & 0.771 & $-0.0075$ & 0.738 \\ [0.7ex]
 0.27 & $+0.0000$ & 0.711 & $+0.0000$ & 0.681 \\
      & $-0.0068$ & 0.750 & $-0.0122$ & 0.750 \\
      & $-0.0075$ & 0.755 & $-0.0075$ & 0.723 \\
\hline
\end{tabular}
\end{table}

For the most part, there is good consistency between the observed period of V8
and the periods based on the HB tracks for $Y_0 = 0.25$.  Some reduction of the
inferred $\teff$\ of this star seems to be necessary, but this could be due, in
part, to the assumption of a reddening that is somewhat too high or to the
adoption of a distance modulus that is too low.  Another possibility is that the
$\langle B\rangle - \langle V\rangle$ and $\langle V\rangle - \langle I_C\rangle$
colors should be increased somewhat to better represent the equivalent static
colors. Indeed, \citet{bcs95} predict that $(B-V)_S = \langle B\rangle -
\langle V\rangle + 0.008$\ in the case of an $ab$-type variable with a pulsation
amplitude $A(B) = 1.06$ mag, as measured for V8.  If we were to adopt this
estimate of $(B-V)_S$, the predicted period would have been 0.745~d, assuming
$Y=0.25$ and $\delta\log\teff = 0.0$, instead of 0.727~d; see the first row of
numbers in Table~\ref{tab:v8}. (Unfortunately, Bono et al.~do not consider $V-I_C$\ colors.)

The remaining entries in Table~\ref{tab:v8} list the calculated periods of V8
when its mass, luminosity, and temperature have been obtained via interpolations
in the HB tracks for $Y_0 = 0.27$.  (If the assumed helium abundance is increased
by $\Delta\,Y_0 = 0.02$, $Z$ must be reduced to $7.42 \times 10^{-4}$ in order for
the resultant mixture to have [Fe/H] $= -1.55$.)  However, the models will
necessarily predict nearly identical values of $\teff$\ and $M_{\rm bol}$ for
V8, independently of the assumed He abundance, because these properties depend
almost entirely on the adopted reddening and distance modulus (which are
unchanged).  Consequently, the main effect of higher $Y$ is to increase the
predicted mass of V8, and thereby to reduce the predicted period.  This explains
the $\approx 0.015$~d reduction in the period, compared to the corresponding
results for $Y_0 = 0.25$, if $\delta\log\teff = 0.0$ or $-0.0075$.  Alternatively, 
to compensate for this effect, cooler temperatures must be adopted if the models
for $Y_0 = 0.27$ are to reproduce the observed period.  We conclude from the 
similarity of the results given in Table~\ref{tab:v8} for $Y_0 = 0.25$ and 0.27
that the helium abundance of V8 could have any value within the range $0.25
\lta Y \lta 0.27$.  (As already discussed, higher values of $Y$ are ruled out.)

It turns out that the periods of the RRc stars in M\,13 are considerably more
difficult to explain, which is unexpected given that we were quite successful
in matching the periods of the $c$-type RR Lyrae in M\,3 (see Paper I) using the
same HB tracks as those employed in the present study.  To be specific, periods
derived for the M\,13 variables are uniformly too high if, as in our analysis of
M\,3, no adjustments are applied to the inferred $\teff$ values.  As shown in
Table~\ref{tab:rrc}, the differences in the periods (in the sense ``predicted
minus observed") are as much as $\sim 0.11$~d.  The tabulated results clearly
show that the assumed He abundance cannot be the main cause of the discrepancies
because it has a relatively small influence on computed periods; higher helium
by $\delta\,Y_0 = 0.035$ implies reduced periods by only $\sim 0.015$~d.  Hence,
either appreciably higher temperatures and/or the adoption of a significantly
smaller distance modulus, implying lower luminosities, would seem to be the most
likely ways of obtaining reduced periods that are in better agreement with the
observed ones.

\begin{table}[ht]

\centering
\caption{Differences Between the Periods of the RRc Stars in M\,13, as Predicted
 by HB Models for $Y_0 = 0.25$, 0.27, and 0.285, and their Observed Periods}
\label{tab:rrc}
\setlength{\tabcolsep}{1.7pt}
\begin{tabular}{cccccccccc}
\hline
\hline
Var. & $P_{\rm obs}$ & & $\Delta P_{Y25}$ & $\Delta P_{Y27}$ & 
 $\Delta P_{Y285}$ & & $\Delta P_{Y25}$ &
 $\Delta P_{Y27}$ & $\Delta P_{Y285}$ \\
 & [days] & & \multicolumn{3}{c}{[from $\langle B-V\rangle$]} & &
     \multicolumn{3}{c}{[from $\langle V-I_C\rangle$]} \\ [0.5ex]
\hline
V5  & 0.382 & & 0.101 & 0.094 & 0.086 & & 0.105 & 0.098 & 0.091 \\
V7  & 0.313 & & 0.042 & 0.034 & 0.026 & & 0.035 & 0.027 & 0.019 \\
V9  & 0.393 & & 0.074 & 0.066 & 0.057 & & 0.059 & 0.051 & 0.043 \\
V34 & 0.389 & & 0.093 & 0.084 & 0.074 & & 0.064 & 0.056 & 0.047 \\
V36 & 0.316 & & 0.108 & 0.101 & 0.093 & & 0.077 & 0.070 & 0.064 \\
\hline
\end{tabular}
\end{table}

\begin{figure}[ht]
\plotone{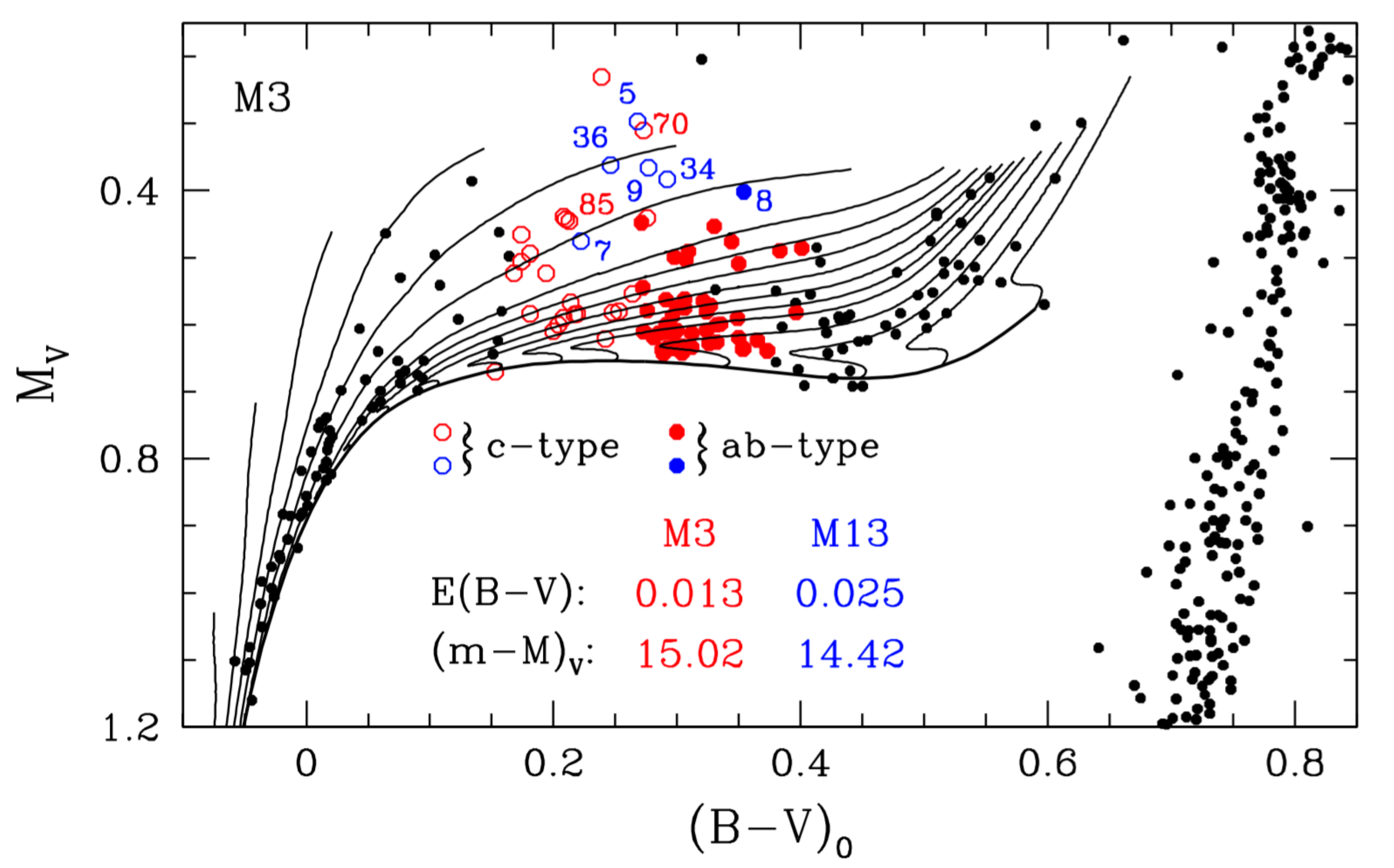}
\caption{Overlay of a ZAHB and several HB tracks for [Fe/H] $= -1.55$,
[$\alpha$/Fe] $= 0.4$, and $Y_0 = 0.25$ onto the CMD for the HB and RGB populations
of M\,3 that have $0.15 \le M_V \le 1.2$.  Non-variable HB stars and giants
have been plotted as small black filled circles, while filled and open circles,
in red, represent the cluster $ab$- and $c$-type RR Lyrae, respectively.  Two
of the latter, V70 and V85, are discussed in the text.  The locations of the
M\,13 variables, which are all identified by their ``V" numbers, are similarly
plotted, but in blue.  The assumed reddenings and distance moduli are as
indicated.}
\label{fig:f17}
\end{figure}

Figure~\ref{fig:f17} highlights some of the puzzling differences between
the M\,3 and M\,13 RR Lyrae.
It reproduces the fit of a ZAHB and several tracks to the HB 
population of M\,3 that was given in Fig.~5 of Paper I, except that (i) the
assumed value of $(m-M)_V$ has been reduced by 0.02 mag and (ii) the locations
of the M\,13 variables have been plotted (on the assumption of the indicated
reddening and distance modulus).  Whereas small black filled circles mark the
CMD locations of non-variable HB stars and giants in M\,3, filled and open
circles (in red for M\,3, and in blue for M\,13) represent the cluster RRab and
RRc stars, respectively.  All of the M\,13 RR Lyrae have been identified, but
only V70 and V85 in M\,3.  The reason for focusing some attention on these two
variables is that they are located very close to two of the M\,13 RRc stars (V5
and V7).  Due to their proximity, one would expect V5 and V70 to have nearly the
same pulsation periods, but they actually differ by $\approx 0.10$~d.  The
measured period of V70 is 0.486~d, in excellent agreement with the predicted
period (0.485~d if $Y_0 = 0.25$), but that of V5 is 0.382~d (as compared with
predictions that are $\sim 0.1$~d higher, depending on the assumed helium
abundance; see Table~\ref{tab:rrc}).

Somewhat better consistency is found for V7 in M\,13 and V85 in M\,3.  Models
for $Y_0 = 0.25$ reproduce the observed period of V85 (0.356~d) to within
0.003~d, whereas the difference between the predicted and observed periods of
V7 is 0.02--0.04~d, according the results given in Table~\ref{tab:rrc}.  In
fact, V7 is the least problematic of the M\,13 RRc stars.  Models for $Y_0 = 0.285$
predict a period within 0.03~d of the observed value, and those for lower $Y$
would do so as well if V7 had a slightly bluer static color and/or the models
predicted a somewhat higher $\teff$\ for this star.  In fact, M\,3 has two
other $c$-type variables at nearly the same location as V85 in 
Fig.~\ref{fig:f17} (namely, V29 and V140), and their observed periods are
about 0.02~d less than those predicted by models for $Y_0 = 0.25$, in acceptable
agreement with the period of V7 in M\,13.  (Higher helium would tend to reduce
these differences, and since our HB simulations support only a small variation of
$Y$ in M\,3, it is possible that V29 and V140 are more helium-rich than V85.) 

Near the beginning of this section, we expressed some concerns about the light
curves of V5 and V36 as possibly being affected by the presence of companions
or by the Blazhko effect, which may explain the large discrepancies between the
predicted and observed periods of these two variables.  The differences are
within acceptable limits in the case of V7 and V8, and in fact, we would predict
periods that are within 0.03~d of those observed for V9 and V34 if they have
$Y \gta 0.28$ and their temperatures were increased by only $\delta\log\teff
\sim 0.007$ (or less, if the adopted static colors are somewhat too red).  A
small zero-point error in the model $\teff$\ scale or in the observed colors is
suggested by the fact that all of the predicted periods for the RRc
stars are too high.

Curiously, the distribution in M\,3 of the first-overtone pulsators as a
function of $B-V$ color is quite different from that seen in M\,13.  As shown
in Fig.~\ref{fig:f17}, M\,3 has many RRc stars with $(B-V)_0 \lta 0.22$,
whereas none of the M\,13 variables have such blue colors.  It is conceivable
that this is simply the consequence of small number statistics, or perhaps this
difference (and the difficulties discussed above) are telling us that the
RR Lyrae in M\,13 differ in some fundamental, yet unrecognized way from those
in M\,3. The fact that most of the M\,13 variables are located very close
to the boundary between RRab and RRc stars (see Fig.~\ref{fig:f17}) may be
relevant to an understanding of their apparently anomalous properties.

\subsection{47\,Tuc}
\label{subsec:tucrr}

According to \citet{csw93}, there appears to be only one bona fide RR Lyrae
variable in 47\,Tuc, as other candidates that lie inside or just outside the
tidal radius have been found to be metallicity, proper-motion, or 
radial-velocity nonmembers.  Observations of V9 obtained by these investigators
have yielded $\langle V\rangle = 13.725$, $\langle B-V\rangle
= 0.359$, $(B-V)_{\rm mag} = 0.396$, and a period of 0.736852 days.  If 47\,Tuc
has $(m-M)_V \approx 13.3$, as our HB simulations suggest, V9 has $M_V \approx
0.4$, which makes it much brighter than nearly all of the core He-burning stars
in this cluster, in addition to being much bluer.

\begin{figure}[t]
\plotone{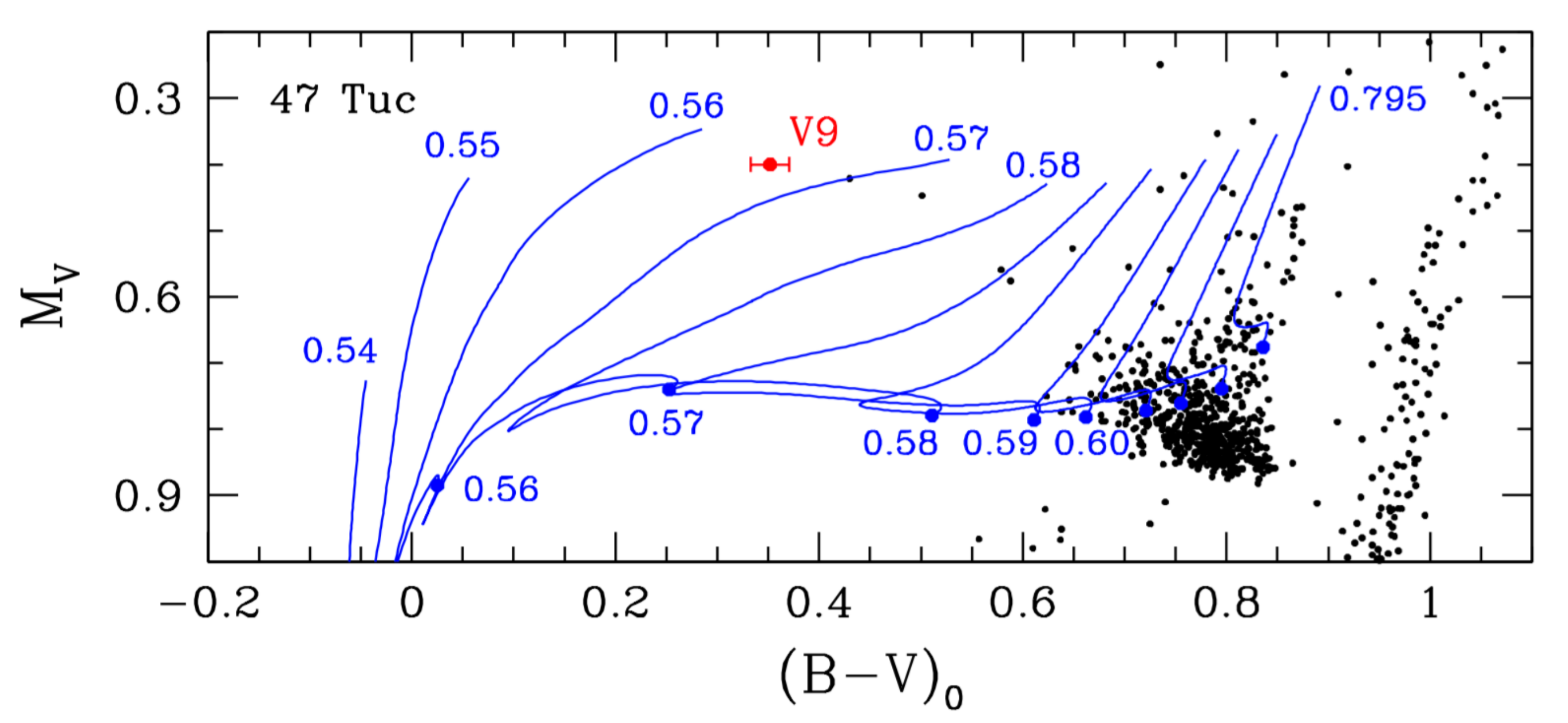}
\caption{Superposition of a grid of HB tracks for [Fe/H] $= -0.70$, 
[$\alpha$/Fe] $= 0.4$, and $Y_0 = 0.287$ onto the CMD of 47\,Tuc in the magnitude
range $0.2 \le M_V \le 1.0$ (assuming $E(B-V) = 0.026$ and $(m-M)_V = 13.27$).
Small black filled circles represent cluster HB stars and giants and the red
filled circle indicates the location of the $ab$-type variable V9 (see the
text).  Blue filled circles mark the ZAHB locations of the HB tracks that have
been plotted.  The four reddest tracks were computed for masses of 0.795, 0.68,
0.64, and $0.62\,\msol$; the assumed masses of the bluer tracks are as
indicated.}
\label{fig:f18}
\end{figure}

Since a variation in $Y \sim 0.03$ is implied by simulations of the observed
morphology of the HB in 47\,Tuc, the simplest explanation of V9 is that it is a
member of the highest-$Y$ population (in order to explain its high luminosity),
with a mass that is sufficiently low for its evolutionary track to enter the
instability strip.  In fact, this suggestion seems to work out remarkably well.
In Figure~\ref{fig:f18}, several evolutionary tracks for [Fe/H] $= -0.70$,
[$\alpha$/Fe] $= 0.4$, $Y_0 = 0.287$, and masses in the range $0.54 \le
{\cal M}/\msol\ \le 0.795$ have been overlaid onto the part of the 47\,Tuc CMD
that contains its HB population and adjacent RGB stars (from
\citealt{bs09}\footnote{\href\protect{http://www.cadc.hia.nrc-cnrc.gc.ca/en/community/\hfil\break STETSON/homogeneous/}}).
The location of V9 is indicated by the red filled circle.  For its static color,
we have simply adopted the mean of the published $\langle B-V\rangle$
and $(B-V)$ colors, as it is not clear which of these colors provides
the best approximation.  \citet{bcs95} suggest that both possibilities will be
bluer than $(B-V)_S$ by $\lta 0.01$ mag, which seems to be at odds with a
difference of 0.037 mag in the measured values of $\langle B-V\rangle$
and $(B-V)$.  (This difference has been plotted as the horizontal
error bar that is attached to the red filled circle to represent the probable
uncertainty in the color of V9.)

\begin{figure*}[t]
\plotone{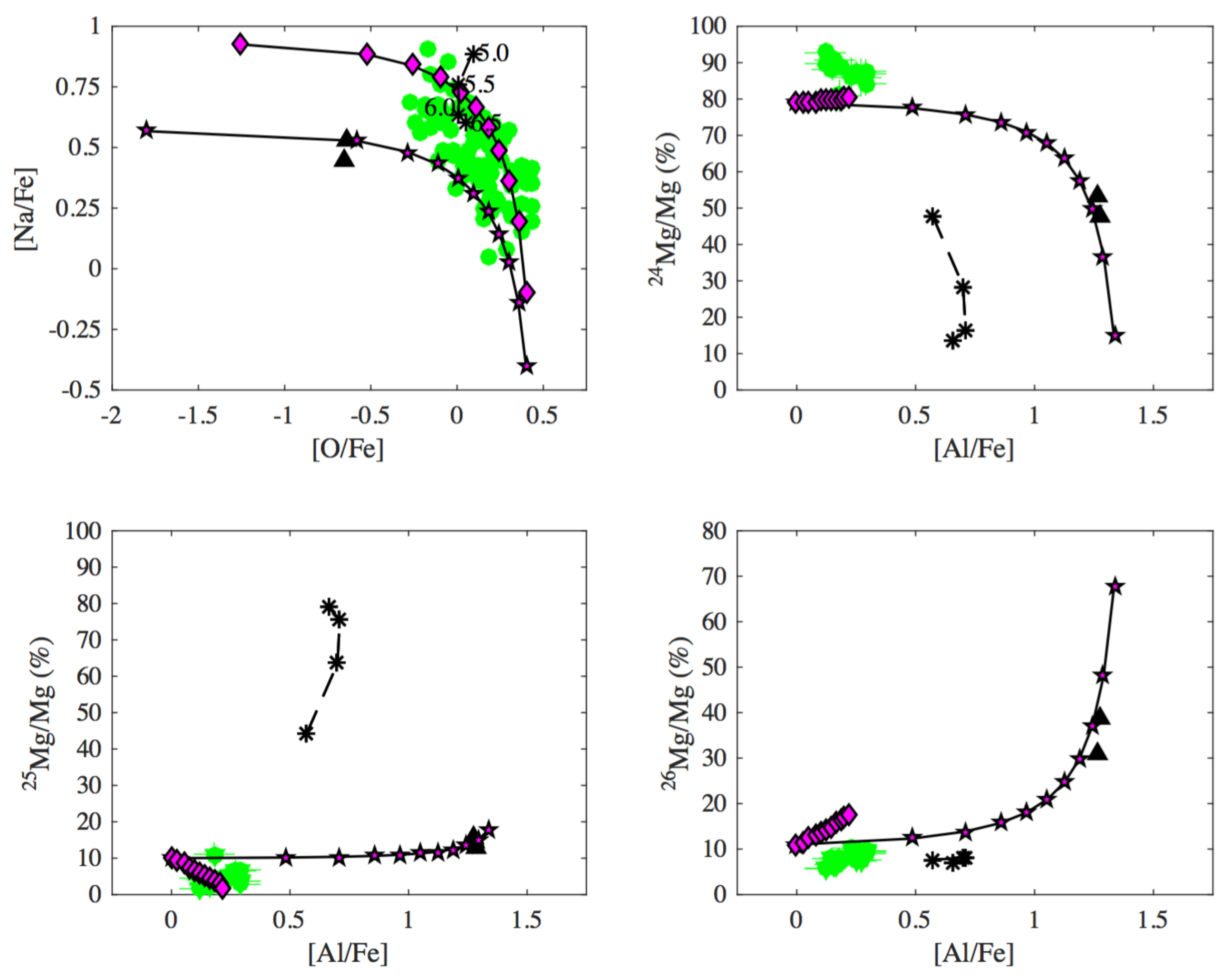}
\caption{The $p$-capture abundance anomalies in 47\,Tuc (green circles are the data from \citealt{gratton:13} (top-left panel) and
         \citealt{thygesen:16} (other panels)) modeled with a $10,000\,\msol$ 
         MS star (the magenta diamonds) as described by \cite{denissenkov:15}, but using the \cite{ags09} initial mix of elements and
         isotopes instead of that of \cite{gs98}. The black triangles and magenta stars show the observational data of \cite{johnson:12}
         and \cite{dacosta:13} for the two M\,13 RGB stars with the most extreme abundance anomalies and their reproduction with H-burning
         yields from a $42,500\,\msol$ MS model with the As09 mix. For comparison, the abundances predicted by \cite{ventura:09} for AGB stars 
         with $Z=0.004$ are also shown.}
\label{fig:f19}
\end{figure*}

Fig.~\ref{fig:f18} illustrates the well-known prediction that, at relatively
high metallicities, masses ``pile up" at the red end of a ZAHB, which causes
the HB in most metal-rich clusters to have a clump-like appearance.  In the
case of 47\,Tuc, nearly the entire horizontal branch is contained within the
color range $0.6 \lta (B-V)_0 \lta 0.83$, despite having masses that vary by
$\gta 0.2\,\msol$, according to our HB tracks.  However, as the mass decreases
below $\sim 0.6\,\msol$, each successive $0.01\,\msol$\ change in mass results
in an ever larger shift to hotter temperatures and bluer colors.  Thus, only a
small decrease in mass ($\sim 0.035\,\msol$) from the mass of the bluest stars
in 47\,Tuc's HB clump, $\approx 0.60\,\msol$, is needed to reach sufficiently
blue colors where post-ZAHB tracks pass close to the location of V9.

Interpolations in the HB tracks, which have $Z = 5.099 \times 10^{-3}$, yield
$M_{\rm bol} = 0.384$, $\log\teff = 3.814$, and a mass of $0.565\,\msol$ for
V9.  Using equation (\ref{eq:pab}), the predicted period of this variable is 0.717~d,
which is within 0.02~d of the observed period.  This surprisingly good agreement
provides further support for our interpretation of the 47\,Tuc CMD, particularly
the results based on our HB simulations.

\section{A supermassive star as a source of the $p$-capture abundance anomalies in 47\,Tuc}
\label{sec:47TucSMS}

\cite{denissenkov:14} have proposed that the ubiquitous abundance anomalies of the proton-capture elements and isotopes
in GCs were produced by MS supermassive stars with ${\cal M}\sim 10^4\,{\cal M}_\odot$, which could have formed
in the process of a runaway merger of massive stars during the first few million years of their lives. 
With the initial chemical composition scaled to the required value of [Fe/H]
from the solar abundance mix of \cite{gs98}, \cite{denissenkov:15} have shown that the O-Na anticorrelation in M\,13 as well as
the correlations of the [Al/Fe] enhancement with the Mg isotopic ratios in M\,13 and in four other GCs can all
be reproduced by mixing pristine gas with core H-burning nucleosynthesis yields from MS stars with ${\cal M}\approx 6\times 10^4\,{\cal M}_\odot$.
Since that publication, the Mg isotopic ratios have been measured along with the non-LTE-corrected [Al/Fe] ratios, 
also in the GC 47\,Tuc, by \cite{thygesen:16}. They are plotted together with the O-Na anticorrelation in 47\,Tuc
RGB and HB stars from \cite{gratton:13} in Fig.~\ref{fig:f19} (the green circles). By comparison,
the black triangles represent the two M\,13 RGB stars with the most extreme abundance anomalies.

Similar success in matching the observed abundances can be obtained if models for supermassive stars assume the
metals mixture given by \cite{ags09}, as adopted in this study, but for lower masses.  
As shown in Fig.~\ref{fig:f19}, the O-Na anticorrelation and the correlations of
the [Al/Fe] enhancement with the Mg isotopic ratios can be reproduced quite well
by the H-burning yields from MS stars with
${\cal M}\approx 4.25\times 10^4\,{\cal M}_\odot$ for M\,13 and ${\cal M}\approx 10^4\,{\cal M}_\odot$ for  47\,Tuc 
(the magenta stars and diamonds, respectively).
The reduction in mass for M\,13 is caused by the lower mass-fraction abundance of the heavy elements in the
As09 mix compared with that given by the GS98 mix at the same [Fe/H] value,
which results in a higher central temperature of the core H burning at the same stellar mass. 
As before, the black asterisks show the theoretical results obtained with the yields from the massive AGB star models of
\cite{ventura:09}, with the initial masses of the stars indicated in the top-left panel. 
The variations of the Mg isotopic ratios with [Al/Fe] clearly present severe problems for the AGB scenario.

The theoretical solid black lines in Fig.~\ref{fig:f19} are dilution curves, along which the relative contributions to
the predicted chemical composition from the pristine gas and supermassive star change from 100\% to 0\%
for the former and in the opposite direction for the latter, beginning with the assumed initial abundances.
The symbols (diamonds and stars) on the curves show the sequence of the dilution factor decreasing from 100\% to 0\% with the step of 10\%.
The good fits to the observed abundance correlations have been obtained assuming that H burning in the supermassive star models continued 
until the He mass fraction in these fully convective objects has reached the value of $Y=0.4$ in the case of M\,13 and $Y=0.3$ for 47\,Tuc.
The difference in the final value of $Y$ for these GCs can be attributed to the fact that the star-to-star variation of $Y_0$ in M\,13,
$\Delta Y_0\approx 0.08$, is found to be larger than $\Delta Y_0\approx 0.03$ in 47\,Tuc. In the case of 47\,Tuc, the mixture of the pristine gas with 
the supermassive star yields reproduces really well not only its O-Na anti-correlation and the correlations between the Mg isotopic ratios 
and Al abundance (especially the observed range of [Al/Fe], small shifts to the assumed initial abundances of Al, $^{24}$Mg and $^{26}$Mg are probably required),
but also our estimated maximum value of $Y_0\approx 0.287$ in 47\,Tuc HB stars. Indeed, the 47\,Tuc stars with the lowest [O/Fe] abundances
should have the dilution factor (the pristine gas contribution with $Y_0\approx 0.257$) $\sim 20\%$ (the top-left panel in Fig.~\ref{fig:f19}), 
which translates into $Y_0\approx 29$ for them when we use $Y=0.3$ for the supermassive star yileds. 
This is very close to $Y_0\approx 0.287$ for the extreme population of stars in 47\,Tuc.
The positions of the two M\,13 stars relative to the predicted abundances show that approximately the same value, $\sim 10$--$20$\%,
of the dilution factor can be used to explain all of their abundance anomalies.

It is interesting that
the possible presence of an intermediate-mass black hole with ${\cal M}\sim 2300\,{\cal M}_\odot$ in 47\,Tuc has recently
been claimed by \cite{kiziltan:17} based on their analysis of the dynamical state of pulsars in this GC.
This could be a remnant of a supermassive star whose short life left an imprint in the stellar populations that currently
reside in this GC in the form of the $p$-capture abundance anomalies.

\section{Summary and Discussion}
\label{sec:concl}

We have generated a set of new ZAHB and HB evolutionary
models\footnote{The ZAHB and HB models used in this work are available 
at \href\protect{http://apps.canfar.net/storage/list/nugrid/data/projects/HB}} using
revision 7624 of the MESA stellar evolution code supplemented with the new OPAL
and low-temperature opacity tables for the \cite{ags09} mix with
[$\alpha$/Fe]\,$=0.4$ that have specially been prepared for this project. 
As shown in Paper I, stellar models produced by the MESA and Victoria codes
are in excellent agreement when very close to the same physics is assumed.
There are slight differences in the Victoria code that was used for this
test and the one employed by \cite{vbf14} to generate the isochrones
that we have fitted to the cluster CMDs.  However, we examined how well ZAHBs based
on the latter code are able to reproduce MESA computations, finding differences
in $M_\mathrm{bol} \lta 0.01$\ mag at all $\log\teff \lta 4.12$, which shows that
the respective models are highly consistent with one another.
(Note that the best agreement is obtained when MESA employs cubic interpolation
of the opacities with respect to $Z$ rather than the default quadratic option.)
Importantly, we have chosen the same initial mix of elements from \cite{ags09}
for both the MESA and Victoria models. 

In the computations of the HB evolutionary tracks, we have implemented a prescription for the convective He core
mixing similar to that used in the preferred ``maximal overshoot'' scheme by \cite{constantino:15} with which they obtained
the best agreement between the predicted and observed periods of non-radial pulsations of the Kepler field HB stars.
We have also taken care that the convective He core does not experience the so-called ``breathing pulses'' in our HB models,
in the accordance with the results of the AGB to RGB star ratio counts in a large number of GCs recently reported by \cite{constantino:16}.

The new HB evolutionary tracks have been used as input data in our HB population synthesis tool to simulate the distributions of
HB stars in the GCs 47\,Tuc, M\,13, and M\,3. We have chosen these three clusters to test our new HB models and HB population
synthesis tool because they have been extensively studied by other researchers who used similar methods. The results of our
comparison of the observed and simulated distributions of HB stars in these GCs are presented 
in Figs.~\ref{fig:f6}--\ref{fig:f8} and in Table~\ref{tab:fit_par}. Neither the HB morphology in 47\,Tuc nor
the one in M\,13 can be explained without assuming a spread in the initial He mass fraction that is found to be $\Delta Y_0\approx 0.03$
in 47\,Tuc and as large as $\Delta Y_0\approx 0.08$ in M\,13. On the other hand, in spite of being as massive as the first two GCs,
M\,3 has a distribution of HB stars that is reproduced assuming a very small variation of $\Delta Y_0\approx 0.01$.
It is also surprising that, unlike 47\,Tuc and M\,13, which both require a mean RGB mass-loss $\Delta {\cal M}\approx 0.2\,{\cal M}_\odot$,
the red HB stars in M\,3 are distributed as if they had lost only $\Delta {\cal M}\approx 0.13\,{\cal M}_\odot$ on the RGB.
These results are not new. They confirm the previous findings by \cite{salaris:16}, \cite{dicriscienzo:10}, \cite{caloi:05},
\cite{dalessandro:13}, \cite{valcarce:16}, and \cite{mcdonald:15}.

The inclusion of the gravitational settling of He in our MS models results in 
lower luminosity HBs by $\sim 0.2$\ mag (to be more specific, our HB models are fainter because
they have lower envelope He abundances).  Therefore, our distance moduli, which
are based on best fits of the predicted distributions of HB stars to those
observed, are systematically shorter, notably for GCs with substantial populations
of stars along the horizontal part of the HB redward of its ``knee'', than those
estimated using MS models without the diffusion of He. Our distance moduli
determinations seem to be very robust, because even a very small change
(e.g., $\sim 0.02$\ mag) in the best-fit value of $(m-M)_V$ leads to a significant
drop in the K-S probability when the simulated and observed magnitude histograms
are compared. In principle, the same is true for the reddening, except
that the best-fit value of $E(B-V)$ turns out to depend on the selected color.
For example, our best-fit value of $E(B-V)=0.026$ for 47\,Tuc changes to $0.039$
and $0.058$ when we switch from the {\it HST ACS} color $(m_\mathrm{F606W}-m_\mathrm{F814W})_0$
to $(V-I)_0$ and $(B-V)_0$, respectively. Zero-point errors in the
photometry or in the adopted color transformations could well be responsible
for this lack of consistency.

The ages of M\,3 and M\,13 estimated using Victoria-Regina isochrones with the
distance moduli best-fitted to the distributions of HB stars in these GCs are
12.6 Gyr and 12.9 Gyr, respectively (see Figs.~\ref{fig:f11}--\ref{fig:f12}).
A very similar age, 13.0 Gyr, was found for 47\,Tuc; i.e., all three clusters
appear to be nearly coeval.  Note that these age determinations would be reduced
by $\sim 0.6$\ Gyr if the clusters have higher oxygen abundances by $\approx 0.2$
dex than we have assumed.  (Clearly, the relative ages of these GCs depend quite
sensitively on whether or not there are cluster-to-cluster differences in [O/Fe],
or more generally, [CNO/Fe].)  In fact, the recent study by \cite{brogaard:17}
of the detached eclipsing binary in 47\,Tuc, V69, found that stellar models for the same
chemical abundances that we have adopted would provide a better match to the properties
of V69 if they had increased [O/Fe] (or higher [Fe/H], or reduced $Y_0$, or some combination
of these changes).  Unfortunately, because suitable low-temperature opacities for enhanced O
are not yet available, we are unable to produce synthetic HB populations to check into the
implications of this possibility.  (Although low-$T$ opacities for [O/Fe] $= 0.4$ can be
used to generate stellar models for [O/Fe] $> 0.4$ at low metallicities, following the
strategy described in Section~\ref{sec:MESA}, this approach cannot be used to model stars that
are as cool and as metal-rich as those found in the HB of 47\,Tuc.)

Nevertheless, the apparent distance modulus of 47\,Tuc that was derived by
\cite{brogaard:17}, $(m-M)_V = 13.30 \pm 0.06$(random) $\pm 0.03$(systematic), which is
based solely on the binary, is in very good agreement with our determination, $(m-M)_V =
13.27$.  Encouragingly, \cite{zoccali:01} obtained exactly the same modulus
as we did, but with a relatively high uncertainty ($\pm 0.14$\ mag), from a fit of
the cluster white dwarfs (WDs) to a sequence of field WDs with measured trigonometric
parallaxes.  This gives us considerable confidence in our HB models for [Fe/H] $\sim -0.7$,
just as the success of our simulations and the apparent consistency of the inferred
distances with those implied by the RR Lyrae standard candle (see Paper I) supports our
predictions for lower metallicities.  Admittedly, synthetic HBs at relatively high [Fe/H]
values do depend on the assumed mass loss to some extent, because the part of the ZAHB
that is populated has a significant slope.  For instance, if the RGB stars in 47\,Tuc
lost the same amount of mass as M\,3 ($\Delta {\cal M}_1 = 0.13\,\msol$, instead of
$0.20\,\msol$), we would obtain $(m-M)_V \approx 13.33$.  However, HB tracks for
higher masses have shorter blue loops, making it difficult to reproduce the locations
of the stars along the blue edge of the observed HB.  The assumption of higher $Y_0$
would not help to alleviate this problem because the models for higher helium abundances
are more luminous, so the blueward extent of the tracks would shifted to brighter absolute
magnitudes, thereby missing the faintest stars with the bluest colors.  Based on these
considerations, we consider it unlikely that our determination of the distance modulus
of 47 Tuc is in error by more than $\pm 0.05$\ mag.

In view of the small difference in the ages of M\,3
and M\,13, and the results of our analysis presented in Section~\ref{sec:evol_HB},
we conclude that, at least for this pair
of GCs, the striking difference in the extension of their HBs to blue colors is
caused by the larger {\it spread} of $Y_0$ in M\,13 than in M\,3. According
to our simulations, there are significant populations of stars ($\gta 40$\%) in
both clusters that have $Y_0 \approx 0.250$, but the distribution of He abundances is
much narrower in M\,3 ($\Delta Y_0 \sim 0.01$) than in M\,13 ($\Delta Y_0 \sim 0.08$).
However, given that M\,3 has an unusually long extension of its HB to red colors,
a difference in the RGB mass-loss efficiency must play a role here as well.
Therefore, we conclude that the main second parameters relevant to these GCs are
$\Delta Y_0$ and mass loss along the giant branch.

In fact, these are among the few second-parameter candidates that are viable
possibilities for M\,3 and M\,13 since, as noted by many others \citep[e.g.,][]{ferraro:97},
both clusters have similar masses and structural parameters (such
as the concentration and central stellar density), as well as similar C$+$N$+$O
abundances \citep{smith:96,cohen:05} and ages (this study and VBLC13).  As pointed
out by \cite{johnson:98}, the different morphologies of the M\,3 and M\,13
CMDs between the turnoff and the lower RGB are much easier to explain in terms
of a difference in helium abundance (they suggested $\Delta Y \sim 0.05$) instead
of age.   On the other hand, it has been known for a long time that these clusters contain
stars with normal oxygen abundances, both on their respective giant branches
\citep{kraft:92} and their HBs \citep{peterson:95}; so, as argued by Ferraro
et al., it cannot be the case that {\it all} of the stars in M\,13 are helium
enhanced.  Thus, $\Delta Y_0$ rather than $Y_0$ seems to be the relevant parameter,
though a cluster with a larger spread in the initial helium content will generally
also have a higher value of $\langle Y_0\rangle$.  As \cite{dantona:02}
proposed more than 15 years ago, a variation in $Y_0$ caused by some mechanism of
self-enrichment could explain the extremely blue HBs in such GCs as M\,13 (they
suggested pollution by the ejecta of massive AGB stars, but recent studies of
Mg isotopic abundance ratios are problematic for this scenario; see Section~\ref{sec:47TucSMS}
and Denissenkov et al.~2015).  This idea is now generally accepted, but precisely how 
this enrichment, and the accompanying light element correlations and anti-correlations,
occurred is not yet understood.

\begin{figure}[t]
\plotone{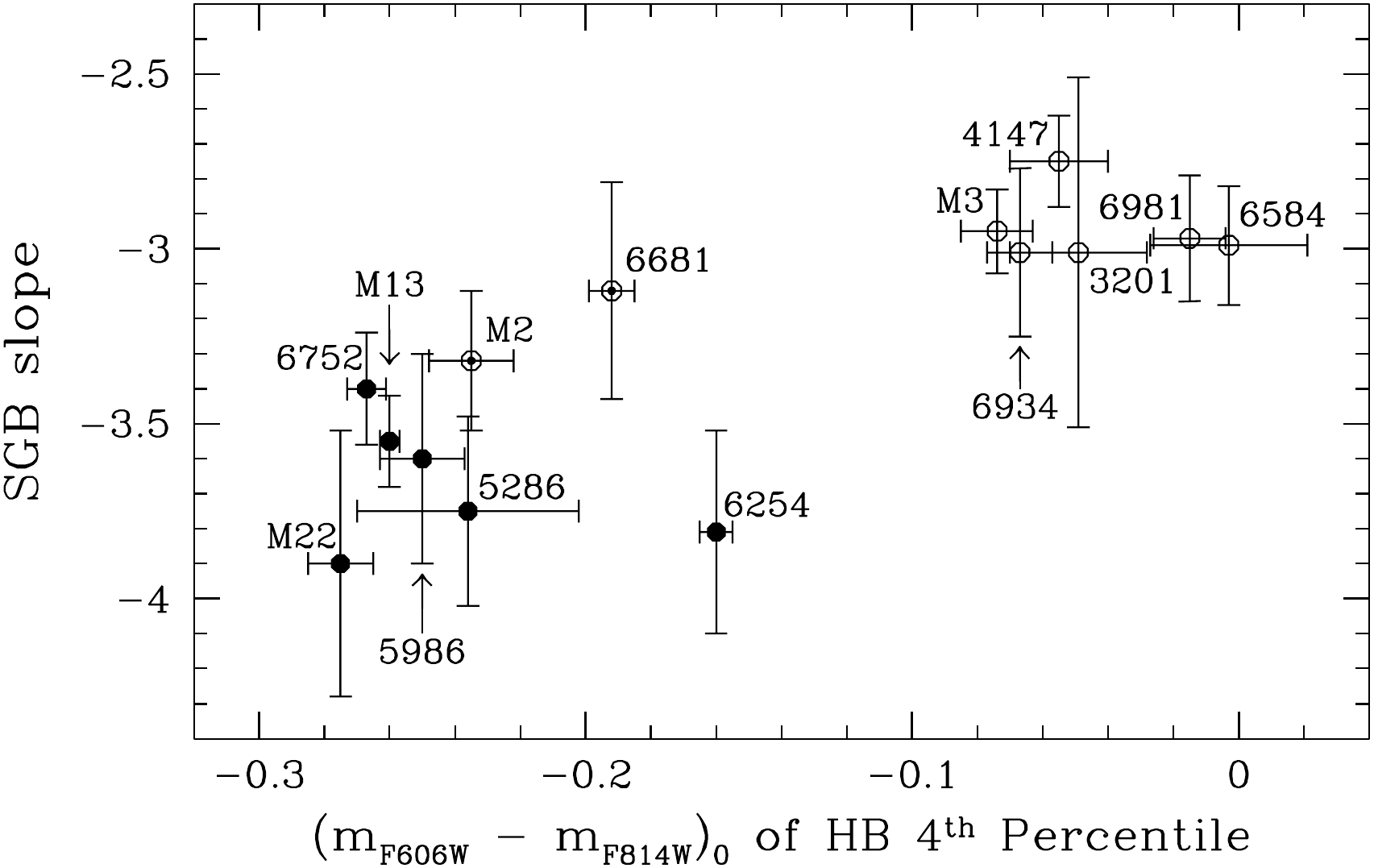}
\caption{Plot of the SGB slopes of a sample of GCs with $-1.78 \le$ [Fe/H] $\le -1.48$
as a function of the intrinsic color corresponding to the fourth percentile of the
color distribution of their HB stars (see the text for the sources of the data).
The clusters are identified by their Messier or NGC numbers.}
\label{fig:f20}
\end{figure} 

In their analysis of the {\it HST} photometric data for 55 of the GCs that were
observed by \cite{sarajedini:07}, VBLC13 noticed that a large fraction of the 
most metal-deficient clusters could be divided into two groups based on differences
in their SGB morphologies.  They found that clusters with somewhat shallower SGB
slopes, like that of M\,3, tended to have fainter absolute magnitudes (and hence
lower masses) and to have larger Galactocentric distances than those with steeper,
M\,13-like SGB slopes.  Although a possible connection with HB type was not considered,
it turns out that, as shown in Figure~\ref{fig:f20}, the SGB slope is correlated
with the blueward extent of an observed HB.  

In this plot, the SGB slopes derived by VBLC13 (from their Table 1\footnote{We took
this opportunity to check the published results and found that, in the case of NGC\,6681,
the objectively derived slope (i.e., from a linear least-squares fit to the
subgiant observations) was not as steep as an eye-estimated fit to the data would suggest.
By varying the color range over which the data were fitted, a visually much more
agreeable slope was obtained; specifically, $-3.10 \pm 0.31$, which has been
adopted for NGC\,6681 instead of the published value. (The small number of subgiants
in the sample together with fluctuations in their distribution with color are
likely responsible for this difficulty.)  The revised slope and its large uncertainty
have the consequence that it is not possible to assign NGC\,6681 to either the
M\,3-like or the M\,13-like group of clusters.  Our examination also suggested that the
SGB slopes of M\,22 and NGC\,6254 are probably closer to $-3.5$ than to the published
values, but as this would not alter their identification as M\,13-like GCs, we opted to
plot the published slopes for both clusters.}) for GCs with $-1.78 \le$ [Fe/H] $\le -1.48$ 
(according to \citealt{cbg09b}) have been plotted as a function of the intrinsic
color corresponding to the fourth percentile of the color distribution of the
cluster HB stars.  To obtain the numerical values along the abscissa, we have simply 
subtracted the reddenings adopted by VBLC13 from the observed colors of the
fourth percentile given by \citet[][i.e., the $P_A$ values listed in their Table 1]{milone:14}.  
For most of the clusters, these reddenings are equivalent to the
dust-map values of $E(B-V)$ from \cite{sfd98}, though some minor adjustments
were needed in the case of highly reddened clusters in order to obtain satisfactory
fits of ZAHB models to the bluest HB stars.  Note that, as shown by VBLC13 (see their Fig.~29),
there is some dependence of the SGB slope on metallicity, and it is for this reason that
we have considered only those GCs with similar metallicities (specifically, the ones with
$-1.78 \le$ [Fe/H] $\le -1.48$).

\begin{figure}[t]
\plotone{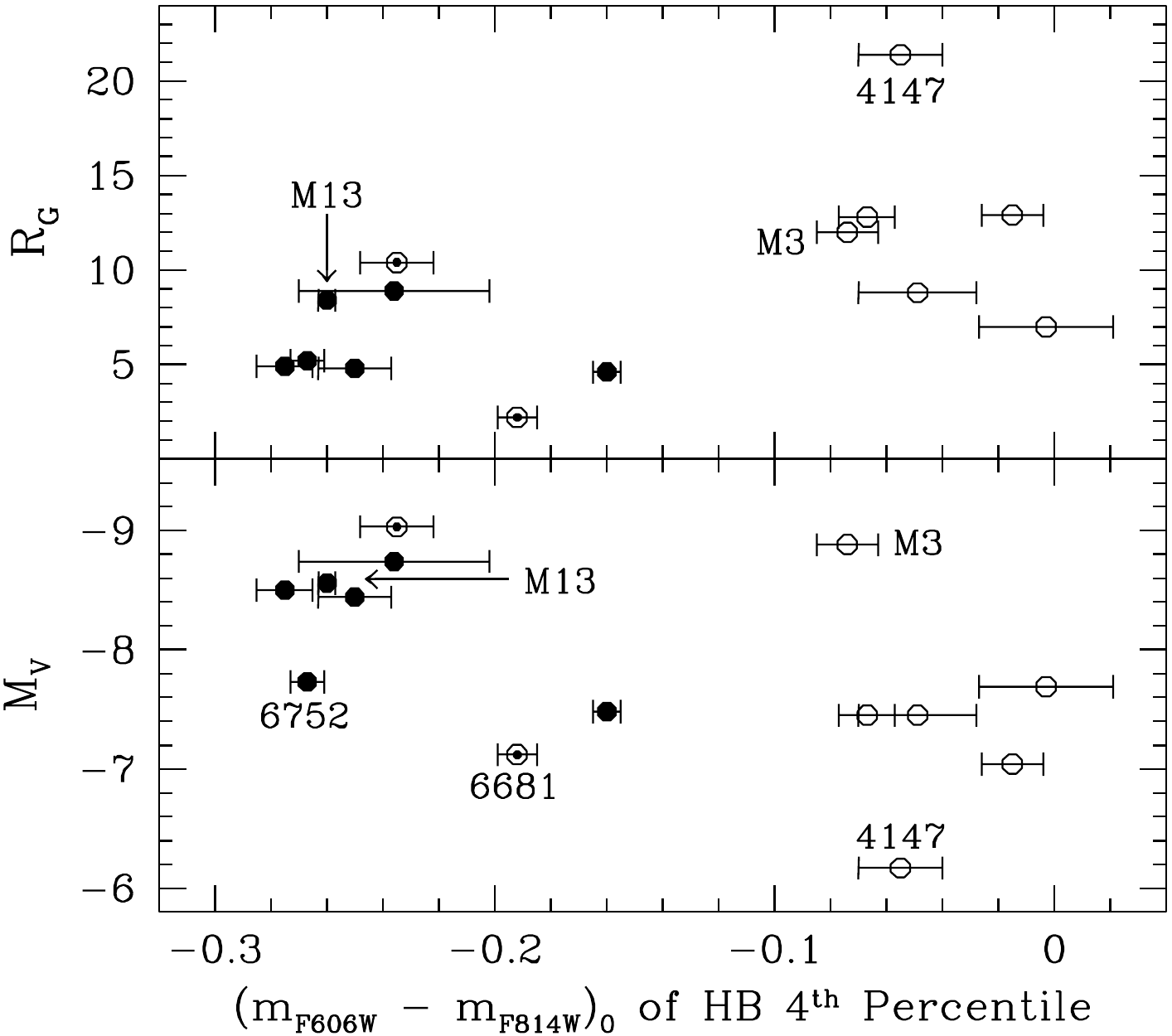}
\caption{As in the previous figure, except that the Galactocentric distances
and their absolute visual magnitudes are plotted as a function of the selected
color on the cluster HBs.  Only a few cluster identifications are provided, but
the others may be deduced from their respective locations in the previous figure.}
\label{fig:f21}
\end{figure}

Although the differences in the cluster SGB slopes are quite subtle, they are
sufficiently different (except in the case of M\,2 and NGC\,6681, which have
intermediate slopes) that it is possible to separate them into two groups that,
as it turns out, have very different HB morphologies.  Figure~\ref{fig:f21}
shows that the M\,13-like GCs tend to be inner-halo, intrinsically luminous (and
hence massive) systems whereas the M\,3-like clusters are typically lower mass
objects located at larger Galactocentric distances, though with some overlap.
Recent work has established that M\,2 \citep{yong:14,milone:15b},
M\,13 \citep{johnson:12}, M\,22 \citep{marino:09}, NGC\,6752 \citep{milone:13},
and NGC\,5286 \citep{marino:15b} are among the relatively few GCs
that show wide variations in the abundances of the light elements (O, Na, etc.).
Since these are the same clusters that have extremely blue HBs, it is a safe bet
that their bluest HB stars have high He abundances and a substantial range in
$Y_0$, in accordance with our simulations for M\,13.  Similarly, we suspect
that all of the clusters that are located close to M\,3 in Figs.~\ref{fig:f20}
and~\ref{fig:f21} will be found to have $Y_0 \sim 0.25$ and a small range in
$\Delta Y_0$, just as we have determined for M\,3. Further work to test this
prediction is strongly encouraged.

Many others have deduced that mass loss differences are partly responsible for
the dissimilarity of the M\,3 and M\,13 HBs (e.g., see \citealt{fusipecci:93,catelan:95},
and the review by \citealt{catelan:09}), and it seems
likely that this is a distinguishing characteristic of the two groups of clusters
that have been plotted as filled and open circles in Figs~\ref{fig:f20} and~\ref{fig:f21}.
Why this should be the case is far from clear.  Perhaps, as noted by \cite{ferraro:97}
(also see \citealt{fusipecci:75}), differences in rotational
velocities affect the morphology of the HB since \cite{peterson:95} have shown
that a substantial fraction of the HB stars in M\,13 rotate rapidly, whereas no rapid
rotators have been found in M\,3 (one of the few distinguishing characteristics
of these two GCs).  In fact, the data tabulated by \cite{vdbergh:11} indicate
that the M\,13-like group of clusters tend to have higher ellipticities than those
that have M\,3-like HBs, but with some exceptions.

We cannot help but wonder whether stellar rotation may help to explain why
the observed periods of some of the $c$-type RR Lyrae in M\,13 are shorter (by as
much as $\sim 0.1$~d) than our models predict.  In this investigation, we have
presented new determinations of intensity-weighted mean $B$, $V$, and $I_C$
magnitudes for the only known RRab star, and 5 of the 7 known RRc variables, in M\,13.
Our models for $Y \sim 0.25$ predict very close to the observed period of the
fundamental-mode pulsator, and models for higher $Y$ can potentially explain the
periods of one (perhaps two) of the first-overtone pulsators.  One of the remaining
RRc stars seems to be multi-periodic, but we are unable to explain why the observed
periods of the rest are in such poor agreement with the model predictions.  What is
particularly disconcerting is that the observed periods do not agree at all well with
those of variables in M\,3 that seem to have nearly the same intrinsic colors and
absolute magnitudes.  Presumably, this discrepancy is a reflection of some fundamental
difference in the properties of these particular variables.  Unfortunately, further
studies of the M\,13 RR Lyrae will be needed to obtain a resolution of this issue.

Returning to the second-parameter problem: age differences appear to be much too small
to be the primary explanation of the diversity in HB types, though they may well be a
contributing factor.  Over the years, members of the astronomical community have
either reached a similar conclusion \citep[see, e.g.][]{fusipecci:93,stetson:96,
buonanno:97,vandenberg:00,vbl13}, or else they have argued in
support of age being the dominant second parameter \citep[e.g.,][]{lee:94,
gratton:10,dotter:10,milone:14}. Insofar as the clusters
that are identified in Fig.~\ref{fig:f20} are concerned, the mean age of the M\,13-like
GCs, including M\,2 and NGC\,6681 as members of this group given that their properties
seem to resemble those of M\,13 more so than M\,3, is approximately 0.5 Gyr greater than
the mean age of the M\,3-like systems --- if the absolute ages derived by VBLC13 are adopted.
This is much less than the 2.6 Gyr age difference that, according to our simulations (see
\S~4.2.3), would be needed to explain the different HB morphologies of M\,3 and M\,13
on the assumption that age is the only distinguishing property of these two GCs.

Similarly, age seems to play no more than a minor role in explaining the wide variations
in the HBs of M\,5, M\,12, NGC\,288, and NGC\,362, for which the metallicities tabulated
by \cite{cbg09a} lie within the range $-1.33 \le$ [Fe/H] $\le -1.30$.  VBLC13
found that NGC\,362 is $\approx 0.75$\ Gyr younger than M\,5, which is in the right sense
to explain why the former has a significantly redder HB than the latter.  However,
using the horizontal method of determining relative GC ages (VandenBerg et al.~1990),
VBLC13 (see their Fig.~24) concluded that M\,5, which has an M\,3-like HB, is coeval
with NGC\,288, which has an M\,13-like HB.  Interestingly, the same figure indicates
that M\,12 is much older than M\,5, and therefore NGC\,288, even though M\,12 and
NGC\,288 both have extremely blue HBs.  Other factors besides age need to be
considered for these, and quite possibly most, GCs.  
(One would reach the same conclusion if, for the clusters highlighted
in this discussion, we had adopted the relative GC ages derived by
\citealt{marinfranch:09}.) In addition to the other
second-parameter candidates mentioned above, the apparent correlation of chemical
properties with mass \citep{recioblanco:06} and with $\sigma_0\,v_{e,0}$, where
$\sigma_0$\ is the surface density of stars at the cluster center and $v_{e,0}$
is the central escape velocity, which are related to the capability of clusters to
retain gas and to resist ram-pressure stripping (see VBLC13) should be kept in mind.

In closing this investigation, it is worth mentioning that, so far, we
have not considered more than three stellar populations with different values of
$Y_{0,i}$ in our HB population synthesis models.  However, this number is just
a free parameter that can be increased to a larger value if
required \cite[e.g., see][who used seven populations
in their model of $\omega$\,Cen]{tailo:16}. In fact, all populations with $Y_{0,i} > Y_{0,1}$ can be considered as
the second of just two generations of stars in GCs. This interpretation
depends on the criterion that is used to distinguish multiple populations. For example,
the SUMO project has revealed three populations in 47\,Tuc and M\,13 with the help of the photometric index 
$C_{U,B,I} = (U-B)-(B-I)$ \citep{monelli:13}, while the ``chromosome maps'' of GCs, that use a more complex search criterion
based on combinations of {\it HST ACS} colors, only distingush the fractions of the first generation stars \citep{milone:17}.
Therefore, it is not yet very meaningful
to compare the fractions of stars belonging to different stellar populations and the number of the populations themselves
in individual GCs that are obtained using different observational and theoretical methods.

The next paper in this series will focus on M\,55 and NGC\,6362, making use
of both their RR Lyrae populations and member eclipsing binary stars to
provide especially tight constraints on their absolute ages.

\acknowledgements

We are very grateful to Santi Cassisi for providing the color transformations that
we have used for hot, horizontal-branch models.
Financial support from the Natural Sciences and Engineering
Research Council via a Discovery Grant to D.A.V.~is acknowledged with
gratitude.  
The work of P.A.D.~was also supported in part by the National Science Foundation under Grant No. PHY-1430152 
(JINA Center for the Evolution of the Elements).

\bibliography{ms}

\end{document}